\documentclass[twocolumn, twocolappendix]{aastex631}

\usepackage{CJKutf8}
\usepackage{graphicx}	
\usepackage{amssymb}	
\usepackage{amsmath}	
\usepackage{url}
\usepackage{makecell}
\usepackage{multirow}
\usepackage[version=4]{mhchem} 
\usepackage{booktabs}
\usepackage{arydshln}
\usepackage{tablefootnote}

\newcommand{\htworegion}{H{\sc ii}}
\newcommand{\velouni}{$\rm \, km\ s^{-1}$}
\newcommand{\arcsecuni}{$\rm ^{\prime\prime}$}
\newcommand{\arcminuni}{$\rm ^{\prime}$}

\begin{document}

\title{The Structure of an 80\,pc Long Massive Filament}

\begin{CJK}{UTF8}{gbsn} 
\author[0000-0002-5077-9599]{Qian-Ru He (何茜茹)}
\affiliation{Purple Mountain Observatory, Chinese Academy of Sciences, No. 10 Yuanhua Road, Nanjing 210023, PR China}
\affiliation{School of Astronomy and Space Sciences, University of Science and Technology of China, Hefei 230026, PR China}
\affiliation{Jodrell Bank Centre for Astrophysics, School of Physics and Astronomy, University of Manchester, \\Oxford Road, Manchester M13 9PL, UK}

\author[0000-0003-0364-6715]{Won-Ju Kim}
\affiliation{I. Physikalisches Institut, University of Cologne, Z\"ulpicher Str. 77, 50937 K\"oln, Germany}
\affiliation{Max-Planck-Institut f\"{u}r Radioastronomie, Auf dem H\"{u}gel 69, 53121 Bonn, Germany}

\author[0000-0001-8509-1818]{Gary A. Fuller}
\affiliation{Jodrell Bank Centre for Astrophysics, School of Physics and Astronomy, University of Manchester, \\Oxford Road, Manchester M13 9PL, UK}
\affiliation{Intituto de Astrof\'isica de Andalucia (CSIC), Glorieta de al Astronomia s/n E-18008, Granada, Spain}
\affiliation{I. Physikalisches Institut, University of Cologne, Z\"ulpicher Str. 77, 50937 K\"oln, Germany}

\author[0000-0003-1665-6402]{Alessio Traficante}
\affiliation{IAPS-INAF, Via Fosso del Cavaliere, 100, I-00133 Rome, Italy}

\author[0000-0001-9751-4603]{Seamus D. Clarke}
\altaffiliation{Institute of Astronomy and Astrophysics, Academia Sinica, No.1, Section 4, Roosevelt Road, Taipei 10617, Taiwan}
\affiliation{Department of Physics, National Cheng Kung University, Tainan, Taiwan}

\author[0000-0003-0007-2197]{Yu Gao (高煜)}
\affiliation{Purple Mountain Observatory, Chinese Academy of Sciences, No. 10 Yuanhua Road, Nanjing 210023, PR China}
\affiliation{Department of Astronomy, Xiamen University, No. 422 Siming South Road, Xiamen, Fujian 361005, PR China}

\author[0000-0003-3151-8964]{Xue-Peng Chen (陈学鹏)}
\affiliation{Purple Mountain Observatory, Chinese Academy of Sciences, No. 10 Yuanhua Road, Nanjing 210023, PR China}

\author[0000-0001-8060-1321]{Min Fang (房敏)}
\affiliation{Purple Mountain Observatory, Chinese Academy of Sciences, No. 10 Yuanhua Road, Nanjing 210023, PR China}

\author[0000-0002-7237-3856]{Ke Wang (王科)}
\affiliation{Kavli Institute for Astronomy and Astrophysics, Peking University, 5 Yiheyuan Road, Haidian District, Beijing 100871, PR China}

\author[0009-0002-8449-1734]{En Chen (陈恩)}
\affiliation{Center for Astrophysics, Guangzhou University, Guangzhou, 510006, PR China}
\affiliation{Purple Mountain Observatory, Chinese Academy of Sciences, No. 10 Yuanhua Road, Nanjing 210023, PR China}

\author[0000-0003-0295-6586]{Tapas Baug}
\affiliation{S. N. Bose National Centre for Basic Sciences, Sector-III, Salt Lake, Kolkata 700106, India}

\author[0000-0003-2536-3142]{Xiao-Long Wang (王小龙)}
\altaffiliation{Physics Postdoctoral Research Station at Hebei Normal University}
\affiliation{Department of Physics, Hebei Normal University, Shijiazhuang 050024, PR China}
\affiliation{Guo Shoujing Institute for Astronomy, Hebei Normal University, Shijiazhuang 050024, PR China}
\affiliation{Purple Mountain Observatory, Chinese Academy of Sciences, No. 10 Yuanhua Road, Nanjing 210023, PR China}

\author[0000-0001-8923-7757]{Chen Wang (王晨)}
\affiliation{National Astronomical Observatories, Chinese Academy of Sciences, 20A Datun Road, Chaoyang District, Beijing 100101, PR China}

\author[0000-0002-8809-2725]{Yong-Xiong Wang (王永雄)}
\affiliation{Jodrell Bank Centre for Astrophysics, School of Physics and Astronomy, University of Manchester, \\Oxford Road, Manchester M13 9PL, UK}

\correspondingauthor{Qian-Ru He}
\email{heqr@pmo.ac.cn}
\correspondingauthor{Won-Ju Kim}
\email{wjkim@mpifr-bonn.mpg.de}
\correspondingauthor{Gary A. Fuller}
\email{gary.a.fuller@manchester.ac.uk}
\correspondingauthor{Xue-Peng Chen}
\email{xpchen@pmo.ac.cn}



\begin{abstract}
Using new Institut de Radioastronomie Millimétrique (IRAM) 30~m telescope \ce{N2H+}, \ce{C^18O} $J=1-0$ and Atacama Pathfinder Experiment (APEX) telescope  \ce{^13CO} and  \ce{C^18O} $J=2-1$ maps
together with archival far-infrared continuum data, and \ce{^12CO}, and \ce{^13CO} $J=1-0$ data, 
we present a comprehensive analysis of the massive filament CFG024.00$+$0.48 (G24) across clump-to-cloud scales. Our results show that G24 is an $\sim$80\,pc giant filament with a total mass of $\sim$$10^5$\,M$_{\odot}$.
In the different tracers the filament width is measured to be about $\sim$2 times the beam size of the observations, as expected for power-law density distributions, giving beam-deconvolved widths in the range from 0.8 to 2.8\,pc. We determine a line-of-sight thickness of $\sim$2.2\,pc demonstrating that G24 is not an edge-on, flatten structure. 
The virial parameter obtained from line mass ($\alpha_{\rm line,vir}=M_{\rm line,vir}/M_{\rm line}$) from the \ce{C^18O}\,(1--0) data is 0.85, and that obtained from $Herschel$-based \ce{H2} column density is 0.52, suggesting G24 is globally close to virial equilibrium. 
The distribution of the 40 dust clumps appears to have a ``two-tier'' fragmentation pattern. For the clump groups, the separation, with a mean/median of 3.68/3.46\,pc, is very close to expected length associated with the maximum fragmentation growth rate of $\lambda_{\rm max}=3.55 \pm0.32$\,pc estimated for the dust.  
However, the longitudinal  centroid velocity profiles of \ce{C^18O} and \ce{N2H+}  show oscillation patterns with wavelengths of 9.8$\pm$0.1\,pc and 9.9$\pm$0.1\,pc, respectively. 
This is $\sim2$ times larger than the  corresponding values of $\lambda_{\rm max}$ of 4.96$\pm$0.63\,pc and 4.65$\pm$1.34\,pc, respectively. 
This suggests that the velocity structure is not dominated by flows directly associated with the fragmentation seen in the dust emission.

\end{abstract}



\section{Introduction} 
\label{sec: Introduction}
\end{CJK}
Several Galactic observational surveys at various wavelengths have shown that Galactic molecular clouds frequently appear as elongated filamentary structures throughout the Milky Way (MW) \citep[e.g.,][]{Schneider_1979ApJS...41...87S,Ungerechts_1987ApJS...63..645U,Molinari_2010PASP..122..314M,André_2010AnA...518L.102A,Wang_2015MNRAS.450.4043W,Wang_2016ApJS..226....9W,Wang_2024AnA...686L..11W,Li_2016AnA...591A...5L,Mattern_2018AnA...619A.166M, Jackson_2006ApJS..163..145J, Su_2019ApJS..240....9S}.
In addition, a significant fraction of filaments (typical lengths of $\sim$1--10\,pc) enclose dense, compact dust/molecular clumps that are often directly associated with ongoing star-forming activities \citep[e.g.,][]{Molinari_2010AnA...518L.100M, André_2010AnA...518L.102A, André_2014prpl.conf...27A, Polychroni_2013ApJ...777L..33P, Könyves_2015AnA...584A..91K, Schisano_2020MNRAS.492.5420S, Zhou_2022MNRAS.514.6038Z}. Massive star clusters seem to be particularly formed at high-density filaments ($\rm > 10^5\ cm^{-3}\ over\ \sim 5\ pc^3$) \citep[e.g.,][]{Wang_2011ApJ...735...64W, Wang_2012ApJ...745L..30W, Wang_2014MNRAS.439.3275W, Motte_2018ARAnA..56...41M, Schneider_2010AnA...520A..49S, Nguyen_Luong_2011AnA...535A..76N, Hennemann_2012AnA...543L...3H, Traficante_2020MNRAS.491.4310T}. For example, junctions of multiple filaments, known as a hub-filament system (HFS; \citealt{Myers_2009ApJ...700.1609M}), are considered to be a potential birthplace of massive stars ($> 8\,\rm M_{\odot}$) and stellar clusters (e.g., \citealt{Liu_2012ApJ...745...61L,Schneider_2012AnA...540L..11S, Peretto_2013AnA...555A.112P, Peretto_2014AnA...561A..83P, Motte_2018ARAnA..56...41M, Yuan_2018ApJ...852...12Y, Kumar_2020AnA...642A..87K, Anderson_2021MNRAS.508.2964A,Zhou_2022MNRAS.514.6038Z}). 
A significant proportion of the filamentary structures are also identified as infrared dark clouds (IRDCs), which are dense molecular clouds seen in silhouette against the bright infrared emission of the Galactic plane. Their cold ($T\,<$\,20\,K), high column density ($N\rm(H_2) > 10^{22}\,cm^{-2}$ ), and high-mass ($10^2$--$10^4\,\rm M_{\odot}$) nature demonstrates that IRDCs are an ideal laboratory for understanding the formation of massive stars and clusters  (e.g., \citealt{Peretto_2009AnA...505..405P, Battersby_2014ApJ...787..113B, Xie_2021SCPMA..6479511X}). 

Notably, some filaments are significantly longer (typical lengths of 11--269\,pc; e.g., \citealt{Wang_2015MNRAS.450.4043W,Zucker_2018ApJ...864..152Z,Wang_2024AnA...686L..11W}) and more massive (up to $10^6 \rm M_{\odot}$) than typical filaments found towards nearby star-forming regions, and are referred to as ``giant filaments'' \citep[GFs;][]{Hacar_2023ASPC..534..153H}. They are considered to be the bones of the MW. 
The characteristics of GFs show a wide range of physical properties such as masses ($\rm \sim 10-10^6\,M_{\odot}$), aspect ratios ($\rm \sim 4-100$), and line masses ($\rm \sim 1000\,M_{\odot\,pc^{-1}}$) \citep{Hacar_2023ASPC..534..153H}. 
However, the formation mechanism and evolution of these giant structures, and their fragmentation and star formation process, remain unclear and require further investigation. 
Since the discovery of the first ``bone'' of the MW, the $\sim$80\,pc long IRDC ``Nessie'' \citep{Jackson_2010ApJ...719L.185J}, the scale of GFs has extended from tens to hundreds of parsecs. For example, 
subsequent work proposed that Nessie is part of a much longer bone-like structure that traces the Scutum-Centaurus spiral arm and could extend up to five times longer \citep{Goodman_2014ApJ...797...53G}. Several studies have also been done towards filaments with lengths comparable to Nessie, for example, G32.02$+$0.06 \citep{Battersby_2014ASSP...36..417B, Battersby_2014ApJ...787..113B},  
IRDC 182223 \citep{Beuther_2010AnA...518L..78B, Tackenberg_2013AnA...550A.116T, Tackenberg_2014AnA...565A.101T}, a $\geqslant$500\,pc long filamentary gas wisp \citep{Li_2013AnA...559A..34L}, and a kpc-scale molecular wave \citep{Veena_2021ApJ...921L..42V}. 
In addition, a series of systematic Galactic censuses of GFs have been carried out in several studies (e.g., \citealt{Ragan_2014AnA...568A..73R, Zucker_2015ApJ...815...23Z, Wang_2015MNRAS.450.4043W, Wang_2016ApJS..226....9W, Abreu-Vicente_2016AnA...590A.131A, Li_2016AnA...591A...5L, Zucker_2018ApJ...864..153Z, Colombo_2021AnA...655L...2C, Ge_2022ApJS..259...36G,Ge_2023AnA...675A.119G, Wang_2024AnA...686L..11W}). 

CFG024.00$+$0.48 (hereafter, G24) is identified as a large-scale filament by \cite{Wang_2015MNRAS.450.4043W}, in which they reported nine longest, coldest (17–21 K), and densest (1.7--9.3\,$\times\,10^4\ \rm M_{\odot}$) filaments with lengths spanning from 37 to 99\,pc and widths of 0.6--3.0\,pc in the MW within $15^{\circ} < l < 56^{\circ}$, based on the dust continuum data of the Herschel infrared Galactic Plane (Hi-GAL) survey and the \ce{^13CO}\,(1--0) data from the Boston University FCRAO Galactic Ring Survey (GRS). 
G24 encloses a 70-\micron\ dark massive clump SDC 24.013+0.488 (hereafter, SDC24), which has a high mass ($1957\pm559\,\rm M_{\odot}$) within a radius of $0.81\pm0.08$\,pc \citep{Traficante_2017MNRAS.470.3882T} and contains two star clusters \citep{Traficante_2023MNRAS.520.2306T}. The adopted kinematic distance of G24 is  $5.2^{+0.18}_{-0.19}$\,kpc \citep{Wang_2015MNRAS.450.4043W}, and thus its physical size is 82\,pc\,$\times$\,1.9\,pc. 
Interestingly, G24 is located slightly off by $\rm \sim 44\,pc$ above the Galactic plane, and thus, it is probably not associated with any particular spiral arms. 
Therefore, G24 is a perfect candidate to investigate how such a GF is formed in the isolated environment, and, furthermore, the general star formation process in such circumstances. 
 \begin{table*}[ht!]
	\centering
	\caption{\label{tab: mol_info}Molecular transitions used in this work.}
	\begin{tabular}{l c c c c c c c c c}
	\toprule[1 pt]
    Molecule & Transition & \multicolumn{1}{c}{Rest frequency} & Telescope & Ref. & \multicolumn{1}{c}{$\theta_{\rm HPBW}$} & \multicolumn{1}{c}{$\Delta\upsilon$} & \multicolumn{1}{c}{$E_{\rm u}/k$} & \multicolumn{1}{c}{$E_{\rm l}/k$} & \multicolumn{1}{c}{$A_{\rm ul}$} \\
        &   & \multicolumn{1}{c}{(MHz)} & & & \multicolumn{1}{c}{($''$)} & \velouni & \multicolumn{1}{c}{(K)} & \multicolumn{1}{c}{(K)} & \multicolumn{1}{c}{(s$^{-1}$)} \\
    \midrule[0.5 pt]
    \ce{^13CO} & 2--1 & 220398.684 & APEX 12~m & This work &30 & 0.10 & 15.9 & 5.3 & $6.08\times10^{-7}$ \\ 
    \ce{C^18O} & 1--0 & 109782.173 & IRAM 30~m & This work &24 & 0.13 & 5.3 & 0.0 & $6.27\times10^{-8}$ \\ 
    \ce{C^18O} & 2--1 & 219560.354 & APEX 12~m & This work &30 & 0.10 & 15.8 & 3.7 & $6.01\times10^{-7}$ \\ 
    \ce{N2H+} & 1--0 & 93173.398 & IRAM 30~m & This work &28 & 0.16 & 4.5 & 0.0 &  $3.63\times10^{-5}$ \\ 
    \ce{^12CO} & 1--0 & 115271.202 & PMO 13.7\,m & MWISP & 55 & 0.16 & 5.5 & 0.0 & $7.20\times10^{-8}$ \\ 
    \ce{^13CO} & 1--0 & 110201.354 & PMO 13.7\,m & MWISP & 55 & 0.17 & 5.3 & 0.0 & $6.33\times10^{-8}$ \\ 
	\bottomrule[1 pt]
    \end{tabular}
 \end{table*}

This paper presents results from the first mapping observations of molecular line emissions towards G24 with the Institut de Radioastronomie Millim\'etrique (IRAM) 30~m telescope and the Atacama Pathfinder EXperiment (APEX) 12~m telescope.We characterized the properties of G24 and its role in star formation in an environment isolated from the galactic spiral arms. The paper is arranged as follows. The observations and data reduction are presented in Section~\ref{sec: IRAM 30m Observation and Data Reduction}. The archival data are enumerated in Section~\ref{sec: Archival Data}. Section~\ref{sec: Analysis} shows the results of the data analysis, determining column density and clump identification. Section~\ref{sec: General Properties of G24} presents the general physical properties of filament G24 determined in this study, including the width, the size along the line of sight, the line mass, and longitudinal profiles. In this section, we also discuss the gravitational stability and possible fragmentation within G24. Lastly, we summarise the findings of this study and the perspective for further studies in Section~\ref{sec: Summary and Conclusion}.

\section{Observation and Data Reduction} 
\label{sec: IRAM 30m Observation and Data Reduction}
\subsection{\texorpdfstring{\ce{N2H+}}{N2H+} and \texorpdfstring{\ce{C^18O} $J=1-0$}{C18O} Data from IRAM~30~m}
\begin{figure*}[ht!]
  \centering 
  \includegraphics[width=0.90\textwidth]{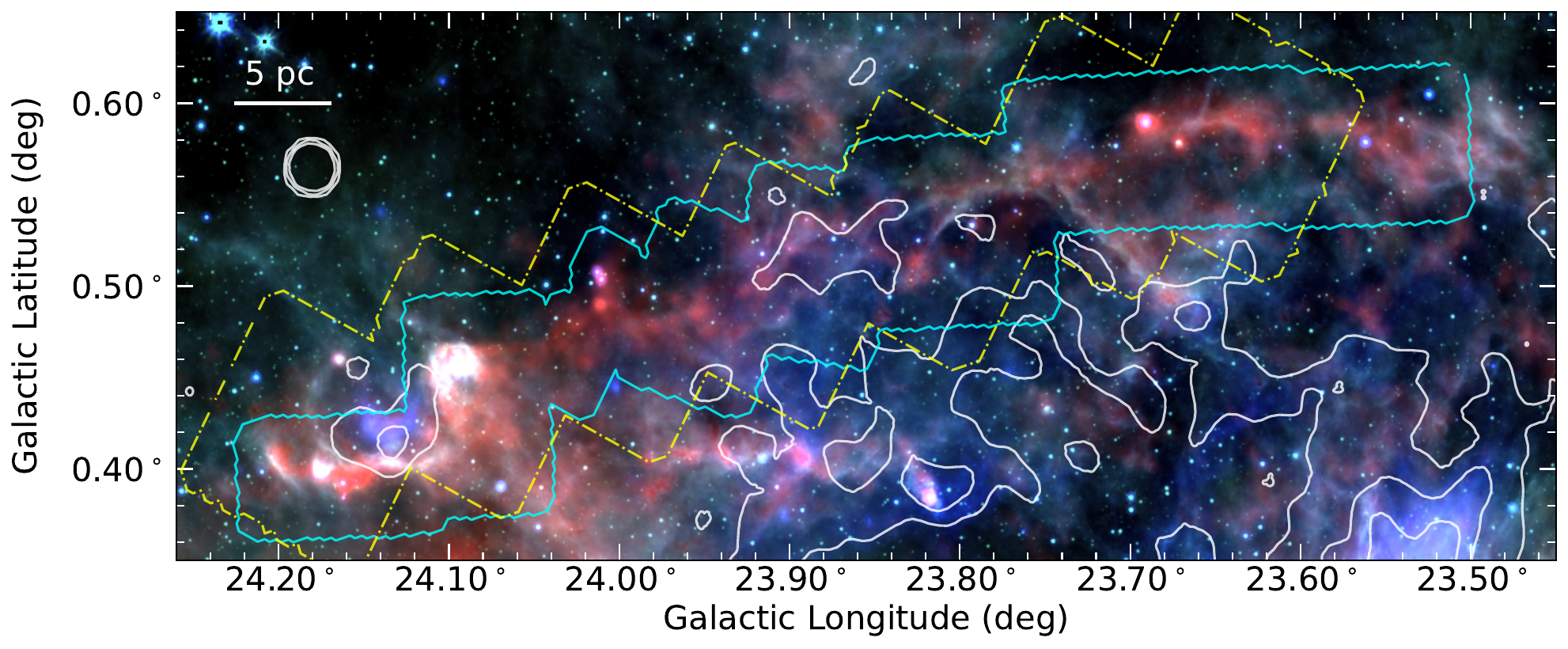}
  \caption{Three colour-composite map of the G24 filament. Red represents 250\,\micron\ continuum emission obtained from the \textit{Herschel} Hi-GAL survey, green indicates the $Spitzer$-MIPS 24\,\micron\ continuum emission, and blue is the $Spitzer$-IRAC 8\,\micron\ emission. The white contours superimposed on the map are the 21\,cm radio continuum emission from VGPS, with levels of [20.5, 25, 30]\,K. The cyan contour is the region observed by the IRAM 30~m telescope, and the yellow dotted-dashed contour is the region observed by the APEX 12~m telescope. The region observed by the PMO 13.7\,m millimetre telescope is larger than the FOV of the presented map and is not shown. The 5\,pc scale indicator on the upper left corner is estimated by adopting a distance of 5.2\,kpc.}
  \label{img: RGB}
 \end{figure*}

The observations of the entire filament with the IRAM~30~m telescope were performed from January to May 2021 and June 2021 (Project IDs: 093-20 and 003-21 and PI: Qian-Ru He). The target region was mapped by the On-The-Fly (OTF) method with the position switching (PSW) mode, using the eight mixer receiver (EMIR, \citealt{Carter_2012AnA...538A..89C}) on the 3~mm wavelength band (E090) tuned at the \ce{N2H+}\,(1--0) rest frequency ($\nu =$ 93173.77\,MHz). The total mapped area is 176\,arcmin$^2$ marked with the cyan contours in Figure~\ref{img: RGB}. For spectroscopic setups, we used the Versatile SPectrometer Array (VESPA) mode and Fourier Transform Spectrometer (FTS) 50\,kHz mode in parallel. The FTS 50\,kHz mode ``offers'' the full 8\,GHz EMIR bandwidth, simultaneously covering \ce{C^18O}\,(1--0) transition. The information on the molecular transitions studied in this work is listed in Table~\ref{tab: mol_info}.  The beam sizes (half-power beam width, HPBW), $\theta_{\rm HPBW}$, for the \ce{C^18O}\,(1--0) and \ce{N2H+}\,(1--0) data are $\rm \sim$24\arcsecuni\ and $\rm \sim$28\arcsecuni, respectively. And the velocity resolutions are $\rm \sim$0.13\,\velouni\ and $\rm \sim$0.16\,\velouni, respectively. The main beam and forward efficiencies ($\eta_{\rm MB}$ and $\eta_f$) are 0.81 and 0.95, respectively, which are used to convert the antenna temperature $T_{\rm A}^*$ to the main beam temperature $T_{\rm MB}$ using the equation $T_{\rm MB}=T_{\rm A}^*\times\frac{\eta_{f}}{\eta_{\rm MB}}$. The data reduction was performed with the Continuum and Line Analysis Single-dish Software (CLASS) of the Grenoble Image and Line Data Analysis Software (GILDAS\footnote{\url{http://www.iram.fr/IRAMFR/GILDAS}}, \citealt{Pety_2005sf2a.conf..721P}). Spectral lines of the \ce{N2H+} and \ce{C^18O} data cube were extracted over the velocity range of $\pm$50\,\velouni\ centred on the centroid velocity of G24 ($\sim$94.5\,\velouni). The baseline of each spectrum was fitted by a 1st order polynomial towards windows of emission-free channels, which was defined by masking out the velocity range of $94.5\pm30$\,\velouni. Then the noise (standard deviation) of the spectrum was obtained within the same windows. The final rms noise of the entire filament, $\sigma_{\rm rms}$, is the median of the dataset composed of rms from all spectra in a data cube. The obtained $\sigma_{\rm rms}$ is $\simeq$135\,mK for \ce{C^18O} and $\simeq$95\,mK for \ce{N2H+} on the $T_{\rm MB}$ scale.  

\subsection{\texorpdfstring{\ce{^13CO}}{13CO} and \texorpdfstring{\ce{C^18O} $J=1-0$}{C18O} Data from APEX~12~m}
The \ce{^13CO}\,(2--1) and \ce{C^18O}\,(2--1) molecular lines were observed with the receiver PI230 tuned to $\sim$220\,GHz on the APEX 12~m telescope (\citealt{Güsten_2006AnA...454L..13G}) with the OTF mode (Project ID: 095.C-0905(A) and PI: G. A. Fuller). The map consists of seven sub-maps with the field of view of $\rm 5^{\prime}\,\times\,5^{\prime}$, covering a total mapping area of 175 arcmin$^2$, as indicated by the yellow dashed contours in Figure~\ref{img: RGB}. 
The spectral resolution for both data cube is $\sim$0.104\,\velouni, and the $\theta_{\rm HPBW}$ is $\sim$30\arcsecuni. The data reduction was performed with CLASS, including baseline subtraction and line extraction. A main beam efficiency of $\eta_{\rm mb}=0.75$ was used to convert from the antenna temperature to the main beam temperature. 
The rms noise ($\sigma_{\rm rms}$) was estimated in the same way as for the IRAM 30~m data, and is $\simeq \rm 257\,mK$ for \ce{^13CO}\,(2--1) and $\simeq \rm 215\,mK$ for \ce{C^18O}\,(2--1).

\section{Archival Data}
\label{sec: Archival Data}

\subsection{Archival PMO \texorpdfstring{\ce{^12CO}}{12CO} and \texorpdfstring{\ce{^13CO} $J=1-0$}{13CO} Data}
\label{subsec: PMO13.7tele}
To investigate the relatively extended, low-density molecular gas surrounding the filament, we used the archival data of the ground state transitions of \ce{^12CO} and \ce{^13CO} obtained from the Milky Way Imaging Scroll Painting (MWISP\footnote{\url{http://english.dlh.pmo.cas.cn/ic/in/}}, \citealt{Su_2019ApJS..240....9S, Sun_2020ApJS..246....7S}) survey towards the northern Galactic plane. The survey was observed with the 13.7\,m millimetre telescope, equipped with a 3$\times$3 multi-beam sideband-separating Superconducting SpectroScopic Array Receiver (SSAR) system \citep{Shan_2012ITTST...2..593S}. The spectral resolution is 61\,kHz, translating to a velocity resolution of $\sim$0.158\,\velouni\ at 115\,GHz and $\sim$0.167\,\velouni\ at 110\,GHz. The main beam efficiency is $\rm \eta_{MB}$ = 0.443 at 115\,GHz and 0.475 at 110\,GHz according to the status report for the observing season\footnote{\url{http://www.radioast.nsdc.cn/zhuangtaibaogao.php}} when the observations were carried out. The size of the map is 1.5 $\times$ 1 deg$\rm ^2$ centred at (l, b)\,=\,(23.74$^{\circ}$, 0.46$^{\circ}$), with $\theta_{\rm HPBW}$ of 55$^{\prime\prime}$ at $\sim$110\,GHz and 49$^{\prime\prime}$ at $\sim$115\,GHz. The raw data were reduced by the MWISP working group with CLASS. A first-order baseline was fitted to the spectra, and the data was regridded into a pixel scale of 30\arcsecuni. The typical rms noise level on the $T_{\rm MB}$ scale is 0.45\,K for \ce{^12CO} and 0.26\,K for \ce{^13CO} \citep{Sun_2020ApJS..246....7S}. 
The summary of this molecular archival data is listed in Table~\ref{tab: mol_info}.

\subsection{Infrared and Radio Continuum Data}

G24 Filament was observed as part of the Hi-GAL survey. The Photodetector Array Camera \& Spectrometer (PACS, \citealt{Poglitsch_2010AnA...518L...2P}) and the Spectral and Photometric Imaging Receiver (SPIRE, \citealt{Griffin_2010AnA...518L...3G}) are equipped. The former carried out observation at 70, 160\,\micron\, while the latter at 250, 350, and 500\,\micron\. The $\theta_{\rm HPBW}$ are $\sim$6\arcsecuni, 12\arcsecuni, 18\arcsecuni, 25\arcsecuni, and 36\arcsecuni, respectively, for the five wavelengths. The level 2.5 data of PACS and level 3.0 data of SPIRE reduced by Herschel Interactive Processing Environment (HIPE, \citealt{Ott_2010ASPC..434..139O}) were obtained from Herschel science archive data website\footnote{\url{http://archives.esac.esa.int/hsa/whsa/}}. In addition, the radio continuum data of the VLA Galactic Plane Survey (VGPS, \citealt{Stil_2006AJ....132.1158S}) at 21\,cm (equivalent to 1.4\,GHz) was used for detecting ionised gas such as free-free emission from \htworegion\ region, and the $\theta_{\rm HPBW}$ is 1\arcminuni. 

 \begin{figure*}[ht!]
  \centering 
  \includegraphics[width=1\textwidth]{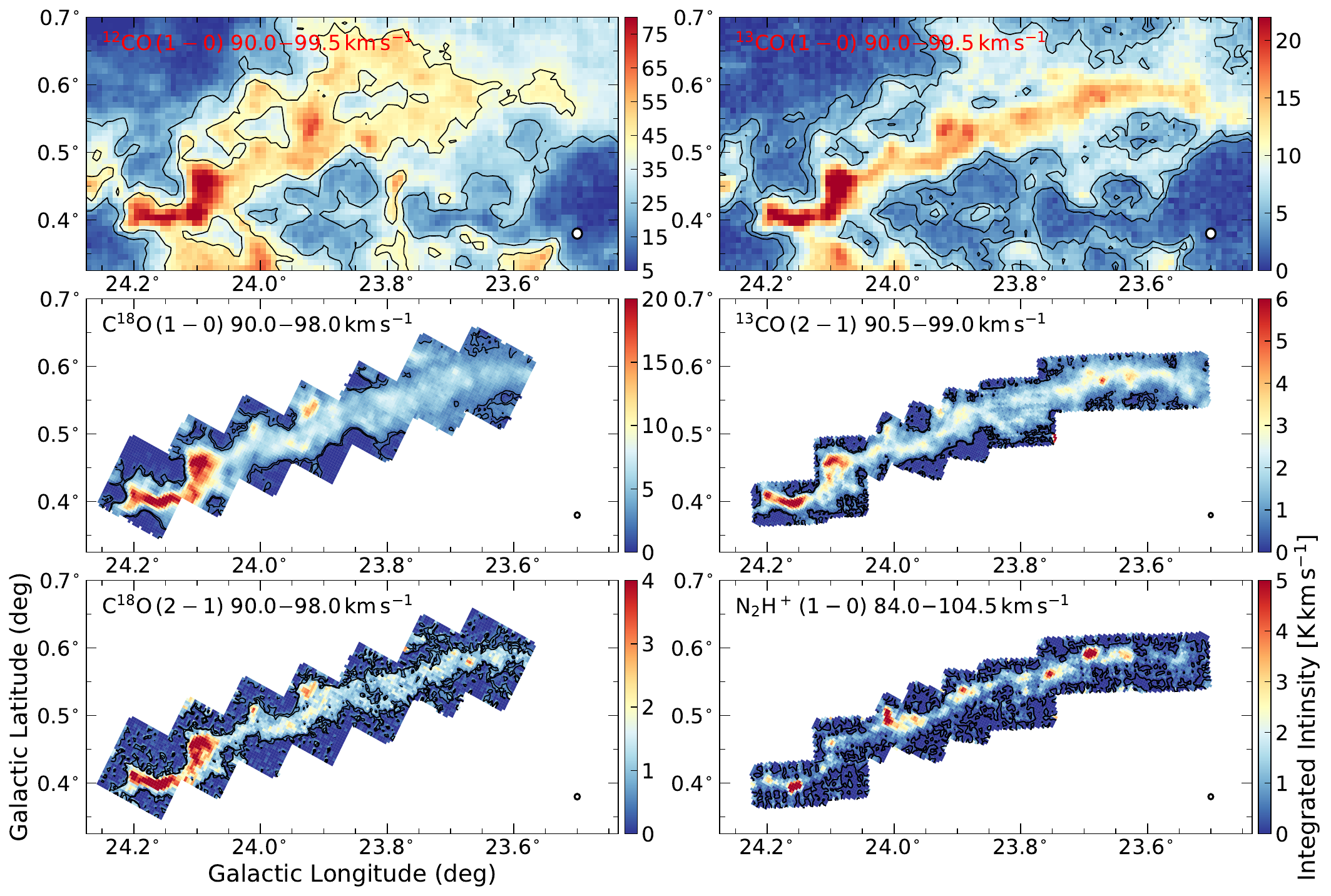}
  \caption{An overview of the integrated intensity maps of molecules in Table~\ref{tab: mol_info}. The velocity range for integration is shown in the upper left corner of each panel, and the beam size ($\theta_{\rm HPBW}$) is shown in the lower right corner. The contours depict the intensity of 3 and 5 times the noise level. 
  }
  \label{img: G24_molecular_data_show}
 \end{figure*}

\section{Analysis}
\label{sec: Analysis}
Figure~\ref{img: RGB} presents the three-colour infrared composite map of G24 with the filed of view (FOV) of $\rm \sim 50^{\prime} \times 20^{\prime}$. The 250\,\micron\ emission, obtained from the Hi-GAL survey, represents the cold dust emission seen in red. The 24\,\micron\ and 8\,\micron\ emission are taken from the \textit{Spitzer}-MIPSE/MIPSEGAL survey \citep{Carey_2009PASP..121...76C} and the \textit{Spitzer}-IRAC/GLIMPSE survey \citep{Benjamin_2003PASP..115..953B} respectively, and reflect the hot dust emission and are coloured in blue and green. Towards the east end of G24, an arch-shaped radio emission (white contours) is associated with an H{\sc ii} region \citep{Liu_2015ApJ...798...30L}. Its bright 8.0\,\micron\ and 24\,\micron\ emission trace polycyclic aromatic hydrocarbons (PAHs) and hot dust heated by high-energy ionising photons, respectively, which together point to active massive star formation. To provide an overview of the spatial distribution of emission intensities of different molecules, the integrated intensity images of the \ce{^12CO} $J$=1--0, \ce{^13CO} $J$=1--0 and 2--1, \ce{C^18O} $J$=1--0 and 2--1 and \ce{N2H+} $J$=1--0 emissions are shown in Figure~\ref{img: G24_molecular_data_show}. The emission distributions of these lines adequately present the filamentary structure of G24 and clumpy features like clumps, which are most prominent in \ce{N2H+} emission.

\subsection{Column Density of G24}
\label{subsec: Clmndens}

To estimate the mass and other properties of the G24 filament, we constructed several column density maps based on the \ce{^13CO}, \ce{C^18O} and \ce{N2H+} data, along with the $Herschel$ dust continuum data.

\subsubsection{Column Density of \texorpdfstring{\ce{^13CO}}{13CO}, \texorpdfstring{\ce{C^18O}}{C18O} and \texorpdfstring{\ce{N2H+}}{N2H+}}
\label{subsubsec: Clmndens_molec} 
Assuming local thermodynamic equilibrium (LTE), the \ce{^13CO} column density, $N$(\ce{^13CO}), can be estimated by the following equation \citep{Garden_1991ApJ...374..540G, Bourke_1997ApJ...476..781B}, 
\begin{equation}
N(\ce{^13CO}) = 2.42 \times 10^{14}\frac{T_{\rm ex}+0.88}{1-{\rm exp}(-5.29/T_{\rm ex})}\int\tau_{\ce{^13CO}}~d\upsilon.
\label{eq:Clmn_13CO}
\end{equation}

The column density of \ce{C^18O} under the LTE condition is estimated following the equation \citep{Mangum-Shirley_2016PASP..128b9201M,Dewangan_2019ApJ...877....1D},
\begin{equation}
\begin{aligned}
N(\ce{C^18O}) = &  2.48 \times 10^{14}\frac{T_{\rm ex}+0.88}{1-{\rm exp}(-5.27/T_{\rm ex})}\int\tau_{\ce{C^18O}}~d\upsilon.
\label{eq:Clmn_C18O}
\end{aligned}   
\end{equation}

The optical depths, $\tau$, of both \ce{^13CO}\,(1--0) and \ce{C^18O}\,(1--0) are derived from the following equation,
\begin{equation}
  \tau = -{\rm ln}\left[ 1-\frac{T_{\rm MB}}{J(T_{\rm ex}) - J(T_{\rm bg})} \right], 
  \label{eq:Tau}
 \end{equation}
 where $J$($T$) is the Rayleigh-Jeans temperature defined as $J(T) = \frac{h\nu/k}{{\rm exp}(h\nu/kT)-1}$, and $T_{\rm bg}$ is the cosmic background radiation temperature (i.e., 2.7\,K). 

The range of \ce{^13CO}\,(1--0) optical depth within the boundary of G24 (white thick contour in Figure~\ref{img: radial_fil_profile_13co1-0}) is 0.16--0.84 with both the mean and median values of 0.47. The range of $\tau_{\ce{C^18O}}$ is 0.02--0.48, and the mean and median values are 0.11 and 0.10, respectively. Although the $\tau_{\ce{^13CO}}$ values are less optically thin compared with those of $\tau_{\ce{C^18O}}$, we determined column densities of \ce{^13CO} and \ce{C^18O} as optically thin cases, where $T_{\rm MB} \approx f[J(T_{\rm ex}) - J(T_{\rm bg})]\tau$. The main beam filling factor, $f$, is assumed to be 1, because \ce{^13CO}\,(1--0) and \ce{C^18O}\,(1--0) emission are extended toward G24 and filling the main beam. 

\begin{figure*}[ht!]
\centering 
\includegraphics[width=1.0\textwidth]{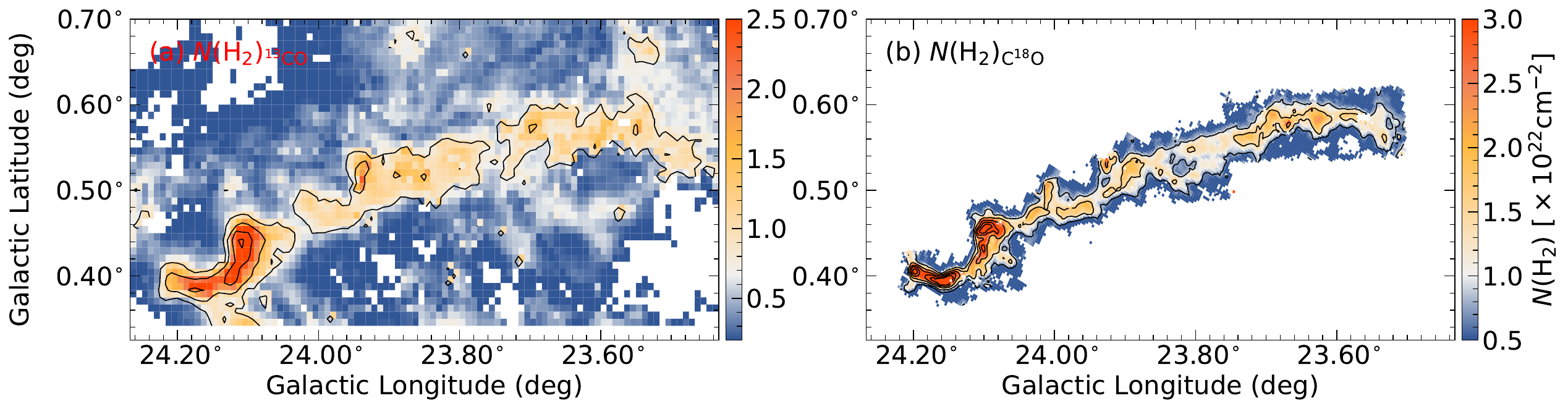}
\includegraphics[width=0.86\textwidth]{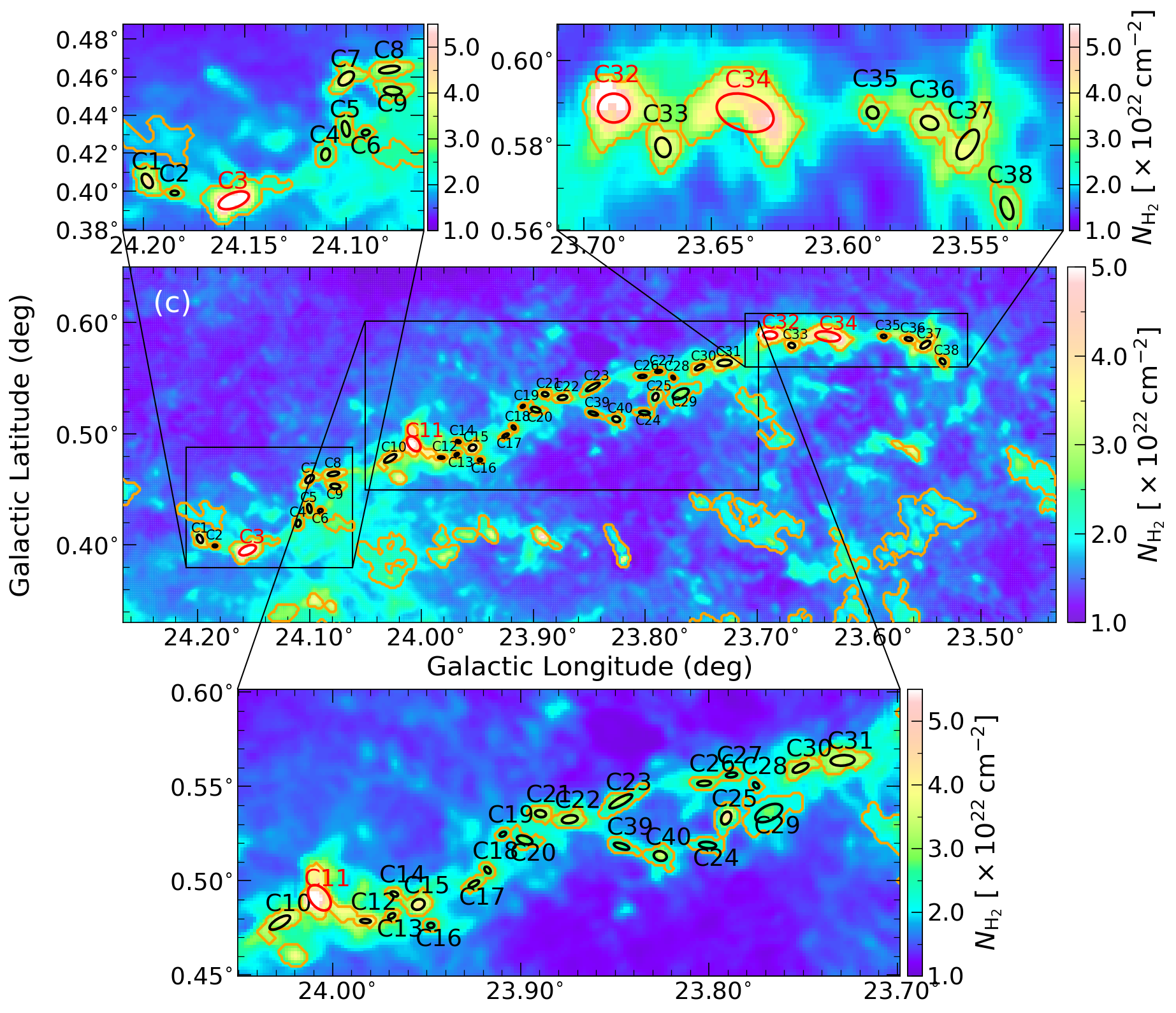}
\caption{
(a) \ce{H2} column density map derived from the \ce{^13CO}\,(1--0) data with contours of [0.8, 1.5, 2.5, 3.5, 4.5]$\rm \times 10^{22}$\,$\rm cm^{-2}$. (b) \ce{H2} column density map derived from the \ce{C^18O}\,(1--0) data with the same contours as panel\,(a).
(c) The \textit{Herschel}-based \ce{H2} column density ($N$(\ce{H2})$_{Herschel}$) map at a resolution of $\sim$18.2\arcsecuni , superimposed 40 clumps (``leaves") identified by \texttt{astrodendro}. The $N$(\ce{H2})$_{Herschel}$ map displayed here is the original map with background. The identified clumps are marked with orange polygons, and the black ellipses are projections. The number of the clumps is labelled in black. Four massive clumps with high surface densities (C3, C11, C32, and C34) are labelled in red. Panel (c) also displays zoomed-in views of the clumps to show their morphology. }
\label{img: column_density1}
\end{figure*}

Assuming that \ce{^12CO}\,(1--0) is optically thick, we derived the excitation temperature, based on which the column densities of \ce{^13CO} and \ce{C^18O} are derived. $T_{\rm ex}$ is calculated with the following equation \citep{Garden_1991ApJ...374..540G, Xu_2018AnA...609A..43X},
\begin{equation}
  T_{\rm ex} = \frac{h\nu}{k}\left\{{\rm ln}[1+\left(\frac{kT_{\rm MB}}{h\nu}+\frac{1}{e^{h\nu/kT_{\rm bg}}-1}\right)^{-1}]\right\}^{-1},
  \label{eq:Tex}
\end{equation}
where $\frac{h\nu}{k} = 5.53$ for \ce{^12CO}\,(1--0), $k$ is Boltzmann constant, $h$ and $\nu$ are Planck constant and the rest frequency of \ce{^12CO}\,(1--0), respectively. 
$T_{\rm MB}$ is the main-beam brightness temperature, which was then obtained by fitting \ce{^{13}CO}\,(1--0) and \ce{C^{18}O}\,(1--0) spectral lines with a multiple-component Gaussian profile. 
The details of the fitting procedure are described in Appendix~\ref{appendix: Spectral Line Fitting}. 
The column density is then derived based on the above quantities. To derive molecular hydrogen column densities ($N(\ce{H2})$) from the column densities of \ce{^13CO} and \ce{C^18O}, we applied an empirical abundance of $N$(\ce{^{13}CO})/$N$(\ce{H2})$~\simeq~1.4\times10^{5}$ and $N$(\ce{C^{18}O})/$N$(\ce{H2})$~\simeq~1.7\times10^{7}$ \citep{Ferking_1982ApJ...262..590F}. The results are presented in panels (a) and (b) of Figure~\ref{img: column_density1} and Table~\ref{tab: Tex_tau_13co}. 
Within the boundary of G24, the $N$(\ce{H2}) ranges estimated based on the \ce{^13CO}\,(1--0) and \ce{C^18O}\,(1--0) data are similar ($\rm \sim 2.2 \times 10^{21}$--$\rm 3.9 \times 10^{22}\,cm^{-2}$ and $\rm \sim 2.1 \times 10^{21}$--$\rm 6.5 \times 10^{22}\,cm^{-2}$, respectively), but the higher mean and median values of the latter ( 10.0/8.8\,$\rm \times 10^{21}\,cm^{-2}$ vs. 1.3/1.1\,$\rm \times 10^{22}\,cm^{-2}$) indicates \ce{C^18O}\,(1--0) traces regions of higher density.

  \begin{table*}
	\centering
	\caption{Derived physical properties using \ce{^13CO}, \ce{C^18O}, \ce{N2H+} $J$=1--0, and \textit{Herschel} dust thermal emission.}
	\label{tab: Tex_tau_13co}
    \renewcommand{\arraystretch}{1.2}
    \setlength{\tabcolsep}{1.0 mm}{
	\begin{tabular}{lccccccccc} 
		\toprule[0.8 pt]
        & $T_{\rm ex}$ & \multicolumn{2}{c}{$\tau$} & & \multicolumn{5}{c}{$N(\ce{H2})_{Herschel}$ (cm$^{-2}$)} \\
        \cline{3-4} \cline{6-10}
        & (K) & \ce{^13CO} & \ce{C^18O} & &\ce{^13CO} & \ce{C^18O} & \ce{N2H+} & Dust (with background) & Dust (without background)  \\
		\midrule[0.5 pt]
        Min & 7.0 & 0.16 & 0.02 & & 2.2$\times 10^{21}$ & 2.1$\times 10^{21}$ & 3.3$\times 10^{21}$ & 1.4$\times 10^{22}$ & 3.5$\times 10^{19}$  \\
        Max & 21.4 & 0.84 & 0.48 & & 3.9$\times 10^{22}$ & 6.5$\times 10^{22}$ & 4.9$\times 10^{23}$ & 8.9$\times 10^{22}$ & 7.3$\times 10^{22}$  \\
        Mean & 10.6 & 0.47 & 0.11 & & 1.0$\times 10^{22}$ & 1.3$\times 10^{22}$ & 3.2$\times 10^{22}$ & 2.3$\times 10^{22}$ & 7.7$\times 10^{21}$  \\
        Median & 10.1 & 0.47 & 0.10 & & 8.8$\times 10^{21}$ & 1.1$\times 10^{22}$ & 2.6$\times 10^{22}$ & 2.1$\times 10^{22}$ & 5.9$\times 10^{21}$  \\
		\bottomrule[1 pt]
	\end{tabular}}
\end{table*}

We also determined column density of \ce{N2H+} with the following equation \citep{Caselli_2002ApJ...565..344C, Zhang_2020MNRAS.497..793Z}, 
 \begin{equation}
  \begin{aligned}
  N({\rm \ce{N2H+}}) & = I_{\rm tot}\frac{8\pi}{\lambda^3A_{\rm ul}}\frac{g_{\rm l}}{g_{\rm u}}\times\frac{1}{J(T_{\rm ex})-J(T_{\rm bg})}\times \\
  & \frac{1}{1-{\rm exp}(-h\nu/(k_{\rm B}T_{\rm ex}))}\times\frac{Q_{\rm rot}(T_{\rm ex})}{g_{\rm l}\,{\rm exp}(-E_{\rm l}/(k_{\rm B}T_{\rm ex}))}, 
  \end{aligned} 
  \label{eq:column_density_N2HP}
 \end{equation}
where $I_{\rm tot}$ is given by $I_{\rm tot}=\int T_{\rm MB}\,d\upsilon/R_{\rm i}$, and the integral term is the integrated intensity of the isolated hyperfine component $F_1 = 2 \to 1$ with a rest frequency of 93.173770\,GHz, and $R_{\rm i}=5/9$ is the relative statistical weight. $A_{\rm ul}$ is the Einstein coefficient. $g_{\rm u}$ and $g_{\rm l}$ are the degeneracy of the upper and lower states and is defined as $g=2J+1$. 
The term $Q_{\rm rot}(T_{\rm ex})=\sum_{J=0}^{\infty}(2J+1){\rm exp}(-E_J/kT)$ is the partition function at the \ce{N2H+} excitation temperature. The energy of the level $J$, is $E_J=J(J+1)hB$, where $B = 46586.867 \times 10^6$\,Hz is the rotational constant \citep{Caselli_2002ApJ...565..344C}. 
To obtain the excitation temperature $T_{\rm ex, \ce{N2H+}}$, the \ce{N2H+}\,(1--0) spectral line data is fitted by a single component hyperfine structure (HFS) model using the Python library \texttt{pyspeckit}\footnote{\url{https://pyspeckit.readthedocs.io/en/latest/}}. 
The details of the fitting procedure are described in Appendix~\ref{appendix: Spectral Line Fitting}. The \ce{H2} column density is then estimated by multiplying the empirical abundance $X$({\ce{N2H+}})$=N$({\ce{N2H+}})/$N$(\ce{H2}) of $7.5 \times10^{-10}$ \citep{Hacar_2018AnA...610A..77H}. As shown in Figure~\ref{img: column_density2} and listed in Table~\ref{tab: Tex_tau_13co}, the obtained \ce{H2} column density varies between $\rm 3.3 \times 10^{21}\,cm^{-2}$ and $\rm 4.9 \times 10^{23}\,cm^{-2}$ with mean/median of 3.2/2.6\,$\rm \times 10^{22}\,cm^{-2}$. 
It traces dense regions with higher \ce{H2} column densities than \ce{^13CO}\,(1--0) and \ce{C^18O}\,(1--0). 
 
 \begin{table*}[ht!]
 \centering
 \small
  \caption{Properties of the dust continuum clumps.}
  \label{tab: Properties of Clumps}
  \renewcommand{\arraystretch}{0.95}
  \setlength{\tabcolsep}{0.5 mm}{
  \begin{tabular}{lccccccccccccccccc}
  \toprule[1pt]
  idx & $l$ & $b$ & R.A. & Dec. & $\theta_{\rm a}$ & $\theta_{\rm b}$ & P.A & $R_{\rm equ}$ & $M_{\rm clump}$ & $M_{\rm back}$ & $N(\ce{H2})$ & $N(\ce{H2})_{\rm back}$ & $\sum$ & $\sum_{\rm back}$ & $T_{\rm mean}$ & 70 & group \\
    & ($^{\circ}$) & ($^{\circ}$) & ($^{\circ}$) & ($^{\circ}$) & ($''$) & ($''$) & ($^{\circ}$) & (pc) & ($\rm M_{\odot}$) & ($\rm M_{\odot}$) & ($\rm cm^{-2}$) & ($\rm cm^{-2}$) & ($\rm g\,cm^{-2}$) & ($\rm g\,cm^{-2}$) & (K) & \micron &  \\
    (1) & (2) & (3) & (4) & (5) & (6) & (7) & (8) & (9) & (10) & (11) & (12) & (13) & (14) & (15) & (16) & (17) & (18)\\
  \midrule[0.5pt]
  C1 & 24.1981 & 0.4054 & 278.4893 & $-$7.5799 & 14.2 & 8.6 & 155 & 0.59 & 474 & 209 & 1.9$\rm \times 10^{22}$ & 8.4$\rm \times 10^{21}$ & 0.090 & 0.040 & 18.1 & 0  & 0 \\
  C2 & 24.1847 & 0.3991 & 278.4887 & $-$7.5948 & 7.1 & 4.2 & 91 & 0.28 & 83 & 49 & 1.4$\rm \times 10^{22}$ & 8.6$\rm \times 10^{21}$ & 0.069 & 0.041 & 18.7 & 0  & 0 \\
  C3 & 24.1555 & 0.3952 & 278.4786 & $-$7.6225 & 28.5 & 13.8 & 67 & 1.14 & 2180 & 781 & 2.4$\rm \times 10^{22}$ & 8.5$\rm \times 10^{21}$ & 0.112 & 0.040 & 18.0 & 2 & 1 \\
  C4 & 24.1102 & 0.4194 & 278.4359 & $-$7.6515 & 11.3 & 7.5 & $-$170 & 0.49 & 210 & 147 & 1.2$\rm \times 10^{22}$ & 8.6$\rm \times 10^{21}$ & 0.058 & 0.041 & 18.7 & 0 & 2 \\
  C5 & 24.1002 & 0.4328 & 278.4193 & $-$7.6542 & 14.8 & 6.5 & 169  & 0.50  & 256 & 155 & 1.4$\rm \times 10^{22}$ & 8.6$\rm \times 10^{21}$ & 0.067 & 0.041 & 18.9 & 0 & 2 \\
  \midrule[0.5pt]
  C6 & 24.0904 & 0.4308 & 278.4165 & $-$7.6638 & 7.0 & 5.2 & 78 & 0.31 & 84 & 57 & 1.2$\rm \times 10^{22}$ & 8.5$\rm \times 10^{21}$ & 0.059 & 0.040 & 18.9 & 0 & 2 \\
  C7 & 24.1000 & 0.4593 & 278.3955 & $-$7.6422 & 16.2 & 9.7 & 47 & 0.59 & 437 & 211 & 1.8$\rm \times 10^{22}$ & 8.5$\rm \times 10^{21}$ & 0.083 & 0.040 & 19.2 & 0 & 3 \\
  C8 & 24.0787 & 0.4639 & 278.3814 & $-$7.6589 & 18.4 & 6.5 & 81 & 0.57 & 399 & 195 & 1.8$\rm \times 10^{22}$ & 8.6$\rm \times 10^{21}$ & 0.083 & 0.040 & 17.7 & 0 & 4 \\
  C9 & 24.0771 & 0.4528 & 278.3906 & $-$7.6655 & 15.7 & 7.8 & 96 & 0.55 & 325 & 184 & 1.5$\rm \times 10^{22}$ & 8.5$\rm \times 10^{21}$ & 0.071 & 0.040 & 18.4 & 0 & 4 \\
  C10 & 24.0279 & 0.4778 & 278.3453 & $-$7.6976 & 22.2 & 8.4 & 59 & 0.63 & 535 & 241 & 1.9$\rm \times 10^{22}$ & 8.6$\rm \times 10^{21}$ & 0.090 & 0.041 & 16.7 & 0 & 5 \\
  \midrule[0.5pt]
  C11 & 24.0069 & 0.4910 & 278.3237 & $-$7.7101 & 27.6 & 17.5 & 142  & 0.95 & 1876 & 555 & 2.9$\rm \times 10^{22}$ & 8.6$\rm \times 10^{21}$ & 0.137 & 0.041 & 16.1 & 0 & 6 \\
  C12 & 23.9824 & 0.4786 & 278.3234 & $-$7.7376 & 10.0 & 3.7 & 93 & 0.31 & 145 & 58 & 2.2$\rm \times 10^{22}$ & 8.6$\rm \times 10^{21}$ & 0.102 & 0.041 & 16.6 & 0 & 7 \\
  C13 & 23.9685 & 0.4813 & 278.3145 & $-$7.7487 & 7.1 & 4.4 & 62 & 0.28 & 93 & 49 & 1.6$\rm \times 10^{22}$ & 8.6$\rm \times 10^{21}$ & 0.077 & 0.041 & 17.1 & 0 & 7 \\
  C14 & 23.9671 & 0.4930 & 278.3034 & $-$7.7445 & 7.3 & 4.4 & 110 & 0.30  & 106 & 54 & 1.7$\rm \times 10^{22}$ & 8.6$\rm \times 10^{21}$ & 0.081 & 0.041 & 16.8 & 0 & 7 \\
  C15 & 23.9544 & 0.4874 & 278.3025 & $-$7.7584 & 12.4 & 9.7 & 67 & 0.56 & 430 & 191 & 1.9$\rm \times 10^{22}$ & 8.6$\rm \times 10^{21}$ & 0.091 & 0.040 & 17.0 & 0 & 7 \\
  \midrule[0.5pt]
  C16 & 23.9477 & 0.4763 & 278.3093 & $-$7.7694 & 6.6 & 4.9 & 89 & 0.30 & 69 & 54 & 1.1$\rm \times 10^{22}$ & 8.6$\rm \times 10^{21}$ & 0.053 & 0.041 & 18.0 & 0 & 7 \\
  C17 & 23.9249 & 0.4984 & 278.2789 & $-$7.7795 & 10.9 & 4.7 & 61 & 0.37 & 138 & 84 & 1.4$\rm \times 10^{22}$ & 8.5$\rm \times 10^{21}$ & 0.066 & 0.040 & 17.6 & 2 & 8 \\
  C18 & 23.9176 & 0.5057 & 278.2690 & $-$7.7826 & 7.3 & 4.8 & 141 & 0.31 & 91 & 57 & 1.3$\rm \times 10^{22}$ & 8.5$\rm \times 10^{21}$ & 0.064 & 0.040 & 17.9 & 0 & 8 \\
  C19 & 23.9095 & 0.5247 & 278.2482 & $-$7.7810 & 6.5 & 4.0 & 63 & 0.27 & 53 & 44 & 1.0$\rm \times 10^{22}$ & 8.5$\rm \times 10^{21}$ & 0.048 & 0.040 & 18.0 & 0 & 8 \\
  C20 & 23.8979 & 0.5217 & 278.2455 & $-$7.7927 & 15.2 & 7.6 & 107  & 0.48 & 181 & 142 & 1.1$\rm \times 10^{22}$ & 8.6$\rm \times 10^{21}$ & 0.052 & 0.041 & 17.9 & 0 & 8 \\
  \midrule[0.5pt]
  C21 & 23.8894 & 0.5356 & 278.2291 & $-$7.7938 & 10.5 & 6.6 & 100 & 0.43 & 213 & 111 & 1.6$\rm \times 10^{22}$ & 8.6$\rm \times 10^{21}$ & 0.078 & 0.040 & 17.3 & 1 & 8 \\
  C22 & 23.8739 & 0.5326 & 278.2245 & $-$7.8089 & 15.2 & 7.4 & 81 & 0.53 & 319 & 172 & 1.6$\rm \times 10^{22}$ & 8.5$\rm \times 10^{21}$ & 0.075 & 0.040 & 17.1 & 1 & 8 \\
  C23 & 23.8468 & 0.5422 & 278.2034 & $-$7.8285 & 24.5 & 7.0 & 61 & 0.67 & 383 & 263 & 1.2$\rm \times 10^{22}$ & 8.3$\rm \times 10^{21}$ & 0.057 & 0.039 & 17.3 & 0 & 9 \\
  C24 & 23.8007 & 0.5189 & 278.2027 & $-$7.8802 & 16.3 & 6.0 & 95 & 0.50 & 156 & 152 & 8.6$\rm \times 10^{21}$ & 8.4$\rm \times 10^{21}$ & 0.041 & 0.040 & 17.3 & 0 & 10 \\
  C25 & 23.7908 & 0.5333 & 278.1852 & $-$7.8823 & 12.7 & 8.9 & $-$155 & 0.60 & 381 & 222 & 1.5$\rm \times 10^{22}$ & 8.6$\rm \times 10^{21}$ & 0.070 & 0.041 & 16.7 & 0 & 10 \\
  \midrule[0.5pt]
  C26 & 23.8025 & 0.5517 & 278.1742 & $-$7.8634 & 13.0 & 4.7 & 87 & 0.41 & 159 & 102 & 1.3$\rm \times 10^{22}$ & 8.6$\rm \times 10^{21}$ & 0.063 & 0.041 & 17.4 & 0 & 11 \\
  C27 & 23.7880 & 0.5561 & 278.1635 & $-$7.8743 & 10.6 & 4.4 & 83 & 0.34 & 93 & 71 & 1.1$\rm \times 10^{22}$ & 8.5$\rm \times 10^{21}$ & 0.053 & 0.040 & 17.5 & 0 & 11 \\
  C28 & 23.7750 & 0.5506 & 278.1623 & $-$7.8884 & 6.7 & 4.5 & 139 & 0.28 & 49 & 49 & 8.7$\rm \times 10^{21}$ & 8.7$\rm \times 10^{21}$ & 0.041 & 0.041 & 17.7 & 0 & 11 \\
  C29 & 23.7683 & 0.5359 & 278.1723 & $-$7.9011 & 27.6 & 14.0 & 64 & 1.00 & 546 & 603 & 7.8$\rm \times 10^{21}$ & 8.6$\rm \times 10^{21}$ & 0.037 & 0.041 & 17.2 & 0 & 11 \\
  C30 & 23.7513 & 0.5599 & 278.1430 & $-$7.9051 & 16.3 & 6.8 & 65 & 0.51 & 314 & 161 & 1.7$\rm \times 10^{22}$ & 8.6$\rm \times 10^{21}$ & 0.080 & 0.041 & 17.1 & 0 & 12 \\
  \midrule[0.5pt]
  C31 & 23.7290 & 0.5638 & 278.1290 & $-$7.9230 & 22.9 & 10.0 & 86 & 0.75 & 703 & 343 & 1.7$\rm \times 10^{22}$ & 8.5$\rm \times 10^{21}$ & 0.082 & 0.040 & 16.8 & 0 & 13 \\
  C32 & 23.6883 & 0.5888 & 278.0877 & $-$7.9477 & 22.4 & 12.2 & 90 & 0.89 & 1632 & 477 & 2.9$\rm \times 10^{22}$ & 8.5$\rm \times 10^{21}$ & 0.138 & 0.040 & 17.0 & 1 & 14 \\
  C33 & 23.6689 & 0.5795 & 278.0870 & $-$7.9690 & 10.9 & 7.9 & 107 & 0.47 & 302 & 132 & 1.9$\rm \times 10^{22}$ & 8.5$\rm \times 10^{21}$ & 0.092 & 0.040 & 17.3 & 1 & 15 \\
  C34 & 23.6368 & 0.5877 & 278.0647 & $-$7.9939 & 40.4 & 15.5 & 98 & 1.20 & 2317 & 882 & 2.2$\rm \times 10^{22}$ & 8.6$\rm \times 10^{21}$ & 0.106 & 0.040 & 16.7 & 0 & 16 \\
  C35 & 23.5868 & 0.5877 & 278.0413 & $-$8.0382 & 8.4 & 5.1 & 95 & 0.33 & 123 & 65 & 1.6$\rm \times 10^{22}$ & 8.3$\rm \times 10^{21}$ & 0.075 & 0.039 & 17.4  & 0 & 17 \\
  \midrule[0.5pt]
  C36 & 23.5644 & 0.5853 & 278.0331 & $-$8.0591 & 12.3 & 5.7 & 97 & 0.43 & 247 & 112 & 1.9$\rm \times 10^{22}$ & 8.6$\rm \times 10^{21}$ & 0.090 & 0.041 & 16.9 & 0 & 18 \\
  C37 & 23.5495 & 0.5802 & 278.0307 & $-$8.0747 & 18.0 & 9.0 & 56 & 0.57 & 442 & 201 & 1.9$\rm \times 10^{22}$ & 8.6$\rm \times 10^{21}$ & 0.090 & 0.041 & 16.9 & 0 & 18 \\
  C38 & 23.5341 & 0.5652 & 278.0369 & $-$8.0953 & 10.9 & 7.2 & 138  & 0.45 & 202 & 125 & 1.4$\rm \times 10^{22}$ & 8.6$\rm \times 10^{21}$ & 0.066 & 0.041 & 17.7 & 0 & 19 \\
  C39 & 23.8464 & 0.5183 & 278.2245 & $-$7.8399 & 15.1   & 5.3    & 107  & 0.44 & 138 & 119 & 9.9$\rm \times 10^{21}$ & 8.5$\rm \times 10^{21}$ & 0.047 & 0.040 & 17.7 & 0 & 20 \\
  C40 & 23.8259 & 0.5131 & 278.2196 & $-$7.8605 & 12.5 & 9.0 & 104 & 0.55 & 251 & 180 & 1.2$\rm \times 10^{22}$ & 8.5$\rm \times 10^{21}$ & 0.056 & 0.040 & 18.0  & 0 & 21 \\
  \bottomrule[1pt]
 \end{tabular}}
 \raggedright
 \footnotesize
 \textbf{Notes.}The meaning of each column is: (1) Clump ID. (2) Galactic longitude of the ellipse centre obtained from \texttt{astrodendro}. (3) Galactic latitude of the ellipse centre. (4) Right ascension of the ellipse centre. (5) Declination of the ellipse centre. (6) The semi-major axis of the ellipse. (7) The semi-minor axis of the ellipse. (8) The position angle in degrees counter-clockwise from the +x direction. (9) The equivalent radius of the ``leaf'', which is determined as $R_{\rm equ}=\sqrt{A/\pi}$, where A is the area of the ``leaf'' (orange polygon). (10)(11) The mass of the clump (``leaf'') and its background mass. (12) (13) The mean column density of the clump (``leaf'') and its background column density. (14)(15) The mass surface density of the clump (``leaf'') and its background surface density. (16) The mean dust temperature of the clump (``leaf''). (17) The number of 70\,\micron\ point sources match with one clump. (18) The group number classified by MST in Section~\ref{subsec: Fragmentation of G24}.
\end{table*}

\subsubsection{\texorpdfstring{\ce{H2}}{H2} Column Density and Dust Temperature} 
\label{subsubsec: Clmndens&Td}

In addition to molecular data, we also utilized $Herschel$ dust thermal continuum emission independently to determine the \ce{H2} column density ($N(\ce{H2})_{Herschel}$) and dust temperature, $T_{\rm dust}$, towards G24 by fitting a spectral energy distribution\,(SED) with the modified black-body function for each pixel, as described in \cite{Lin_2016ApJ...828...32L}. The 160, 250, 350, and 500\,\micron\ data were included for the SED fitting procedure, but the 70\,\micron\ data were excluded due to the potential contamination by small hot dust grains. To improve the angular resolution of the column density better than the resolution of 500\,\micron\ map ($\sim36$\arcsecuni), we applied the method introduced by \cite{Palmeirim_2013AnA...550A..38P}, as described in their Appendix~A. The detailed description for the SED fitting procedure is explained in Appendix~\ref{appendix: SED Fitting} of this study. The obtained final $N$(\ce{H2}) map with a angular resolution ($\theta_{\rm HPBW}$) of $\sim$18.2\arcsecuni\ is presented in panel (c) of Figure~\ref{img: column_density1}, while the obtained dust temperature map is is shown in Figure~\ref{img: G24_NH2andTemp_1} with the angular resolution of 36\arcsecuni. 
The weak extended emission surrounding the G24 filament is the foreground or background dust emission, and we subtracted the emission by following the method from \cite{Peretto_2010AnA...518L..98P} and the procedure is introduced in Appendix~\ref{appendix: Background Subtraction}.

The general properties of G24 are obtained within the boundary of the filament presented as a white thick contour in Figure~\ref{img: radial_fil_profile_clmn}. The details of the definition of the boundary will be demonstrated in Section~\ref{subsubsec: Width of G24}. The dust temperature within the boundary is in the range of 15.1--20.9\,K with both mean and median of 18.1\,K. The column density before the background subtraction (shown in panel (c) of Figure~\ref{img: column_density1}) is in the range of $\rm 1.4\times 10^{22}$--$\rm 8.9\times 10^{22}\, cm^{-2}$, and with the mean/median of $\rm 2.3 \times 10^{22}/\rm 2.1 \times 10^{22}\, cm^{-2}$. 
While that after the background subtraction (shown in Figure~\ref{img: Bg_sb_map} panel (c)) is within the range of $\rm 3.5\times 10^{19}$--$\rm 7.3\times 10^{22}\, cm^{-2}$, and with the mean/median of $\rm 7.7\times 10^{21}/5.9\times 10^{21}\, cm^{-2}$. 
The dust temperature of G24 is significantly lower than that of the surrounding environment, while the column density increases from the outskirts to the inner region. In addition, several condensations within G24 are clearly visible towards the much colder areas in the temperature map.

\subsection{Clump Identification}
\label{subsec: Clump Identification}
Star-forming filaments or GF often enclose dense, compact structures which are the potential sites of active star formation. 
We identified compact structures in the \textit{Herschel}-based \ce{H2} column density map for its high resolution (18.2\arcsecuni). 
The identification was performed with the algorithm \texttt{Dendrogram} \citep{Rosolowsky_2008ApJ...679.1338R}, which is implemented by the Python library \texttt{astrodendro}\footnote{\url{http://www.dendrograms.org/}}. The required input parameters are \texttt{min\_value}, \texttt{min\_delta}, and \texttt{min\_npix}. To identify the smallest and densest substructure, termed ``leaf'', in the column density map, we set the minimum column density of `leaf'' (\texttt{min\_value}) to be $\rm 1\times 10^{22}\,cm^{-2}$, and the level for merging structures, \texttt{min\_delta}, is limited to $\rm 1\times 10^{20}\,cm^{-2}$. The minimum number of pixels (\texttt{min\_npix}) for a ``leaf'' is set to 10 pixels. 
 
As shown in panel (c) of  Figure~\ref{img: column_density1}, the orange polygons represent the ``leaves'' identified by \texttt{astrodendro} Although the ``leaves'' were identified in the $N_{\rm H_2}$ map, the coherency between the centroid velocity of G24, which is $\sim$94--96\,\velouni, and that of the ``leaves'' were also checked using the \ce{N2H+} data. If the centroid velocity of a ``leaf'' falls out of the range, it is considered incoherent with the G24 filament in velocity and excluded, even if it is spatially associated with G24. Finally, 40 ``leaves'' associated with G24 in both spatial and velocity were obtained, as indicated by black ellipses in panel (c). These black ellipses are the projections of ``leaves'' onto the position-position (PP) plane. The major axis is derived from the intensity-weighted second-order moment map in the direction of greatest elongation in the PP plane, and the minor axis is perpendicular to the major axis. 
 
The derived physical properties of the ``leaves'' are listed in Table~\ref{tab: Properties of Clumps}. The equivalent radius, $R_{\rm equ}$, is determined as $R_{\rm equ}=\sqrt{A/\pi}$, where $A$ is the area of a ``leaf'' (orange polygon). At a distance of 5.2\,kpc, the equivalent radius of ``leaves'' is 0.28--1.2\,pc, corresponding to the scale of clump. The mass ($M_{\rm clump}$) is estimated by the sum of column densities of pixels within a ``leaf". The mass surface density ($\Sigma$) is calculated by dividing the total mass by the area of the orange polygon ($A$). The background mass $M_{\rm back}$ and surface density $\Sigma_{\rm back}$ of each clump are also provided in the table. Studies from \citep{Traficante_2018MNRAS.473.4975T, Traficante_2020MNRAS.491.4310T} towards massive clumps found that they are dynamically active and show evidence of parsec-scale infall motions when their surface density is $\rm \geqslant 0.1\,g\,cm^{-2}$. In our study, four clumps (C3, C11, C32, and C34) out of the 40 clumps are considerably massive based on their surface density ($\Sigma \sim$ 0.10--0.14\,g\,cm$^{-2}$) and mass ($M_{\rm clump}\sim$1632--2317\,$\rm M_{\odot}$). They are marked with red ellipses in Figure~\ref{img: column_density1} panel (c). In addition, five clumps are associated with 70\,\micron\ point sources (70\,\micron -bright, see Column (17) of Tabel~\ref{tab: Properties of Clumps}), indicating active star formation activity. While no significant 70\,\micron\ emission is detected in the remaining 35 clumps (70\,\micron -quiet), indicating they are in the very early stage of star formation. A detailed analysis of the clumps will be presented in a follow-up study. 

 \begin{table*}[ht!]
  \centering
  \caption{Width of the G24 filament measured with different tracers. }
  \label{tab: FWHM_estimation}
  \setlength\tabcolsep{1.3mm}{
  \begin{tabular}{lcccccccccc}
  \toprule[1pt]
  & \multicolumn{2}{c}{Resolution}  &      Integral Range  &  \multicolumn{3}{c}{Plummer(RadFil)} & \multicolumn{2}{c}{Gaussian(RadFil)} & \multicolumn{2}{c}{Gaussian(slice-by-slice)}    \\\cline{2-3}\cline{5-7}\cline{8-9}\cline{10-11}
    &  &  &  & \textit{p} & $R_{\rm flat}$ & Width &  Stddev   & FWHM  & $\rm FWHM_{obs}$ & $\rm FWHM_{deconv}$  \\
    &   (\arcsecuni)  &    (pc)    &    (\velouni)    &    & (pc)     & (pc)    & (pc)     & (pc)    & (pc)    & (pc)       \\
    (1) & (2) & (3) & (4) & (5) & (6) & (7) & (8) & (9) & (10) & (11)   \\
  \midrule[0.5pt]
  \ce{^12CO}\,(1--0) & 54.8 & 1.38 & 90--99.5 & 3.64$\pm$1.39 & 2.47$\pm$0.95 & 4.10$\pm$2.09  & 1.92$\pm$0.09 & 4.52$\pm$0.20 & 3.10$\pm$0.72 & 2.77$\pm$0.81 \\
  \ce{^13CO}\,(1--0) & 54.8 & 1.38 & 90--99.5 & 2.81$\pm$0.72 & 1.39$\pm$0.45 & 2.97$\pm$1.28 & 1.42$\pm$0.06 & 3.33$\pm$0.15 & 2.85$\pm$0.61 & 2.49$\pm$0.71 \\
  \ce{C^18O}\,(1--0) & 23.6 & 0.59 & 90.5--99 & 3.07$\pm$0.28 & 0.63$\pm$0.08 & 1.23$\pm$0.19 & 0.61$\pm$0.01 & 1.44$\pm$0.03 & 1.22$\pm$0.25 & 1.07$\pm$0.29 \\
  \ce{^13CO}\,(2--1) & 30.1 & 0.76 & 90--98 & 2.36$\pm$0.65 & 0.54$\pm$0.24 & 1.44$\pm$0.84 & 0.71$\pm$0.05 & 1.68$\pm$0.11 & 1.38$\pm$0.28 & 1.16$\pm$0.33 \\
  \ce{C^18O}\,(2--1) & 30.1 & 0.76 & 90--98 & 2.39$\pm$0.41 & 0.42$\pm$0.12 & 1.09$\pm$0.41 & 0.59$\pm$0.03 & 1.40$\pm$0.07 & 1.35$\pm$0.26 & 1.12$\pm$0.31 \\
  \ce{N2H+}\,(1--0) & 27.8 & 0.70 & 84.0--104.5 & 3.32$\pm$0.81 & 0.59$\pm$0.17 & 1.06$\pm$0.39 & 0.53$\pm$0.02 & 1.22$\pm$0.06 & 1.15$\pm$0.22 & 0.91$\pm$0.28 \\
  Dust $N_{\rm H_2}$  & 18.2 & 0.46 & --- $\dagger$ & 3.33$\pm$0.21 & 0.45$\pm$0.04 & 0.81$\pm$0.08 & 0.41$\pm$0.01 & 0.96$\pm$0.02& 0.94$\pm$0.21 & 0.82$\pm$0.25  \\
  \bottomrule[1pt]
  \end{tabular}}
  \footnotesize
 \textbf{Notes.} ($\dagger$) We only considered the $N$(\ce{H2})$_{Herschel}$ within the boundary (white thick contour) in Figure~\ref{img: radial_fil_profile_clmn}. 
 \end{table*}

\section{General Properties of G24}
\label{sec: General Properties of G24}
\subsection{Physical Properties of G24}
\begin{figure*}[ht!]
  \centering 
  \includegraphics[width=0.85\textwidth]{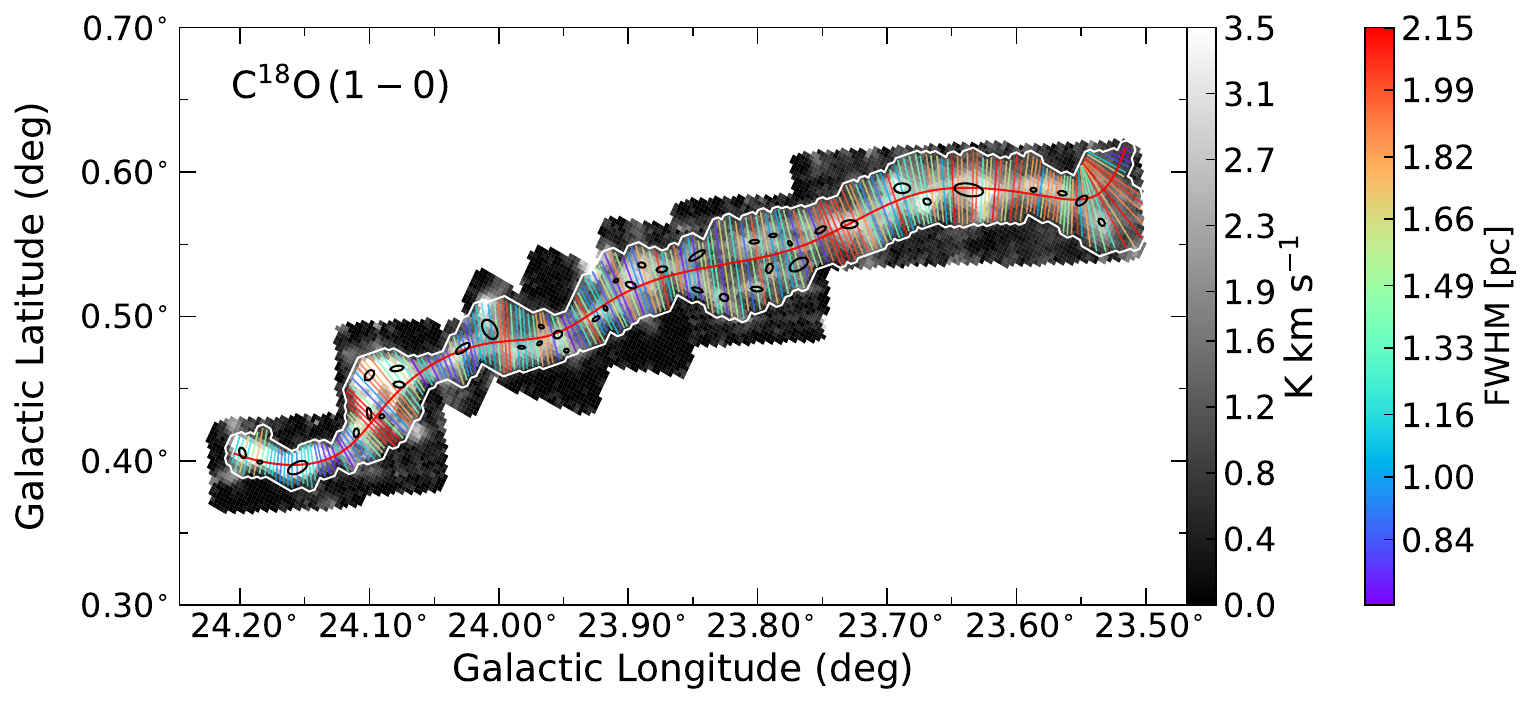}
  \caption{The skeleton (red curve), boundary (white contour) of G24 and paths of transverse slices, colour-coded by the FWHM of the multiple-component Gaussian model. The background is the integrated intensity map of \ce{C^18O}\,(1--0) with the integral range of 90.5--99.0\,\velouni. 
  Clumps are superimposed on the figure and marked by black ellipses.}
  
 \label{img: radial_fil_profile_c18o1-0}
 \end{figure*}
\subsubsection{Filament Width}
\label{subsubsec: Width of G24}

A paramountly important attribute of filamentary molecular clouds is their transverse diameter since the fragmentation properties of quasi-static cylindrical filaments are expected to scale with the filament's diameter \citep[e.g.,][]{Nagasawa_1987PThPh..77..635N, Inutsuka_1992ApJ...388..392I}. Many works have put effort into exploring whether at least nearby molecular filaments have a typical half-power width or a true common scale \citep[e.g.,][]{Arzoumanian_2011AnA...529L...6A, Arzoumanian_2019AnA...621A..42A, Panopoulou_2022AnA...657L..13P}. In this section, we investigated the width of G24.
We firstly determined the length and boundary of G24 on the integrated intensity maps of molecules in Table~\ref{tab: mol_info} and the \textit{Herschel}-based \ce{H2} column density ($N(\ce{H2})_{Herschel}$) map (with background) using the python library \texttt{RadFil}\footnote{\url{https://github.com/catherinezucker/radfil}} \citep{Zucker_2018ApJ...864..153Z}. 
The velocity ranges for integration are listed in Table~\ref{tab: FWHM_estimation}. 
To estimate the width of a filament, \texttt{RadFil} first identifies the skeleton of the filament and extracts the transverse slices (radial intensity profiles) perpendicular to the local skeleton at a fixed interval along the skeleton. 
The radial profiles are then stacked into a single profile and fitted by the Gaussian and Plummer-like model provided in \texttt{RadFil} to obtain the width. 
\begin{figure}
  \centering 
  \includegraphics[width=1.0\columnwidth]{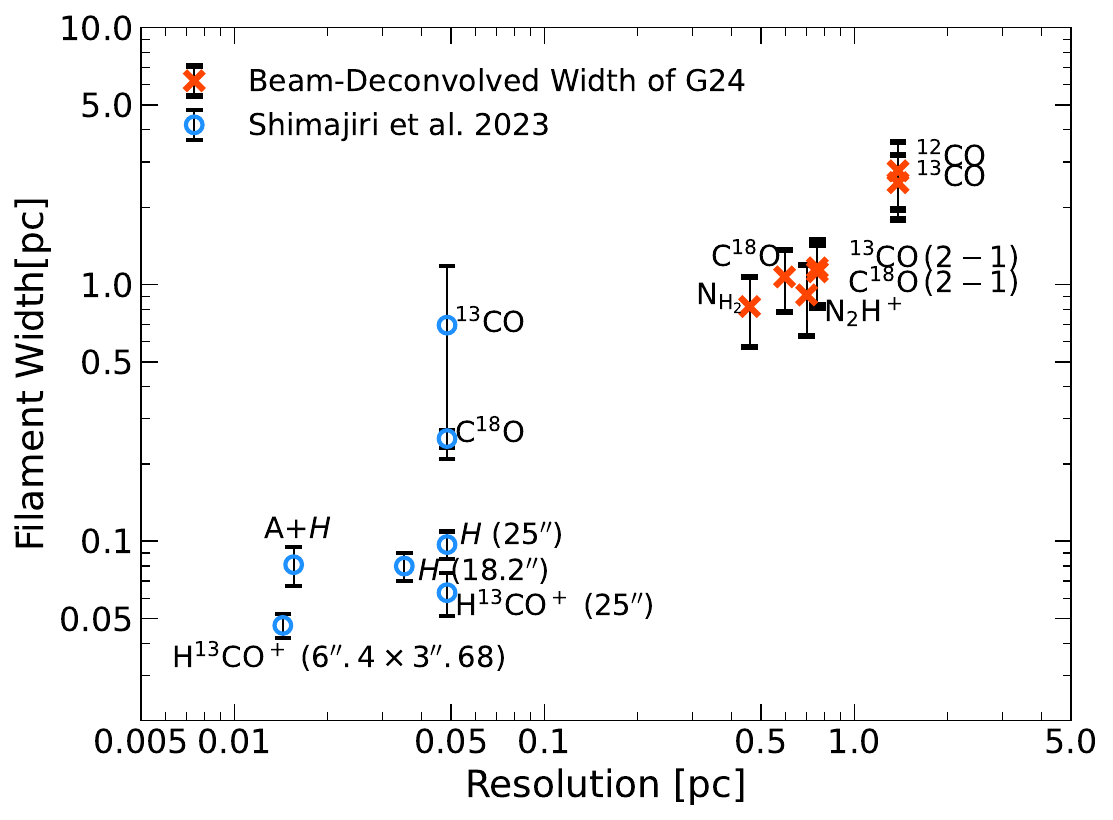}
  \caption{The spatial resolution of the molecular data versus beam-deconvolved widths. The red crosses are the data from this study (G24), and blue circles indicate the data from \cite{Shimajiri_2023AnA...672A.133S} (NGC~2024). The spatial resolution and filament width are in pc units by adopting the distance of 5.2\,kpc for G24 and 400\,pc for NGC~2024. The data from \cite{Shimajiri_2023AnA...672A.133S} are shown with abbreviations for ease of presentation, where $H$ denotes $Herschel$ and A denotes ArT\'eMiS. } 

 \label{img: Resolution_vs_Width}
 \end{figure}

For the Gaussian model, the width of a filament is the full width at half maximum (FWHM) and defined as ${\rm FWHM}={\rm\sqrt{8ln2}}\sigma$, where $\sigma$ is the standard deviation. The Plummer-like model is defined as \citep{Arzoumanian_2011AnA...529L...6A, Cox_2016AnA...590A.110C},
 \begin{equation}
      H(r) = \frac{H_{\rm 0}}{[1+(\frac{r}{R_{\rm flat}})^2]^{(p\rm -1)/\rm 2}},
      \label{Plummer-like model}
 \end{equation}
where $H_0$ is the height at the peak of the profile, $R_{\rm flat}$ is the flattening radius, and $p$ is the power law exponent of the radial profile at radius much larger than $R_{\rm flat}$. The width of the Plummer-like profile is defined as $2(2^{2/(p-1)}-1)^{1/2}R_{\rm flat}$. 
As listed in Table~\ref{tab: FWHM_estimation} Column (7) and (9), the widths obtained by the two models do not differ significantly. 
Figure~\ref{img: radial_fil_profile_c18o1-0} shows the skeleton (red curve) and boundary (white contour) of \ce{C^18O}\,(1--0) integrated intensity defined by \texttt{RadFil}. The corresponding Plummer-like and Gaussian fittings are presented in Figure~\ref{img: radial_fil_profile_c18o1-0_fitting}. 
The skeletons and boundaries determined by other molecules and $N(\ce{H2})_{Herschel}$ are displayed in Figures~\ref{img: radial_fil_profile_13co1-0}--\ref{img: radial_fil_profile_clmn}. 
The specific parameter settings for the radial profile fitting can be found in Appendix~\ref{appendix: The Radial Profile}. 

Some of the widths, obtained by Plummer-like fitting in Column (7) of Table~\ref{tab: FWHM_estimation} using \ce{^13CO}, \ce{C^18O} $J=2-1$, \ce{N2H+}\,(1--0) and $N(\ce{H2})_{Herschel}$, are not spatially resolved. 
\cite{André_2022AnA...667L...1A} compared the widths of filaments from \textit{Herschel} Gould Belt Survey and Hi-GAL survey to synthetic filaments, and found that the measured filament width increases as the spatial resolution worsens and/or the distance to the filament increases. The observed trend was found to be consistent with the beam convolved widths of filaments with Plummer-like density profiles. In addition, they found that when the filament width is less than a factor of $\sim2$ broader than the telescope's HPBW, the width is dominated by the power-law wing of the filament profile, and the flat inner portion of their column density profiles is unresolved. 
The widths fitted by the Plummer-like profiles obtained using the data of aforementioned molecules and $N(\ce{H2})_{Herschel}$ are less than a factor of $\sim2$ broader than the corresponding HPBW, thus are dominated by the power-law wing, which are primarily affected by the convolution of the beams. 

We also measured G24's width by fitting each of the radial intensity profiles with multiple Gaussian components for the following reason. \texttt{RadFil} fits the radial intensity profile stacking from all the transverse slices. However, some of the individual intensity profiles show multiple peaks, which may arise when the radial profile crosses multiple clumps or multiple local filamentary structures (such as the ring structure at the middle of G24 in Figure~\ref{img: radial_fil_profile_c18o1-0}). These profiles cannot be fitted by a single-component Gaussian or Plummer-like model. Therefore, we fitted each of the radial intensity profiles with multiple Gaussian components. 
For a radial profile consisting of multiple intensity components, its amplitude, mean, and standard deviation are obtained by performing a weighted average on the corresponding components of the amplitude, mean, and standard deviation, with the amplitude of each intensity component serving as the weight. 
The process of determining the number of Gaussian components and performing the fitting is identical to that for fitting multiple velocity components of spectral lines (assuming each velocity component follows a Gaussian profile) described in Appendix~\ref{appendix: Spectral Line Fitting}, except that the spectral emission line was replaced with the radial intensity profile. 
The FWMH of each radial profile is shown as the colour-coded segment perpendicular to the skeleton in Figure~\ref{img: radial_fil_profile_c18o1-0}.
The final width of G24 (FWHM$_{\rm obs}$) is the median of the FWHM across all of the radial profiles and is listed in Table~\ref{tab: FWHM_estimation} Column (10). The uncertainty is the standard deviation of the widths of all the radial profiles. The beam-deconvolved FWHM (FWHM$_{\rm deconv}$) is also derived (Column (11)), and ranges from $\sim0.8$ to $\sim2.5$\,pc. 
The FWHM$_{\rm deconv}$ of the \ce{C^18O}\,(1--0) data is narrower ($1.07\pm0.29$\,pc) than that of the \ce{^13CO}\,(1--0) ($2.49\pm0.71$\,pc) and \ce{^12CO}\,(1--0) data ($2.47\pm0.95$\,pc). The same trend also appears in higher rotational energy levels (i.e., $J=2-1$). The dense gas tracer \ce{N2H+}\,(1--0) shows the narrowest FWHM$_{\rm deconv}$ of 0.91$\pm$0.28\,pc, compared with CO lines. 

The pattern that filament widths varying in different gas tracers is also found towards other molecular filaments. 
\cite{Shimajiri_2023AnA...672A.133S} found that the widths of the NGC~2024 filament from denser gas tracers are smaller (0.081\,pc from dust continuum and 0.063\,pc from \ce{H^13CO+}\,(1--0)) than those from tracers of less dense gas  (i.e., 0.25\,pc from \ce{^13CO}\,(1--0) and 0.69\,pc from \ce{C^18O}\,(1--0)). 
In hydrodynamical simulations of a turbulent isothermal molecular clouds \citep{Priestley-Whitworth_2020_MNRAS.499.3728P}, the apparent widths of filaments measured with synthetic line intensities of dense gas tracers (such as \ce{N2H+} and HCN) also tend to be smaller.

Figure~\ref{img: Resolution_vs_Width} shows the distribution of determined filament widths from different density tracers of G24 (Column (11) in Table~\ref{tab: FWHM_estimation}) and \cite{Shimajiri_2023AnA...672A.133S} (NGC~2024). The widths are measured at different spatial resolutions in pc adopting distances of 5.2\,kpc for G24 and 400\,pc for NGC~2024. 
The widths obtained from our work are generally larger. This is possibly due to the large distance of G24, as all the data points of NGC~2024 are at smaller spatial resolutions. 
The beams in our study correspond to a much larger spatial scale than those in NGC~2024. 

The width of filament G24 obtained from the $Herschel$-based \ce{H2} column density is consistent with the typical widths reported by \cite{Schisano_2014ApJ...791...27S}. 
They found that distant filaments with distances $>1.5$\,kpc  have wider deconvolved widths with a larger spread, with a mean of 0.82\,pc and standard deviation of 0.57\,pc. The typical width of those filaments is $1.9\pm0.2$ times the HPBW spatial resolution. The width of G24 measured from $N(\ce{H2})_{Herschel}$ is 0.94\,pc, which is also $\sim2$ times the beam size ($\theta_{\rm HPBW}$). Such a correlation of partially resolved measured size and observation resolution is consistent with the presence of an underlying  power-law density profile \citep{André_2022AnA...667L...1A, Adams_1991ApJ...382..544A,Ladd_1991ApJ...382..555L} and results in weak constraints defining an resolution-dependent intrinsic width of the structure. Also, as a giant filament, it is also possible that G24 contains sub-structures \citep{Hacar_2023ASPC..534..153H} which  would require future  higher-resolution observations to constrain.

We also estimate the thickness of G24 along the LOS using the non-LTE radiative transfer code RADEX \citep{Van_der_Tak_2007AnA...468..627V}. The details are provided in Appendix~\ref{appendix: Line Radiative Transfer}. The simulation obtained a \ce{H2} volume density ($n_{\ce{H2}}$) range of $10^3$--$10^4\,{\rm cm}^{-3}$ for G24. 
Based on the obtained $N(\ce{H2})_{Herschel}$ and $n_{\rm \ce{H2}}$, the mean and median thickness along the LOS are both 2.2\,pc, $\sim2$ times greater than the \ce{C^18O} projected FWHM$_{\rm deconv}$ of 1.07\,pc (see Table~\ref{tab: FWHM_estimation}) and are $\sim40$ times shorter than the length of G24 in the plane of the sky. This rejects the possibility that G24 is a projection of flat structures, such as edge-on sheets, along the LOS. 
 
  \begin{table*}[ht!]
  \centering
  \caption{General physical properties of the G24 filament.}
  \label{tab: filament_properties}
  \begin{tabular}{lccccc}
  \toprule[1 pt]
  Molecule/Line & Length & Mass & $M_{\rm line}$ & $M_{\rm line, vir}$ & mean/max $N_{\ce{H2}}$ \\
  {} & (pc) & ($\rm M_{\odot}$) & ($\rm M_{\odot}\ pc^{-1}$) & ($\rm M_{\odot}\ pc^{-1}$) & ($\rm 10^{22}\ cm^{-2}$) \\
  \midrule[0.5 pt]
  \ce{^13CO}\,(1--0) & 78.8 & $\rm 9.1\times 10^4$ & 1160$\pm$13 & 2212$\pm$77 & 0.94/3.9 \\
  \ce{C^18O}\,(1--0) & 74.8 & $\rm 6.3\times 10^4$ & 837$\pm$2 & 717$\pm$32 & 1.3/6.5 \\
  $N$(\ce{H2})$_{Herschel}$ & 85.8 & $\rm 1.2\times 10^5$ & 1374$\pm$58 & --- & 2.3/8.9 \\
  \cite{Wang_2015MNRAS.450.4043W} & 82 & $\rm 1.1\times 10^5$ & 1362.7 & --- & 0.5/2.6 \\
  \bottomrule[1 pt]
 \end{tabular}
 \end{table*}

\subsubsection{Line Mass}
\label{subsubsec: Line Mass of G24} 

Filaments can be categorized in terms of gravitational stability by the ratio of the virial line mass to the line mass. Therefore, in this section, we derived the line mass for the subsequent discussion on gravitational stability.
We determined the total mass and lengths of the G24 filaments using the \ce{H2} column densities (listed in Table~\ref{tab: Tex_tau_13co}) derived by \ce{^13CO}\,(1--0), \ce{C^18O}\,(1--0), and \textit{Herschel} dust continuum emission. The masses measured within the G24 filament boundaries (white contour in Figure~\ref{img: radial_fil_profile_13co1-0} and \ref{img: radial_fil_profile_c18o1-0}) are $9.1\times 10^4\,\rm M_{\odot}$, $\rm 6.3\times 10^4\,M_{\odot}$ and $1.2\times 10^5\,\rm M_{\odot}$, respectively, for those tracers. 
The skeleton lengths determined with \ce{^13CO}, \ce{C^18O}, and dust continuum emission are 78.8, 74.8, and 85.8\,pc, respectively. 
The line mass for each tracer is derived by the ratio of the total mass to the length ($M\rm _{line}=Mass/Length$), and are $1160\pm13$, $837\pm2$ and $1374\pm 58$\,$\rm M_{\odot}\,pc^{-1}$, respectively. The length and line mass derived from the Herschel dust continuum emission is consistent with the value reported by \cite{Wang_2015MNRAS.450.4043W}, while the values derived from \ce{^13CO} and  \ce{C^18O} are smaller than \cite{Wang_2015MNRAS.450.4043W}, which may be due to smaller area of profiles identified among the tracers (see Figure~\ref{img: radial_fil_profile_13co1-0} and \ref{img: radial_fil_profile_c18o1-0}).

We also estimated the critical line mass and virial line mass towards G24. The critical line mass is for the case of an infinite, isothermal cylinder in the equilibrium between thermal and gravitational pressure \citep{Ostriker_1964ApJ...140.1056O}. \cite{Inutsuka_1992ApJ...388..392I} found that when the line mass $M_{\rm line}$ is larger than the critical value for the equilibrium, $M_{\rm line,crit} = 2c_{\rm s}^2/G$, an unmagnetised filament collapses radially, and the fragmentation occurs once the interstellar gas becomes non-isothermal. The $c_{\rm s}$ represents the isothermal sound speed and is defined as $c_{\rm s} = (kT_{\rm kin}/\mu m_{\rm H})^{1/2}$, where $\mu$ is the mean molecular weight per free particle of 2.33, $G$ is the gravitational constant. 
For the kinetic temperature ($T_{\rm kin}$), we used a median dust temperature of 18.1\,K, assuming that the gas and dust are well coupled. The derived critical line mass at the given temperature is $M_{\rm line,crit}\sim25\,\rm M_{\odot}\,pc^{-1}$, and is significantly smaller than those lines masses derived from the \ce{^13CO}, \ce{C^18O}\,(1--0) and \textit{Herschel} data. 
The lower critical mass implies that G24 is thermally supercritical and gravitationally unstable to radial contraction.

The virial line mass includes the contribution of the non-thermal gas motions when estimating the critical line mass, which is defined as $M_{\rm line,vir}=2\sigma_{\rm tot}^2/G$ \citep{André_2014prpl.conf...27A, Kainulainen_2016AnA...586A..27K, Fiege_2000MNRAS.311..105F}, where $\sigma_{\rm tot}=\sqrt{c_{\rm s}^2+\sigma_{\rm NT}^2}$ is the total velocity dispersion including thermal and non-thermal gas motions. The virial line mass can then be written as \citep{Dewangan_2019ApJ...877....1D},
\begin{equation}
  M_{\rm line,vir} = \left[1+\left(\frac{\sigma_{\rm NT}}{c_{\rm s}}\right)^2\right]\times\left[16\,{\rm M_{\odot}}\,{\rm pc}^{-1}\times\left(\frac{T}{10{\,\rm K}}\right)\right],
  \label{eq:virial_line_mass}
 \end{equation}
 where the non-thermal velocity dispersion $\sigma_{\rm NT}$ is defined by
 \begin{equation}
 \sigma_{\rm NT} = \sqrt{\frac{{\rm \Delta} v^2}{\rm 8ln2} - \sigma_{\rm T}^2} = \sqrt{\sigma_{\rm obs}^2-\frac{kT_{\rm kin}}{\mu m_{\rm H}}}.
  \label{eq:NT_velocity}
 \end{equation}
The $\sigma_{\rm obs}$ is the observed velocity dispersion and  $\sigma_{\rm T}$ is the thermal broadening defined as $(kT_{\rm kin}/\mu m_{\rm H})^{1/2}$, and $\mu$ is the  molecular weight ($\mu=29$ for \ce{^13CO}, and $\mu=30$ for \ce{C^18O}). 
The virial line masses estimated using the \ce{^13CO}\,(1--0) and \ce{C^18O}\,(1--0) data within the boundary of G24 are 2212$\pm$77 and 717$\pm$32\, $\rm M_{\odot}\,\rm pc^{-1}$, respectively. The uncertainties of the line masses is derived from the propagation of the computational uncertainties of $T_{\rm ex}$, $\tau_{\rm 13}$, and velocity dispersion. We will discuss the gravitational stability in Section~\ref{subsec: Gravitational Stability of G24}.

\subsection{Longitudinal Profiles of G24}
\label{subsec: Longitudinal_profile} 

\begin{figure*}[ht!]
  \centering 
   \includegraphics[width=0.7\textwidth]{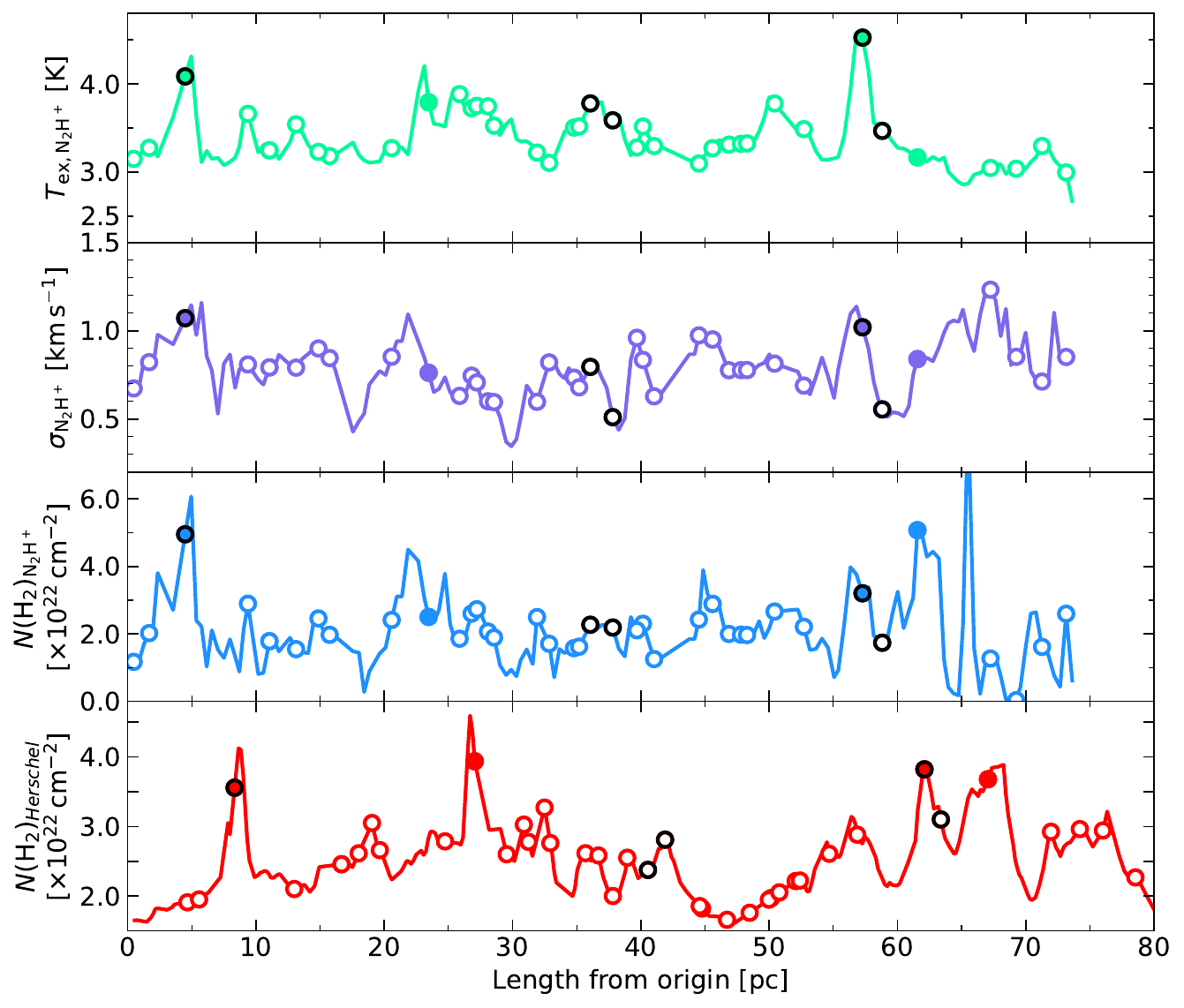}
  \caption{Longitudinal profiles of filament properties based on the \ce{N_2H^+}\,(1--0) data along the G24 skeleton. 
  From top to bottom, the first three panels show the excitation temperature ($T_{\rm ex}$), velocity dispersion ($\sigma$), and  \ce{H2} column density ($N(\ce{H2})_{\ce{N2H+}}$). 
  The \textit{Herschel}-based \ce{H2} column density profile is also presented in the fourth panel for comparison. Empty circles indicate the positions of the 70\,\micron-quiet clumps in the same colour as the lines, while circles with black edges are the 70\,\micron-bright clumps. Four massive clumps with high surface densities (C3, C11, C32, C34) are indicated by solid coloured circles. 
  The galactic longitude decreases from the origin, along the length of the filament. }
  \label{img: longitudinal_fil_profile_N2HP}
 \end{figure*}

\begin{figure}[ht!]
  \centering 
  \includegraphics[width=0.9\columnwidth]{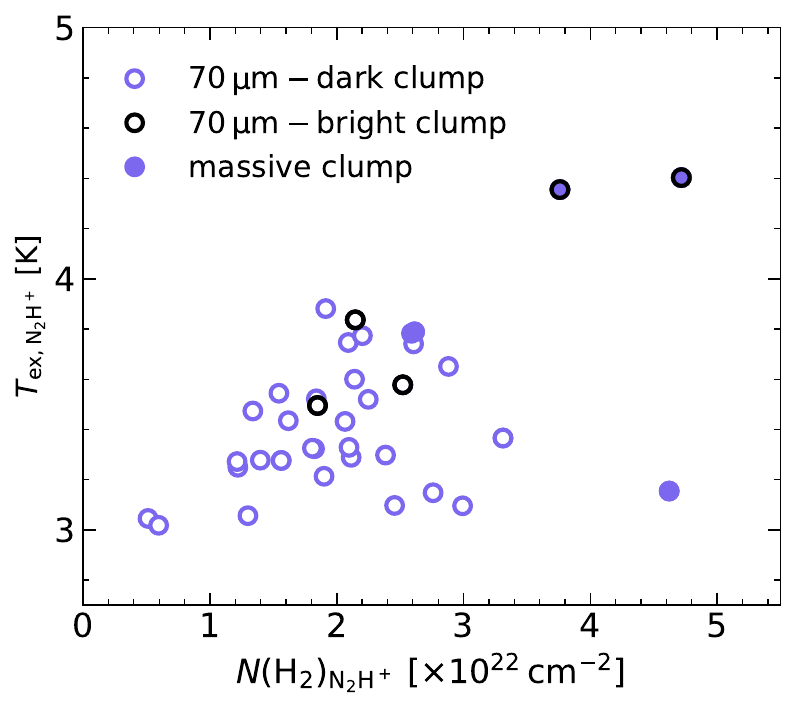}
   \caption{Scatterplot of \ce{N2H+}-based \ce{H2} column density and \ce{N2H+} excitation temperature. Empty circles indicate the positions of the 70\,\micron-quiet clumps, while circles with black edges indicate the 70\,\micron-bright clumps. Four massive clumps with high surface densities (C3, C11, C32, C34) are marked by solid coloured circles.}
  \label{img: longitudinal_fil_profile_N2HP_scatter}
 \end{figure}

 \begin{figure*}[ht!]
  \centering 
  \includegraphics[width=0.86\textwidth]{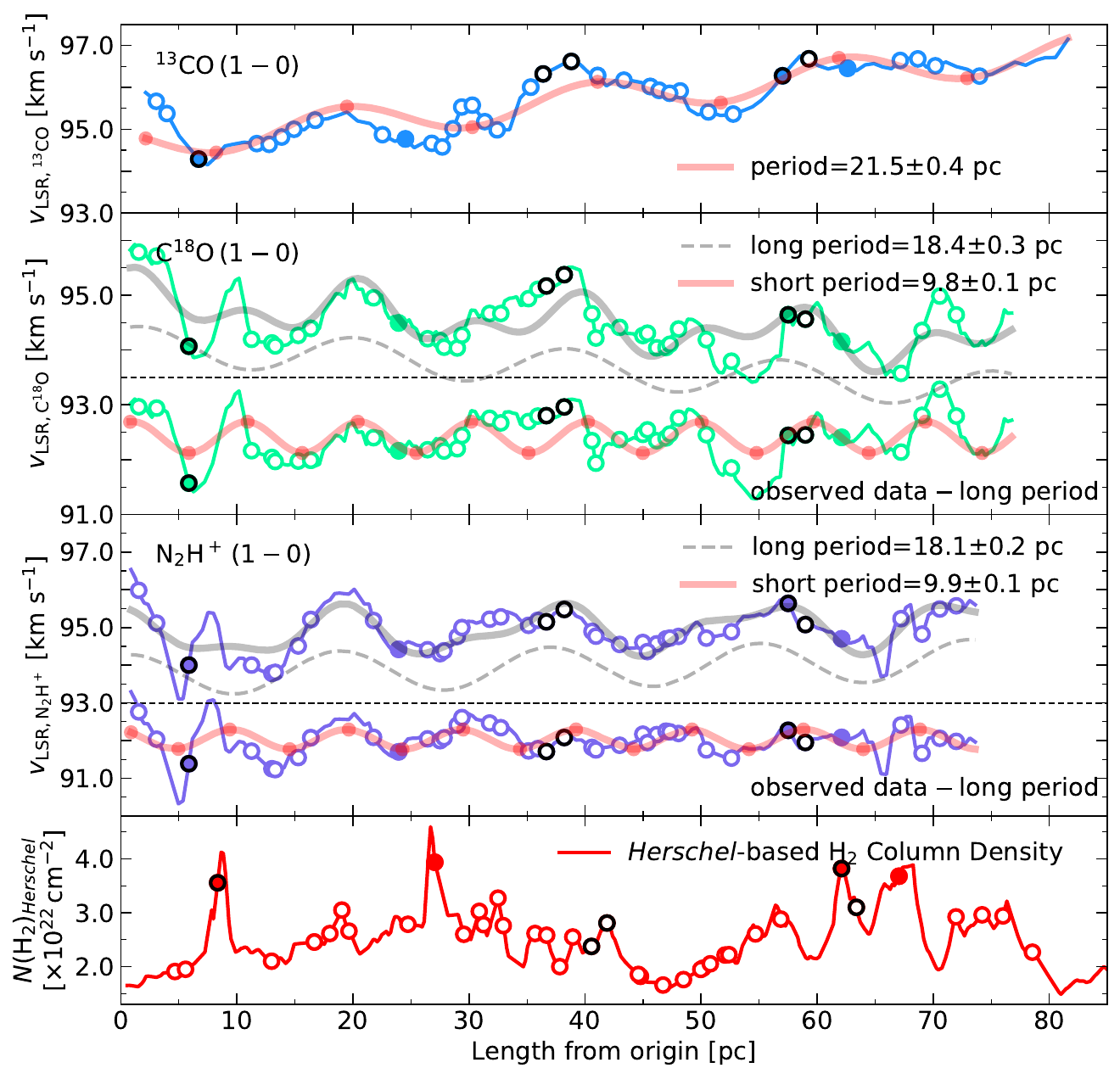}
  \caption{Longitudinal profiles of centroid velocities ($v_{\rm LSR}$) along the skeletons. From top to bottom, the first three panels are $v_{\rm LSR}$ obtained from the \ce{^13CO}, \ce{C^18O} and \ce{N2H+}\,(1--0) data. The \textit{Herschel}-based \ce{H2} column density ($N(\ce{H2})_{Herschel}$) profile is also presented in the 4th panel for comparison. 
 All symbols are the same as in Figure~\ref{img: longitudinal_fil_profile_N2HP}. 
 The $v_{\rm LSR}$ profile of \ce{^13CO} is fitted by a model of single-sinusoidal function, while that of \ce{C^18O} and \ce{N2H+} are fitted by models of double-sinusoidal function, one with a long period and the other with a short period. The translucent grey curve represent the single-sinusoidal model for \ce{^13CO} or the double-sinusoidal model for the \ce{C^18O} and \ce{N2H+} data. In the 2nd and 3rd panels, the grey dashed curve depicts the profile of the long-period component of the double-sinusoidal model shifted lower than the original profile on the y-axis. The $v_{\rm LSR}$ profile with the same colour of each molecule, separated by a horizontal dashed line,  is the residual after removing the long-period component from the original profile. The translucent red curve superimposed on it is the short-period sinusoidal model, with translucent red dots used to delineate oscillation segments. The period of each sinusoidal component is labelled at the upper right corner. }
  \label{img: longitudinal_velocity_gradient_1}
 \end{figure*}

 \begin{figure}[ht!]
  \centering 
  \includegraphics[width=1.0\columnwidth]{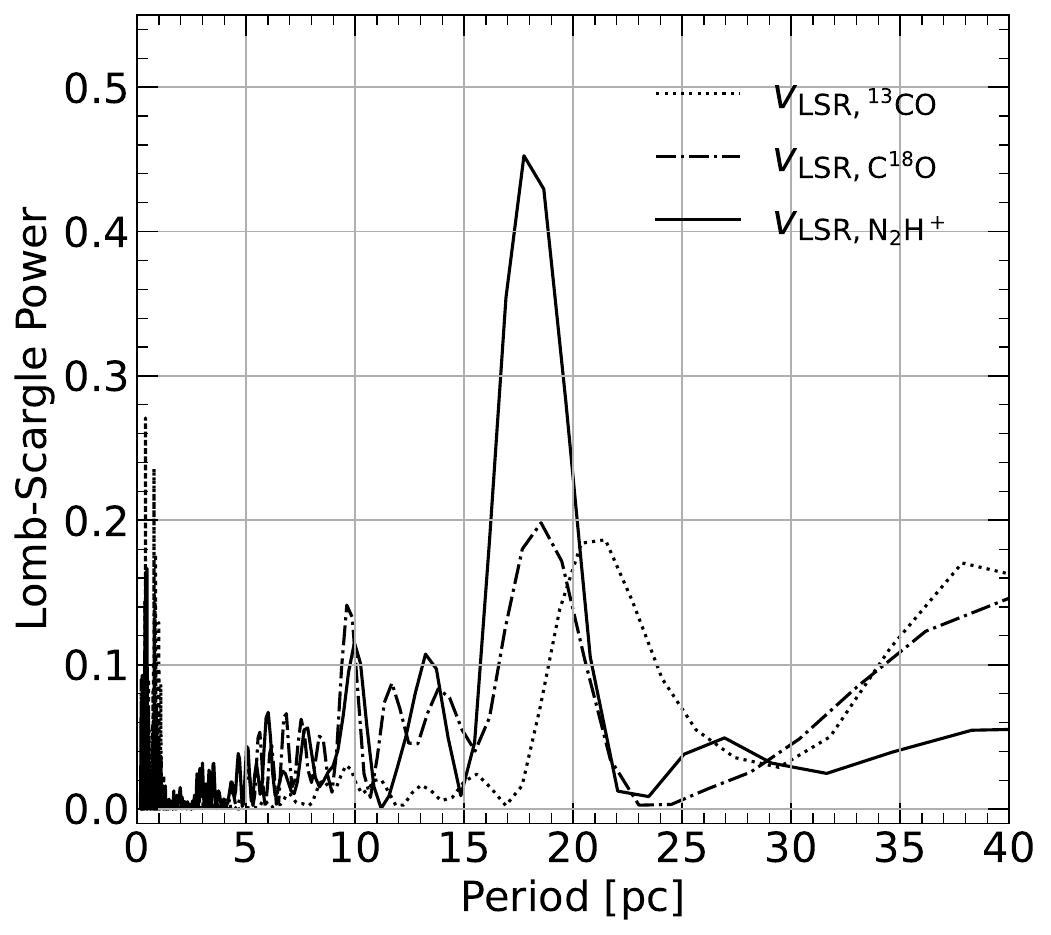}
  \caption{Lomb-Scargle periodogram of longitudinal profile of centroid velocity ($v_{\rm LSR}$). The dotted, dot-dashed and solid line indicate the \ce{^13CO}, \ce{C^18O} and \ce{N2H+}\,(1--0) data, respectively.}
  \label{img: longitudinal_velocity_gradient_2}
 \end{figure}

To investigate the variation of properties of G24 along the longitudinal direction, we extracted the longitudinal profile of the excitation temperature, $T_{\rm ex,\ce{N2H+}}$, the velocity dispersion $\sigma_{\ce{N2H+}}$ and  \ce{H2} column density $N(\ce{H2})_{\ce{N2H+}}$ obtained from the \ce{N2H+}\,(1--0) data, as presented in Figure~\ref{img: longitudinal_fil_profile_N2HP}. The profile of the \textit{Herschel}-based \ce{H2} column density is also presented in the 4th panel for comparative purposes. In each panel, the points are extracted from each transverse radial intensity profile determined by \texttt{RadFil} in Section~\ref{subsubsec: Width of G24}.  
Although the massive clumps (indicated by solid coloured circles) are appear to be located at positions with high \textit{Herschel}-b \ce{H2} column density, there appears to be little correlation between the location of these clumps, the  70\,\micron-bright clumps (circles with black edges), or the other clumps with the \ce{N2H+} column density. However, 
interestingly, more than half of the clumps are around the local maximum velocity dispersion, indicating there may be more non-thermal motions, such as gravitational infall inside the clumps, broadening the velocity dispersion. Also, the 70\,\micron-bright clumps appear to preferentially occur at peaks in the excitation temperature. 

Figure~\ref{img: longitudinal_fil_profile_N2HP_scatter} shows the distribution of G24 clumps with respect to $N(\ce{H2})_{\ce{N2H+}}$ and $T_{\rm ex, \ce{N2H+}}$. The T$_{\rm ex}$ values of the 70\,\micron-bright clumps (circles with black edges) increase as their $N$(\ce{H2})$_{\ce{N2H+}}$ increase. 
In contrast, the 70\,\micron-dark clumps do not show any correlation between these two quantities and have relatively low $T_{\rm ex}$ values compared with the bright clumps.  This trend suggests that the embedded heating sources are contributing to an increase in temperature of the bright sources.

\subsubsection{Velocity profiles}

To investigate the velocity structure along the filament
Figure~\ref{img: longitudinal_velocity_gradient_1} presents the longitudinal profiles of centroid velocity ($\upsilon_{\rm LSR}$) for \ce{^13CO}\,(1--0), \ce{C^18O}\,(1--0), and \ce{N2H+}\,(1--0). The profile of the $N(\ce{H2})_{Herschel}$ is also shown for comparison. 

The longitudinal profiles of the centroid velocity exhibit periodic oscillation patterns. To investigate this structure we have used the Lomb-Scargle method, which is a powerful tool for analysing irregularly sampled time series data to identify frequencies and periodicities. The method constructs a power spectrum using sinusoidal and cosinusoidal functions and evaluates how well these functions fit the data at different frequencies. A frequency with a significantly higher power value indicates the presence of a periodic signal near that frequency. 
As presented in the Figure~\ref{img: longitudinal_velocity_gradient_2}, the power spectrum obtained from the \ce{^13CO}\,(1--0) $\upsilon_{\rm LSR}$ (dotted line) shows one significant peak at 21.5\,pc, while that of the \ce{C^18O}\,(1--0) data shows two peaks, one at 9.6\,pc and a second, higher peak at 18.5\,pc. 
The periodogram of the \ce{N2H+}\,(1--0) data presents similar features as that of the \ce{C^18O} data. The fundamental frequency is at 18.7\,pc, and a weaker secondary peak at 10.0\,pc can also be seen. 

We used false alarm probability (FAP) to assess the statistical significance of the periods (peaks) detected in the periodogram. 
The definition of FAP is such that under the null hypothesis of no periodic signal can be detected in the observed data, the probability that the maximum power generated by random noise is greater than, or equal to, the target observed power \citep{Baluev_2008MNRAS.385.1279B}. 
That is $\rm FAP=P(P_{max, noise}\geqslant P_{max, obs} | H_0)$, where $\rm P_{max, obs}$ is the maximum power in the observed periodogram and $\rm P_{max, noise}$, the maximum power in the periodogram of a noise signal. 
We used the bootstrap method to generate the periodogram of a noise signal. 
The method eliminates potential periodicity by randomly permuting the observed data, while at the same time retaining the original observed time series (longitudinal length in this study). 
Specifically, we sampled the observed longitudinal $v_{\rm LSR}$ data with its original values and length, and then generated a Lomb-Scargle periodogram using the shuffled data. We repeated the step 5000 times and chose the maximum power of each periodogram, and then generated an empirical distribution of the maximum power. 
The FAP of the periods at $\sim19$\,pc for both the \ce{C^18O} and \ce{N2H+} longitudinal $v_{\rm LSR}$ profile are $<2\times10^{-4}$, and that at approximately $\sim10$\,pc for the \ce{C^18O} data is also $<2\times10^{-4}$ while for the \ce{N2H+} data is 0.03. These values are smaller than 0.05, a typical value for rejecting the null hypothesis, indicating the extreme significance of these periods. 
For comparison, the statistical significance at the 6.79\,pc of the \ce{C^18O} periodogram is 0.16, and that at 6.0\,pc of the \ce{N2H+} periodogram is 0.73, which are insignificant because of large FAP and is possibly a spurious data noise. 

According to the periods, the longitudinal profiles of $\upsilon_{\rm LSR}$ were fitted by a model of single- (for \ce{^13CO}) or double-sinusoidal function (for \ce{C^18O} and \ce{N2H+}), and the latter consists of a long period of $\sim20$\,pc and a short period of $\sim10$\,pc. The fitted results are indicated by grey curves in the first three panels in Figure~\ref{img: longitudinal_velocity_gradient_1}. 
In the 2nd and 3rd panels, the long-period component of the double-sinusoidal function model is depicted by a grey dashed curve shifted lower than the original profile on the y-axis. The residual after removing the long-period component from the original profile and the fitted short-period component are also presented. 
The periods obtained from the models are consistent with those in the periodogram. For $\upsilon_{\rm LSR}$ of the \ce{^13CO} data the period is $21.5\pm0.4$\,pc. For $\upsilon_{\rm LSR}$ of the \ce{C^18O} data, two periods were obtained---a long period of $18.4\pm0.3$\,pc and a short period of $9.8\pm0.1$\,pc. The periods for $\upsilon_{\rm LSR}$ of the \ce{N2H+} data closely resemble those for \ce{C^18O}, with the long period being $18.1\pm0.2$\,pc and the short period being $9.9\pm0.1$\,pc. 
The long periods of \ce{C^18O} and \ce{N2H+} are much longer compared with the theoretical length corresponding to the maximum growth rate of a perturbation to a self-gravitating filament 
(Section~\ref{subsec: Fragmentation of G24}) indicating that the velocity field is not dominated by flows directly associated with the fragmentation, but by a larger scale component.

The \ce{^13CO} longitudinal $v_{\rm LSR}$ profile is distinctly different from  that of  \ce{C^18O} and \ce{N2H+}. This is likely due to the lower critical density of the \ce{^13CO} transition resulting in it predominantly tracing the less dense, extended gas surrounding the filament which is subject to different influences that which set its velocity structure compared to the inner part of the filament traced by \ce{C^18O} and \ce{N2H+}.

The formation mechanism of the long periods will be explored and discussed in detail as an independent, dedicated paper. 
Here, the segment of oscillation is determined based on the local maxima and minima of the short period of the double-sinusoidal function model. The beginning and end of each segment are marked with translucent red dots in the 2nd and 3rd panels of Figure~\ref{img: longitudinal_velocity_gradient_1}. 
Finally, we obtained 8 segments of oscillation for the \ce{^13CO}, 16 segments for the \ce{C^18O} and 15 segments for the \ce{N2H+} data.

The velocity along the filament may reflect transport and accretion of material onto clumps. We therefore calculated accretion rates for defined segments by assuming a simple cylindrical model taken from \cite{Kirk_2013ApJ...766..115K}, $\dot{M}=\frac{\triangledown V_{||}M}{\rm tan(\alpha)}$, where $\triangledown V_{||}$ is the velocity gradient of a segment of oscillation, $M$ is the mass of the segment, and $\alpha$ is the inclination angle to the plane of the sky and is assumed to be 45$^{\circ}$. The lengths, velocity differences ($|\upsilon_{\max} - \upsilon_{\rm min}|$), velocity gradients, masses, and accretion rates of these segments are listed in columns (1)--(12) of Table~\ref{tab: Oscillaiton_segments} for the \ce{^13CO} and \ce{C^18O} and Table~\ref{tab: Oscillaiton_segments2} for the \ce{N2H+} data. We note that the length of oscillation segments may vary slightly, as ideal equidistant points (translucent red points in Figure~\ref{img: longitudinal_velocity_gradient_1}) may not correspond to observed data points. In such cases, the nearest observed data points were selected. 

\begin{table*}[ht!]
  \centering
  \small
  \caption{Properties of oscillation segments based on the \ce{^13CO} and \ce{C^18O}\,(1--0) data along G24.}
  \label{tab: Oscillaiton_segments}
  \setlength{\tabcolsep}{0.6mm}{
  \begin{tabular}{clccccccccccc}
  \toprule[1pt]
    Molecule & num & Segment  & Length & $|\upsilon_{\rm max} - \upsilon_{\rm min}|$ & $\triangledown V_{||}$ & Mass & $\dot{M}$ & $T_{\rm dust}$ & $\sigma_{\rm \upsilon}$ & $M_{\rm line}$ & $M_{\rm line,vir}$ & $\alpha_{\rm line,vir}$   \\
     &     & [pc, pc]    &    [pc]    &    [\velouni]    &    [\velouni    &    [$\rm \times 10^3$   & [$\rm \times 10^3\,M_{\odot}$ & [K] & [\velouni] & [$\rm M_{\odot} \,pc^{-2}$] & [$\rm M_{\odot} \,pc^{-2}$]  \\
     &     &     &     &     &  $\rm pc^{-1}$]  &  $\rm M_{\odot}$]    &    $\rm Myr^{-1}$] &   \\
    (1) & (2) & (3) & (4) & (5) & (6) & (7) & (8) & (9) & (10) & (11) & (12) & (13)  \\
  \midrule[0.5pt]
  \ce{^13CO}    & 1   & {[}2.1, 8.2]   & 6.1 & 1.53 & 0.30 & 5.10 & 1.52 & 18.9 & 1.71 & 1078$\pm$107 & 1353$\pm$126 & $1.26^{+1.26}_{-0.63}$       \\
  (1--0)     & 2   & {[}8.2, 19.5] & 11.3   & 1.07  & 0.08 & 18.8 & 2.60 & 18.9 & 1.20 & 1651$\pm$20 & 679$\pm$28 & $0.41^{+0.41}_{-0.21}$       \\
         & 3   & {[}19.5, 30.3]  & 10.8    & 0.16  & 0.003 & 7.56 & 0.02 & 17.4 & 1.24 & 664$\pm$28 & 719$\pm$123 & $1.08^{+1.08}_{-0.54}$       \\
         & 4   & {[}30.3, 41.1] & 10.8    & 0.71  & 0.14  & 12.73 & 1.84 & 18.7 & 1.61 & 1193$\pm$26 & 1193$\pm$54 & $1.00^{+1.00}_{-0.50}$       \\
         & 5   & {[}41.1, 51.7]   & 10.6    & 0.96  & 0.09  & 10.22 & 0.94 & 18.3 & 1.45 & 1051$\pm$17 & 1015$\pm$70 & $0.97^{+0.97}_{-0.48}$       \\
         & 6   & {[}51.7, 61.9]   & 8.2    & 1.18 & 0.15 & 13.32 & 2.03 & 18.0 & 2.50 & 1210$\pm$11 & 2842$\pm$89 & $2.35^{+2.35}_{-1.17}$       \\
         & 7   & {[}61.9, 72.9] & 11.0    & 0.07  & 0.004 & 8.88 & 0.04 & 18.2 & 2.34 & 792$\pm$17 & 2494$\pm$81 & $3.15^{+3.15}_{-1.57}$       \\
         & 8   & {[}72.9, 81.6]  & 8.7    & 0.72 & 0.08  & 6.81 & 0.52 & 19.2 & 2.04 & 647$\pm$9 & 1913$\pm$176 & $2.96^{+2.96}_{-1.48}$       \\
  \midrule[0.5pt]
     \ce{C^18O}     & 1   & {[}0.8, 5.9]  & 5.1    & 1.76   & 0.31 & 5.01 & 1.54 & 18.9 & 0.96 & 875$\pm$11 & 443$\pm$96 & $0.51^{+0.51}_{-0.25}$        \\
    (1--0)   & 2   & {[}5.9, 10.9]  & 5.0    & 0.40 & 0.26 & 2.99 & 0.78 & 19.1 & 1.25 & 846$\pm$12 & 730$\pm$19 & $0.86^{+0.86}_{-0.43}$        \\
         & 3   & {[}10.9, 15.6]  & 4.7    & 0.13 & 0.003 & 5.51 & 0.02 & 19.2 & 0.82 & 1582$\pm$8 & 334$\pm$16 & $0.21^{+0.21}_{-0.11}$        \\
         & 4   & {[}15.6, 20.4] & 4.8    & 0.77  & 0.22 & 5.76 & 1.26 & 18.1 & 0.82 & 665$\pm$7 & 331$\pm$43 & $0.50^{+0.50}_{-0.25}$        \\
         & 5   & {[}20.4, 25.5]  & 5.1    & 0.73 & 0.15 & 3.17 & 0.48 & 16.9 & 0.88 & 764$\pm$5 & 379$\pm$14 & $0.50^{+0.50}_{-0.25}$        \\
         & 6   & {[}25.5, 30.2]  & 4.7   & 0.35  & 0.03 & 3.16 & 0.10 & 17.4 & 0.84 & 702$\pm$16 & 342$\pm$64 & $0.49^{+0.49}_{-0.24}$        \\
         & 7   & {[}30.2, 35.1]   & 4.9    & 0.21 & 0.06  & 3.07 & 0.18 & 18.4 & 0.84 & 644$\pm$7 & 343$\pm$32 & $0.53^{+0.53}_{-0.27}$        \\
         & 8   & {[}35.1, 40.3]   & 5.1    & 0.07 & 0.042 & 4.19 & 0.18 & 18.1 & 0.76 & 845$\pm$3 & 285$\pm$19 & $0.34^{+0.34}_{-0.17}$        \\
         & 9   & {[}40.3, 45.0]   & 4.7    & 0.60 & 0.08 & 3.34 & 0.25 & 18.2 & 0.94 & 858$\pm$4 & 429$\pm$22 & $0.50^{+0.50}_{-0.25}$        \\
         & 10  & {[}45.0, 50.1] & 5.1    & 0.17 & 0.08 & 4.86 & 0.39 & 17.7 & 1.03 & 827$\pm$3 & 505$\pm$33 & $0.61^{+0.61}_{-0.31}$        \\
         & 11  & {[}50.1, 54.8]  & 4.7    & 1.01 & 0.17 & 2.93 & 0.49 & 17.8 & 0.83 & 877$\pm$4 & 339$\pm$22 & $0.39^{+0.39}_{-0.19}$        \\
         & 12  & {[}54.8, 59.7]   & 4.9    & 1.35  & 0.29 & 4.10 & 1.19 & 17.6 & 1.15 & 820$\pm$8 & 626$\pm$24 & $0.76^{+0.76}_{-0.38}$        \\
         & 13  & {[}59.7, 64.5]   & 4.8    & 0.64 & 0.17 & 4.37 & 0.75 & 17.7 & 1.16 & 783$\pm$5 & 637$\pm$26 & $0.81^{+0.81}_{-0.41}$        \\
         & 14  & {[}64.4, 69.3]   & 4.9    & 0.37 & 0.09 & 4.42 & 0.38 & 18.2 & 0.76 & 520$\pm$5 & 290$\pm$61 & $0.56^{+0.56}_{-0.28}$        \\
         & 15  & {[}69.3, 74.2]   & 4.9    & 0.34 & 0.17 & 1.97 & 0.33 & 18.2 & 0.86 & 886$\pm$6 & 359$\pm$22 & $0.41^{+0.41}_{-0.20}$        \\
         & 16  & {[}74.2, 76.8]   & 2.6    & 0.51  & 0.26 & 1.05 & 0.28 & 18.7 & 1.16 & 100$\pm$1 & 638$\pm$49 & $6.39^{+6.39}_{-3.20}$        \\
  \bottomrule[1pt]
  \end{tabular}}
 \raggedright
 \textbf{Notes.}The meaning of each column is: (1) Molecular tracer used for obtaining properties of the oscillation segments. (2) The number of each segment. (3) The length of G24 corresponding to the start point and end point of each oscillation segment. (4) The length of each oscillation segment. (5) The velocity difference between the start point and end point of each oscillation segment. (6) The velocity gradient, which is obtained as the slope of a linear regression fitting of all centroid velocity points within one oscillation segment. (7) The mass of each oscillation segment. (8) The mass accretion rate of each oscillation segment. (9) The excitation temperature $T_{\rm ex}$ of each oscillation segment, replaced by dust temperature. (10) The velocity dispersion of each oscillation segment, obtained by multiple-component Gaussian fitting (introduced in Appendix~\ref{appendix: Spectral Line Fitting}) for the spectral line averaged over all lines within a section. (11) The line mass $M_{\rm line}$ of each oscillation segment. (12) The virial line mass $M_{\rm line, vir}$ of each oscillation segment. (13) The virial parameter defined as $\alpha_{\rm line,vir} = M_{\rm line, vir}/M_{\rm line}$ in Section~\ref{subsec: Gravitational Stability of G24}. The uncertainty is a factor of 2. 
\end{table*}
\begin{table}
  \centering
  \small
  \caption{Properties of oscillation segments based on the \ce{N2H+}\,(1--0) data along G24. }
  \label{tab: Oscillaiton_segments2}
  \setlength{\tabcolsep}{0.2mm}{
  \begin{tabular}{clccccc}
  \toprule[1pt]
 num & Segment & Length & $|\upsilon_{\rm max} - \upsilon_{\rm min}|$  & $\triangledown V_{||}$   & Mass     & $\dot{M}$  \\
     & [pc, pc]    &  [pc]    &    [\velouni]    &    [\velouni    &    $\rm [\times 10^3$   &    [$\rm \times 10^3\,M_{\odot}$    \\
     &     &     &     &  $\rm pc^{-1}$]  &  $\rm M_{\odot}$]    &    $\rm Myr^{-1}$]   \\
    (1) & (2) & (3) & (4) & (5) & (6) & (7) \\
  \midrule[0.5pt]
   1   & {[}0.9, 5.0]  & 4.1    & 3.43   & 0.80  & 1.73 & 1.39    \\
   2   & {[}5.0, 9.4]   & 4.4    & 0.80  & 0.24  & 5.72 & 1.34    \\
   3   & {[}9.4, 14.5]   & 5.2    & 0.54 & 0.007 & 1.63 & 0.01     \\
   4   & {[}14.5, 19.6]  & 5.1    & 1.45 & 0.35 & 2.54 & 0.88  \\
   5   & {[}19.6, 24.2] & 4.6    & 1.36 & 0.40  & 1.28 & 0.51   \\
   6   & {[}24.2, 29.4]  & 5.2    & 0.55 & 0.04 & 6.95 & 0.30     \\
   7   & {[}29.5, 34.4]   & 4.9    & 0.14 & 0.05  & 5.65 & 0.28   \\
   8  & {[}34.4, 39.2]     & 4.8    & 0.11 & 0.07 & 1.60 & 0.12   \\
   9  & {[}39.2, 44.2] & 5.0    & 0.85  & 0.17 & 1.73 & 0.30      \\
   10  & {[}44.2, 49.2]   & 5.0    & 0.58  & 0.10  & 3.84 & 0.39  \\
   11  & {[}49.2, 54.1]    & 4.9    & 0.26 & 0.04  & 2.17 & 0.09 \\
   12  & {[}54.1, 58.8]   & 4.7    & 0.24 & 0.03 & 6.18 & 0.17  \\
   13  & {[}58.8, 64.0]  & 5.2    & 0.28 & 0.09 & 7.25 & 0.69     \\
   14  & {[}64.0, 68.9]   & 4.9    & 0.09  & 0.20 & 5.94 & 1.21  \\
   15  & {[}68.9, 73.7]  & 4.8    & 0.92 & 0.15  & 6.44 & 0.94    \\
  \bottomrule[1pt]
  \end{tabular}}
\end{table}

\begin{figure}
  \centering 
  \includegraphics[width=1.0\columnwidth]{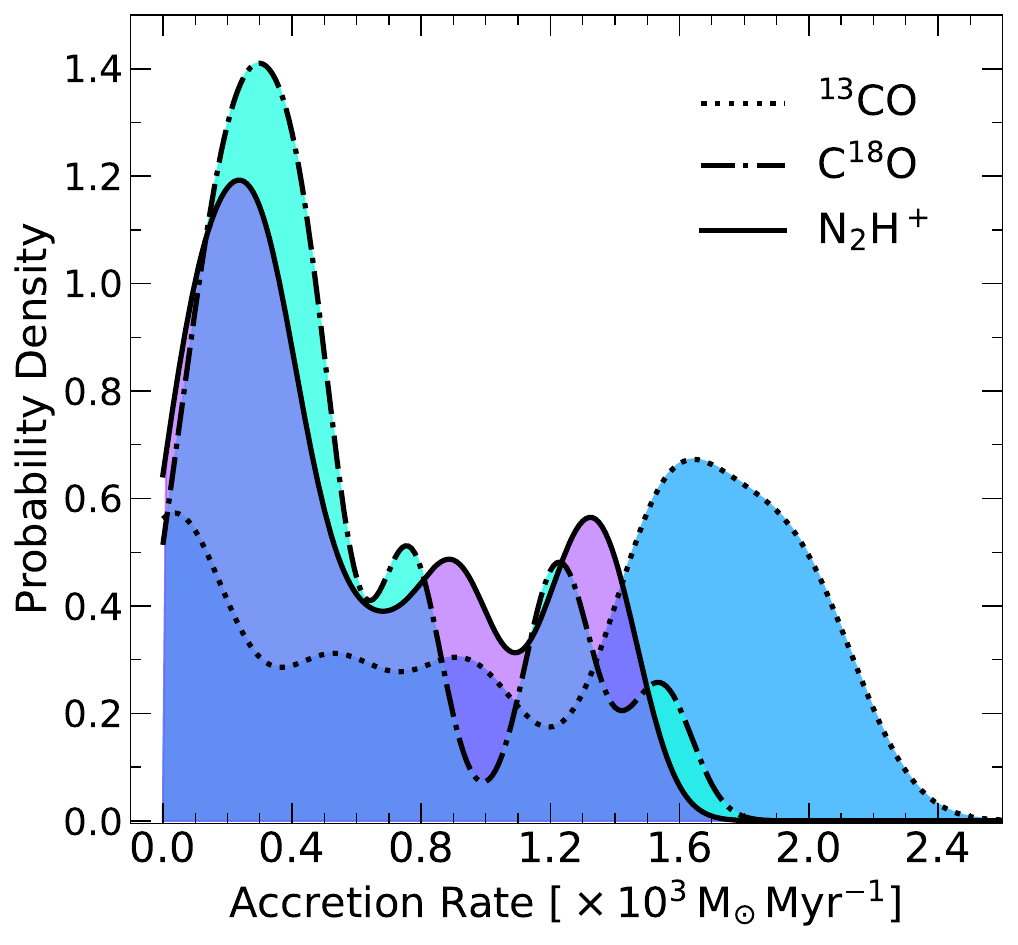}
  \caption{KDEs of the accretion rates generated with the \ce{^13CO} (dotted curve filled with blue), \ce{C^18O} (dotted-dashed curve filled with cyan), and \ce{N2H+} (solid curve filled with purple) $J$=1--0 data.}
  \label{img: longitudinal_velocity_gradient_hist}
 \end{figure}

For the \ce{^13CO} data, the length of the oscillation segment (half of the fitted period) is 10.75$\pm$0.20\,pc. The mean/median observed velocity gradient is 0.11/0.09\,\velouni\,pc$^{-1}$, and the mean/median accretion rate is $1.06/1.23 \times 10^3\,\rm M_{\odot}\,\rm Myr^{-1}$. 
For the \ce{C^18O} data, the length of the segments is 4.88$\pm$0.05\,pc. The mean/median observed velocity gradient is 0.15/0.16\,\velouni\,pc$^{-1}$, and the mean/median accretion rate is $0.54/0.39 \times 10^3\,\rm M_{\odot}\,\rm Myr^{-1}$. 
The velocity gradients and accretion rates of the \ce{N2H+} data show similar ranges to those from the \ce{C^18O} data. Within the similar lengths of the segments (4.94$\pm$0.06\,pc), the mean/median velocity gradient is 0.18/0.10\,\velouni\,pc$^{-1}$. The mean/median rate of accretion is $\rm 0.57/0.39 \times 10^3\,M_{\odot}\,Myr^{-1}$. We remind the reader that the velocity gradient was determined by the slope of a linear regression fit for all centroid velocity points within a segment, rather than the ratio of the velocity difference to the length. 

To present the distribution of accretion rate estimated with different line tracers, we generated the kernel density estimations (KDE) (Figure~\ref{img: longitudinal_velocity_gradient_hist}). We used a Gaussian kernel with a bandwidth of 0.22 for the KDEs. The bandwidth was calculated by cross-validation, which is a statistical method helping in selecting optimal parameters, using \texttt{GridSearchCV} function of the Python \texttt{sklearn} library. The KDEs of the \ce{C^18O} and \ce{N2H+} data show multiple peaks of probability density but with a single dominant one corresponding to the accretion rate of 0.30 and $\rm 0.24 \times 10^3\,M_{\odot}\,Myr^{-1}$ for the \ce{C^18O} and \ce{N2H+} data, respectively. 
While the KDE of the \ce{^13CO} data has two nearly equal probability density peaks, corresponding to the accretion rates of 0.04 and $1.64 \times 10^3\,\rm M_{\odot}\,Myr^{-1}$, respectively. 
The accretion rates of the dominant peaks in the KDEs of the \ce{C^18O} and \ce{N2H+} data are approximately one order of magnitude greater than that found in low-mass star formation regions, e.g., a 0.33\,pc-long filament in Serpens South with an accretion rate of $\rm \sim30\,M_{\odot}\,Myr^{-1}$ \citep{Kirk_2013ApJ...766..115K}, and are several times greater than those found in high-mass star formation regions, e.g., Monoceros R2 with an accretion rate of $\rm \sim100\,M_{\odot}\,Myr^{-1}$ \citep{Trevino-Morales_2019AnA...629A..81T}, and Orion with that of $\rm \sim55\,M_{\odot}\,Myr^{-1}$ for fibers in the NGC1333 proto-cluster \citep{Hacar_2017A&A...602L...2H}, and are also larger than that of some giant molecular filaments, for example, the $\sim72$\,pc California molecular filament with an accretion rate range of $\rm \sim20-101\,M_{\odot}\,Myr^{-1}$ obtained from the \ce{^13CO}\,(1--0) data \citep{Guo_2021ApJ...921...23G}. 
While the accretion rate ($\rm 0.04 \times 10^3\,M_{\odot}\,Myr^{-1}$) of one of the peaks in the KDE of the \ce{^13CO} data is comparable to the values reported in the aforementioned works. 

On the other hand, intriguingly, the KDEs show secondary distinct peaks for the \ce{C^18O} and \ce{N2H+} data at $\rm 1.23\times 10^3\,M_{\odot}\,Myr^{-1}$ and $\rm 1.32\times10^3\,M_{\odot}\,Myr^{-1}$, respectively, which are close to the other peak ($\rm 1.64 \times 10^3\,M_{\odot}\,Myr^{-1}$) of KDE for the \ce{^13CO} data. These accretion rates are much higher than those reported in the aforementioned works. 
Since the velocity differences measured in Table~\ref{tab: Oscillaiton_segments} and \ref{tab: Oscillaiton_segments2} are similar to those reported in the aforementioned works, the high accretion rates exceeding $\rm 1\times 10^3\,M_{\odot}\,Myr^{-1}$ are due to greater masses and higher \ce{H2} column densities within the segments. 
Most of these segments host massive clumps, indicating active accretion activities within them.

 \begin{figure*}
  \centering 
  \includegraphics[width=0.8\textwidth]{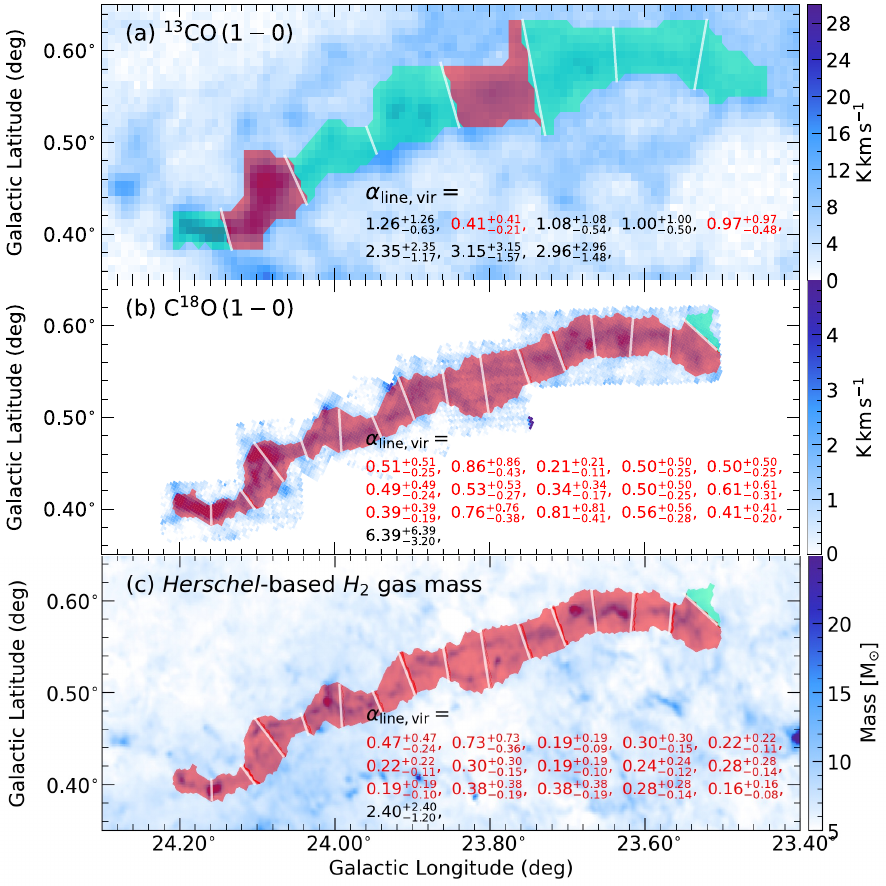}
  \caption{The virial parameter ($\alpha_{\rm line, vir}=M_{\rm line, vir}/M_{\rm line}$) derived from the \ce{^13CO}\,(1--0),  \ce{C^18O}\,(1--0) and \textit{Herschel}-based \ce{H2} column density data. The filament is divided into several segments based on the oscillation of centroid velocity in the 1st and 2nd panels of Figure~\ref{img: longitudinal_velocity_gradient_1}. 
  Taking a factor of 2 as the uncertainty, the segments with $\alpha_{\rm line, vir}\times2<2.0$ are indicating gravitationally bound, and filled in with translucent red, otherwise the segments are gravitationally unbound and filled with translucent green. The $\alpha_{\rm line, vir}$ are shown at the bottom of each panel.
  (a) The $\alpha_{\rm line, vir}$ derived from the \ce{^13CO}\,(1--0) data. The background is the integrated intensity map of \ce{^13CO} over a velocity interval from 90.0\,\velouni\ to 99.5 \velouni. 
  (b) The $\alpha_{\rm line, vir}$ derived from the \ce{C^18O}\,(1--0) data. The background is the integrated intensity map of \ce{C^18O} with the velocity range of 90.5--99.0 \velouni. 
  (c) The $\alpha_{\rm line, vir}$ with the mass of segments derived from \textit{Herschel}-based \ce{H2} gas mass, and the lengths and $M_{\rm line, vir}$ derived from the \ce{C^18O}\,(1--0) data. The background is the \textit{Herschel}-based gas mass map derived from \ce{H2} column density. }
  \label{img: Virial_line_mass}
 \end{figure*}

\subsection{Gravitational Stability of G24} 
\label{subsec: Gravitational Stability of G24}

As mentioned in Section~\ref{subsubsec: Line Mass of G24}, the critical mass of the G24 filament is smaller than the line masses (see Table~\ref{tab: filament_properties}), indicating that the G24 is radially unstable. Thus, we examined whether the G24 filament is truly out of the gravitationally stable condition \citep[e.g.,][]{Arzoumanian_2013AnA...553A.119A} by deriving the virial parameter, $\alpha_{\rm line,vir}$, which is determined by the mass ratio $M_{\rm line,vir}/M_{\rm line}$. A source with $\alpha_{\rm line,vir}>2$ is gravitationally unbound due to strong thermal or turbulent motions preventing gravity collapse, while it is gravitationally bound if $\alpha_{\rm line,vir}<2$. $\alpha_{\rm line,vir}\simeq1$ indicates that a source is in the virial equilibrium state, and $\alpha_{\rm line,vir}<<1$ implies the dominance of gravity. 
The virial line mass ($M_{\rm line, vir}$) obtained with the \ce{^13CO}\,(1--0) data is 2212$\pm$77\,M$_{\rm \odot}\,\rm pc^{-1}$, approximately 2 times higher than the line mass ($M_{\rm line, vir}$) 1160$\pm$13\,M$_{\rm \odot}\,\rm pc^{-1}$, indicating that G24 is gravitationally unbound. However, the $\alpha_{\rm line,vir}$ estimated by the \ce{C^18O}\,(1--0) data presents an opposite result that the $M_{\rm line, vir}$,  717$\pm$32M$_{\rm \odot}\,\rm pc^{-1}$ is nearly equal to the $M_{\rm line, vir}$ of 837$\pm$2\,\,M$_{\rm \odot}\,\rm pc^{-1}$, resulting $\alpha_{\rm line,vir}=0.86\pm0.04$ and being close to virial equilibrium. 
We also estimated the virial parameter with $M_{\rm line}$ obtained from the\textit{Herschel}-based \ce{H2} column density ($N(\ce{H2})_{Herschel}$) map. 
Comparing it with $M_{\rm line, vir}$ obtained from the \ce{C^18O} data leads to $\alpha_{\rm line,vir}=0.52\pm0.03$. 
We remind the reader that the observed line masses are estimated based on the assumption that the angle of inclination to the plane of the sky is zero.  
\cite{Beaumont_2013ApJ...777..173B} conducted a comparative analysis of the intensity structures in synthetic molecular cloud observations in PPP and PPV space, and found that the measurements of the virial parameter have a larger scatter of 0.3\,dex (a factor of two) because of the projection effect. 
Therefore, when the uncertainty is taken into account, the gas traced by \ce{C^18O} is in a state between virial balance and gravitationally bound, while the $\alpha_{\rm line,vir}$ derived from $N(\ce{H2})_{Herschel}$ and the \ce{C^18O} data indicates a virial equilibrium condition.

In order to investigate whether different parts of the G24 filament are locally, gravitationally bound or not, we divided the filament into several segments based on the oscillation segments of centroid velocity in Figure~\ref{img: longitudinal_velocity_gradient_1} and estimated the virial parameter. Panel (a) and (b) of Figure~\ref{img: Virial_line_mass} shows the $\alpha_{\rm line,vir}$ along G24 obtained from the \ce{^13CO} and \ce{C^18O} data, respectively, and Column (13) of Table~\ref{tab: Oscillaiton_segments} lists the values. 
We also estimated the $\alpha_{\rm line,vir}$ with the mass of segments derived from the \textit{Herschel}-based \ce{H2} gas mass, and the lengths and $M_{\rm line, vir}$ derived from the \ce{C^18O} data (panel (c)). 
When a factor of 2 is taken into account for the uncertainty, the gravitationally bound ($\alpha_{\rm line, vir}\times2<2.0$) and unbound ($\alpha_{\rm line, vir}\times2\geqslant2.0$) segments are colour-coded with translucent red and green, respectively.

Taking a factor of 2 as the uncertainty, with the \ce{^13CO} emission (panel (a)), 7 out of 8 segments have $\alpha_{\rm line,vir}\geqslant 2$, and only 1 segment retains  $\alpha_{\rm line,vir}< 2$. This indicates that most of G24 is likely gravitationally unbound based on the \ce{^13CO} data. 
On the other hand, the $\alpha_{\rm line,vir}$ estimated with \ce{C^18O} (panel (b)) shows 15 segments having $\alpha_{\rm line,vir}<2$ when taking a factor of 2 as the uncertainty, and only one segment has $\alpha_{\rm line, vir}>2$  because of its broad line width. In addition, 11 out of the 15 segments are close to 1, considering their uncertainties. Therefore, the most portion of G24 is gravitationally bound and close to being virial equilibrium based on the \ce{C^18O} data, which is similar to most other massive filaments \citep[e.g.,][]{Arzoumanian_2013AnA...553A.119A, Mattern_2018AnA...619A.166M}. 
As for the $\alpha_{\rm line,vir}$ derived from the $N(\ce{H2})_{Herschel}$ and \ce{C^18O} data, 9 segments have $\alpha_{\rm line,vir}>0.25$,  and 7 segments have $\alpha_{\rm line,vir}<0.25$. The latter are still smaller than 0.5 when taking a factor of 2 as the uncertainty. This suggests that a portion of the G24 filament may be dominated by gravity.

\subsection{Fragmentation of G24}
\label{subsec: Fragmentation of G24}
\begin{table*}[ht!]
    \centering
    \caption{Separation of clumps/groups and $p$-values rejecting the null hypothesis that the clumps/groups are randomly located along the filament. The separation was measured by two algorithms: the nearest neighbour (NN) and minimum spanning tree (MST). Four methods are used to estimate the $p$-values: median-interquartile range, mean-standard deviation, Kolmogorov–Smirnov (KS) test and Anderson-Darling (AD) test. The statistical significances of a possible ``two-tier'' fragmentation pattern are calculated by the frequentist approach and the Bayesian approach, and are shown in the last two columns. }
    \label{tab: null hypothesis test}
   \begin{tabular}{cccccccccc}
    \toprule[1pt]
        \multirow{2}{*}{Object} & \multirow{2}{*}{method} & \multicolumn{2}{c}{median-interquartile range} & \multicolumn{2}{c}{mean-standard deviation} & KS test & AD test & Frequentist & Bayesian   \\
    \cline{3-8}
        & & separation [pc] & $p$-value &  separation [pc] & $p$-value & $p$-value & $p$-value & evidence ratio & Bayes factor   \\
    \midrule[0.5 pt]
        \multirow{2}{*}{Clump} & NN & 1.39 & 0.38 & 1.61 & 0.66 & 0.05 & 0.15 & ``Two-tier'' & ``Two-tier''   \\
         & MST & 1.91 & 0.66 & 2.20 & 0.36 & 0.62 & 0.25 & 541 & 11   \\
    \midrule[0.5 pt]
        \multirow{2}{*}{Group} & NN & 2.75 & 0.04 & 2.78 & 0.004 & 0.0001 & 0.001 & ``Single-tier'' & ``Single-tier''   \\
         & MST & 3.46 & 0.05 & 3.68 & 0.04 & 0.02 & 0.05 & 2.4 & 1.2   \\
    \bottomrule[1pt]
    \end{tabular}
\end{table*}

 \begin{figure*}[ht!]
    \centering
	\includegraphics[width=0.9\textwidth]{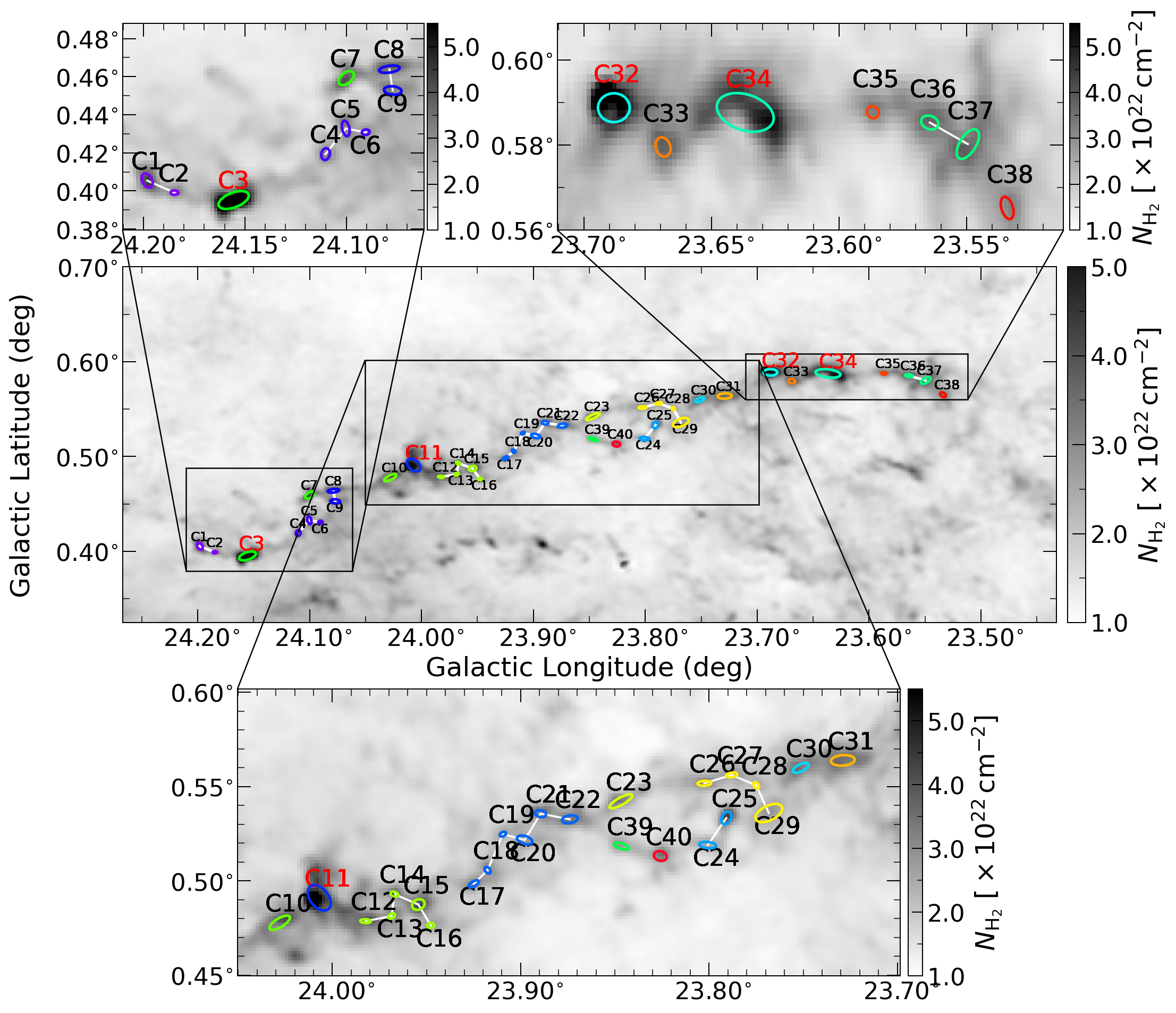}
    \caption{The \textit{Herschel}-based \ce{H2} column density map identical to panel (c) of Figure~\ref{img: column_density1}. Clumps with the same colour are classified into the same group by the MST algorithm. The projected separation distance between two clumps is set to 1.9\,pc as the criterion for MST grouping. 
    The numbers of the clumps are labelled in black. Four massive clumps with high surface densities (C3, C11, C32, and C34) are labelled in red. The figure also displays zoomed-in views of the clumps, facilitating a closer look at the details of their grouping.}
    \label{img: G24_clump&YSO_surf_dence}
 \end{figure*}

A self-gravitating equilibrium isothermal cylinder is unstable to axisymmetric perturbations with wavelengths greater than about 2 times the diameter when its line mass is nearly the same as that for equilibrium. The growth rate of a perturbation has a maximum at a finite wavelength in the dispersion relation, that is, at twice the critical wavelength. It indicates that the self-gravitating filament is expected to fragment, with a typical separation of approximately four times its diameter \citep{Stodólkiewicz_1963AcA....13...30S, Nagasawa_1987PThPh..77..635N,Inutsuka_1992ApJ...388..392I}. Thus, at zero magnetic field intensity, the wavelength corresponding to the maximum growth rate can be written as  
 \begin{equation}
     \lambda_{\rm max}\sim2\times\lambda_{\rm crit},
 \end{equation}
and the critical wavelength is defined as $\lambda_{\rm crit} = 3.94R_{\rm flat}$. We estimated $\lambda_{\rm max}$ for different tracers available in this work and compared with the fitted period length of oscillation segments. Using the flattening radius ($R_{\rm flat}$) listed in Column (6) in Table~\ref{tab: FWHM_estimation}, the $\lambda_{\rm max}$ obtained from the \ce{^13CO}\,(1--0) data is 10.95$\pm$3.55, close to half of the fitted period length of the longitudinal centroid velocity profile of the \ce{^13CO} data (10.75$\pm$0.20). The $\lambda_{\rm max}$ obtained from the \ce{C^18O}\,(1--0) and \ce{N2H+}\,(1--0) data are 4.96$\pm$0.63\,pc and 4.65$\pm$1.34\,pc, respectively. These values are close to half of the fitted short period length of the longitudinal centroid velocity profile of the \ce{C^18O} (4.88$\pm$0.05\,pc) and \ce{N2H+} (4.94$\pm$0.06\,pc) data. 
The factor of two difference between the fitted period length of oscillation segments and the $\lambda_{\rm max}$ may attribute to the presence of additional driving sources for velocity, such as outflows from young stellar objects or gas accretion flows on to the filament. And the mechanism could be the MHD-transverse wave propagating along the filament. Such mechanism can also produce a periodic velocity oscillation which is a superposition of gravitational instability, and was reported both in observations \citep{Stutz_2016AnA...590A...2S, Liu_2019MNRAS.487.1259L} and in simulations \citep{Nakamura_2008ApJ...687..354N}. 

With the $R_{\rm flat}$ obtained from the \textit{Herschel}-based \ce{H2} column density ($N(\ce{H2})_{Herschel}$) data, the $\lambda_{\rm max}$ is 3.55$\pm$0.32\,pc. 
To compare the $\lambda_{\rm max}$ with the separations between the individual dust clumps, we measured the separation using two methods, the nearest neighbour (NN) algorithm and the minimum spanning tree (MST) algorithm, which are provided in the Python library \texttt{FRAGMENT} \citep{Clarke_2019MNRAS.484.4024C}. The mean/median separation calculated by NN is 1.61/1.39\,pc, and by MST is 2.20/1.91\,pc. These values are listed in Table~\ref{tab: null hypothesis test}, and are all smaller than the $\lambda_{\rm max}$. 
 
The inconsistency between the wavelength corresponding to the maximum growth rate and the projected separation of the clumps might be associated with the spatial distribution of clumps along the longitudinal direction of G24. 
First, to explore whether the clumps are randomly placed along the filament, we employed four methods provided in \texttt{FRAGMENT}---median-interquartile range, mean-standard deviation, Kolmogorov–Smirnov (KS) test, and Anderson-Darling (AD) test---to test the null hypothesis that the clumps are randomly placed along the filament. As shown in Table~\ref{tab: null hypothesis test}, the large $p$-values for the four methods can not reject the null hypothesis. In other words, these clumps may be randomly located. 
Meanwhile, to investigate whether these clumps might be the products of hierarchical fragmentation, we calculated the statistical significance of a possible ``two-tier'' fragmentation pattern in G24 using the two model selection methods included in \texttt{FRAGMENT}: the frequentist approach using the Akaike information criterion (AIC), and the Bayesian approach using the odds ratio. 
For both methods, the data are the separation distribution resulting from applying the MST method. 
The frequentist approach obtained an evidence ratio of 541 for the ``two-tier'' fragmentation pattern. The evidence ratio is the weight of the ``best'' model, $w_{\rm max}$, over the weight of the other model, $w_{\rm min}$. In the context of this study, two patterns were compared. If one pattern, for example, the ``two-tier'' fragmentation pattern, is determined by AIC as the ``best'' model, the other pattern, that is, the ``single-tier'' fragmentation pattern, is the other model. The specific determination rules are introduced in \cite{Clarke_2019MNRAS.484.4024C}. The ratio suggests the weight of the ``two-tier'' fragmentation pattern is 541 times more likely than the ``single-tier'' fragmentation pattern, and the hypothesis of the former pattern can not be rejected. 
And the Bayesian approach obtained the odds ratio (the ratio of the marginal likelihood of the ``two-tier'' fragmentation pattern over the ``single-tier'' fragmentation pattern) of 11, indicating positive evidence for the former pattern. 
Both approaches support the ``two-tier'' fragmentation pattern, suggesting it may be taking place in G24.  

The two model selection approaches also proved the best-fit parameters of the distance of clumps' separation. The long and short separations of the ``two-tier'' fragmentation pattern are 4.95\,pc and 1.89\,pc for the frequentist approach, and 3.31 and 1.78\,pc for the Bayesian approach. 
The short separations of the two methods are close to the median separation calculated by the MST method (1.91\,pc). In addition, \cite{Clarke_2019MNRAS.484.4024C} found that the NN technique systematically leads to lower estimates of the characteristic spacing,  while MST is able to recover the underlying spacing distribution better. 
Therefore, we adopted the value 1.9\,pc as the threshold to group the clumps using the MST algorithm. If the distance between two clumps is less than 1.9\,pc, they are considered to belong to the same group (same colour-coded ellipses in Figure~\ref{img: G24_clump&YSO_surf_dence}). Consequently, we identified a total of 22 distinct groups based on this classification, and the grouping number of each clump is shown in Column (18) of Table~\ref{tab: Properties of Clumps}. Furthermore, we determined the projected separations of these groups with the NN and MST algorithm, as done for clumps. 
The centroid coordinates of each group are the weighted average of the coordinates of all clumps in the group, with the weights set as the clumps' masses. 
The separations between the groups are also listed in Table~\ref{tab: null hypothesis test}. The mean/median separation calculated by NN is 2.78/2.75\,pc, and by MST is 3.68/3.46\,pc.  Since MST is able to recover the underlying spacing distribution better than NN \citep{Clarke_2019MNRAS.484.4024C}, the mean/median separation calculated by the MST method may more closely approximate the actual distance, and is quite approximate to the $\lambda_{\rm max}$ of 3.55$\pm$0.32\,pc from the $Herschel$ data, as well as the long separation of the ``two-tire'' fragmentation patter 3.31\,pc obtained from the Bayesian approach. 
The $p$-values of the aforementioned four methods are $\leqslant0.05$, and are able to reject the null hypothesis that the groups are located randomly. 
Both the frequentist and Bayesian approaches preferred the ``single-tier'' fragmentation pattern for the clump groups. The evidence ratio and the Bayes factor can be found in Table~\ref{tab: null hypothesis test}. 
The possible ``two-tier'' fragmentation pattern of clumps is not inconsistent with the possibility indicated by the aforementioned $p$-values that the clumps may be randomly distributed.
The reason is that identifying the typical separation characteristics of the ``second tier'' requires measuring the separation of two clumps within an individual group. However, since each group contains only a few clumps, the statistical data is too sparse to achieve this.

Another possible reason for the inconsistency is that the inclination angle between the long axis of the filament and the plane of the sky is greater than zero, which can lead to an underestimation of the projected separation of the clumps. In addition, the unresolved radius, as discussed in Section~\ref{subsubsec: Width of G24}, $R_{\rm flat}$ can yield a $\lambda_{\rm max}$ value greater than the real one. In other words, a resolved radius, which is smaller than those unresolved $R_{\rm flat}$ listed in Table~\ref{tab: FWHM_estimation}, will lead to a shorter $\lambda_{\rm max}$. 
Consequently, the projected separations within the clump groups are remarkably close to the $\lambda_{\rm max}$ of the \textit{Herschel}-based \ce{H2} column density data. This supports the interpretation that  on the size-scale traced by the dust emission,  G24 is undergoing longitudinal fragmentation consistent with the fragmentation of an isothermal cylinder. However, the factor of two difference between the fitted period length of the \ce{C^18O} and \ce{N2H+} longitudinal centroid velocity profile and the $\lambda_{\rm max}$ of the \ce{C^18O} and \ce{N2H+} data suggests that this is not the case for this gas. At these size scales the fragmentation may be driven by a different mechanism and the  flows directly associated with the fragmentation at the dust  clump groups level do not dominate the velocity structure of the gas traced by these species at the angular resolution of the observations here.

\section{Summary and Conclusion}
\label{sec: Summary and Conclusion}

In this paper, we present the results of mapping observations towards a giant filament CFG024.00+0.48 (G24) using the \ce{C^18O} and \ce{N2H+}\,(1--0) data from the IRAM 30~m telescope and the \ce{^13CO} and \ce{C^18O}\,(2--1) data from the APEX 12~m telescope. Combining with the archival data of CO and \ce{^13CO}\,(1--0) from the PMO 13.7~m millimetre telescope, we investigated the general physical properties of G24 (e.g., width, the size along the LOS, line mass, and longitudinal profiles), gravitational stability, and the possible fragmentation within G24. The results show that G24 is an $\sim$80\,pc-long and massive ($\sim10^5\,\rm M_{\odot}$) giant filament. 
In different tracers the beam-deconvolved width of G24 varies from $\sim$0.8 to 2.8\,pc. 
These deconvolved widths are less than and to approximately 2 times the telescope beam sizes and  the true intrinsic width of the G24 filament could be somewhat smaller. 
The size (thickness) of G24 along the line of sight estimated from the \ce{C^18O} data by RADEX is 2.2\,pc, which is comparable with its width, suggesting that G24 is not a projection of flat structures, such as an edge-on sheet.

We identified 40 clumps associated with G24 from the \textit{Herschel}-based \ce{H2} column density ($N(\ce{H2})_{Herschel}$) map. Four of the 40 clumps are considerably massive based on their surface density ($\Sigma \sim$ 0.10--0.14\,g\,cm$^{-2}$) and mass ($M_{\rm clump}\sim$1632--2317\,$\rm M_{\odot}$). Five of them are associated with 70\,\micron\ point sources (70\,\micron -bright), and the rest are 70\,\micron -quiet. Analyses of the separations between the dust clumps indicate that they appear to have a ``two-tier'' fragmentation pattern, where the short fragmentation length, $\sim$1.9\,pc, corresponds to the separation between two clumps. When grouping the clumps with a threshold of $\sim$1.9\,pc, the separation of the clump groups, with a mean/median of 3.68/3.46\,pc, is very similar to the $\lambda_{\rm max}$, $3.55\pm0.32$\,pc, obtained from the $N(\ce{H2})_{Herschel}$ data. 

The line mass ($M_{\rm line}$) measured from the \ce{C^18O}\,(1--0) data, $\rm 837\pm2\,M_{\odot}\,pc^{-1}$, is comparable to the virial line masses ($M_{\rm line, vir}$) os $\rm 717\pm32\,M_{\odot}\,pc^{-1}$, leading to the virial parameter ($\alpha_{\rm line,vir}=M_{\rm line,vir}/M_{\rm line}$) to be 0.86, indicating gravitationally bound. 
When the filament is separated into several segments, the $\alpha_{\rm line,vir}$ of the majority segments are close to 0.5. Taking a factor of 2 as the uncertainty, the $\alpha_{\rm line,vir}$ of segments indicates that G24 is close to virial equilibrium. 
The $\alpha_{\rm line,vir}$ obtained from the $N(\ce{H2})_{Herschel}$ data is 0.52. 
When the filament is separated into several segments,  9 of them have $\alpha_{\rm line,vir}>0.25$,  and 7 segments have $\alpha_{\rm line,vir}<0.25$. The latter are still smaller than 0.5 when taking a factor of 2 as the uncertainty. This suggests that a portion of the G24 filament may be dominated by gravity. 
 
The longitudinal profiles of centroid velocity show oscillation patterns. The lengths of the oscillation segments are 10.75\,pc for the \ce{^13CO}\,(1--0), 4.88\,pc for the \ce{C^18O}\,(1--0) data and 4.94\,pc for the \ce{N2H+}\,(1--0) data, which are approximately 2 times the wavelength corresponding to the maximum growth rate ($\lambda_{\rm max}$) of $10.95\pm3.55$\,pc, $4.96\pm0.63$\,pc and $4.65\pm1.34$\,pc, respectively, predicted by the model of a self-gravitating equilibrium isothermal cylinder. This suggests that flows directly associated with the fragmentation at the clump group level do not dominate the velocity structure of the gas traced by these species at this angular resolution. 
The accretion rate at the highest probability density of the KDE of these segments is $0.30\times10^3\,\mathrm{M_{\odot}\,Myr^{-1}}$ for the \ce{C^18O}\,(1--0) data and $0.24\times10^3\,\mathrm{M_{\odot}\,Myr^{-1}}$ for the \ce{N2H+}\,(1--0) data. 

In summary, although an infinite isothermal cylinder successfully matches the clump group separations identified in the dust clumps, the overall structure of the G24 filament as traced by molecular line emission and its velocity field point to a more complex fragmentation processes than predicted by an infinite isothermal cylinder.

\section*{Acknowledgements}

We gratefully acknowledge helpful discussions with Patrick Koch, Jia-Wei Wang, Shi-Yu Zhang. This research is supported by the National Science Foundation of China (12041305). This research made use of the data from the Milky Way Imaging Scroll Painting (MWISP) project, which is a multi-line survey in \ce{^12CO}/\ce{^13CO}/\ce{C^18O} along the northern galactic plane with PMO-13.7\,m telescope. We are grateful to all the members of the MWISP working group, particularly the staff members at PMO-13.7m telescope, for their long-term support. MWISP was sponsored by National Key R\&D Program of China with grants 2023YFA1608000 \& 2017YFA0402701 and by CAS Key Research Program of Frontier Sciences with grant QYZDJ-SSW-SLH047. 

Q.-R.He acknowledges 
grants from the STFC and CSC 201804910938, without which, this work would not have been possible.
G.A.Fuller acknowledges support from the Collaborative Research Centre 956, funded by the Deutsche Forschungsgemeinschaft (DFG) project ID 184018867 and from the University of Cologne and its Global Faculty programme.
G.A.Fuller also acknowledges financial support 
from grant PID2020-114461GB-I00, funded by MCIN/AEI/10.13039/5011000110
and through the "Centre of Excellence Severo Ochoa" award for the Instituto de Astrof\'isica de Andalucia
(SEV-2017-0709). 
X.-P.Chen and K.Wang acknowledge supports from the National Science Foundation of China (12041305) and the Tianchi Talent Program of Xinjiang Uygur Autonomous Region.
K.Wang also acknowledges the China-Chile Joint Research Fund (CCJRF No. 2211). CCJRF is provided by Chinese Academy of Sciences South America Center for Astronomy (CASSACA) and established by National Astronomical Observatories, Chinese Academy of Sciences (NAOC) and Chilean Astronomy Society (SOCHIAS) to support China-Chile collaborations in astronomy. 
X.-L.Wang acknowledges the support by the Science Foundation of Hebei Normal University (Grant No. L2024B56).
S.D.Clarke is supported by the National Science and Technology Council (NSTC) in Taiwan through grants NSTC 112-2112-M-001-066 and NSTC 111-2112-M-001-064.


%


\appendix
\restartappendixnumbering 
\section{SED fitting}
\label{appendix: SED Fitting}

Taking SED fitting at a resolution of 500\,\micron\ as an example, the fitting procedure is as follows.
 
First, all $Herschel$ data (160, 250, 350, 500\,\micron) are convolved to the same angular resolution ($\theta_{\rm HPBW}$) as 500\,\micron\ with kernels provided by \cite{Aniano_2011PASP..123.1218A}, which can correct the defects that non-Gaussian beam shape of \textit{Herschel} telescope at the same time. 
 
Second, the data of four bands are re-gridded to the same pixel size as 500\,\micron, and are converted to the same unit of Jy/pixel.  
 
The third step is the SED fitting, following the modified black-body model described in \cite{Lin_2016ApJ...828...32L}:
  \begin{equation}
     S_{\rm \nu}=\Omega_{\rm m} B_{\rm \nu}(T_{\rm d})(1-e^{-\tau_{\rm \nu}}),
 \end{equation}
where $\Omega_{\rm m}$ is the solid angle, $\rm S_{\rm \nu}$ is the observed flux density, $B_{\rm \nu}(T_{\rm d})$ is the Planck function of dust temperature $T_d$
 \begin{equation}
     B_{\rm \nu}(T_{\rm d})=\frac{2h\nu^3}{c^2}\frac{1}{{\rm exp}(h\nu/kT_{\rm d})-1}.
 \end{equation}

The column density is defined as
 \begin{equation}
     N_{\rm H_2}=\frac{\tau_{\rm \nu}}{\kappa_{\rm \nu}\mu m_{\rm H}}\times R_{\rm gd},
 \end{equation}
where $\rm \mu$ is the mean molecular weight and is 2.83 \citep{Kauffmann_2008AnA...487..993K}, $R_{\rm gd}$ is the gas-to-dust mass ratio, and we adopt a ratio of 100. The dust opacity law is 
\begin{equation}
    \kappa_{\rm \nu}=\kappa(\frac{\nu}{\rm 230\,GHz})^{\beta},
\end{equation}
where $\kappa$ is the dust mass absorption coefficient at reference frequency $\nu=230$\,GHz, and is fixed to $\kappa_{\rm 230}\rm =0.5\,cm^2\,g^{-1}$ \citep{Preibisch_1993AnA...279..577P}. Dust opacity index $\beta \rm = 2$ is assumed to be fixed. 
 
\begin{figure}
 \centering 
 \includegraphics[width=1.0\columnwidth]{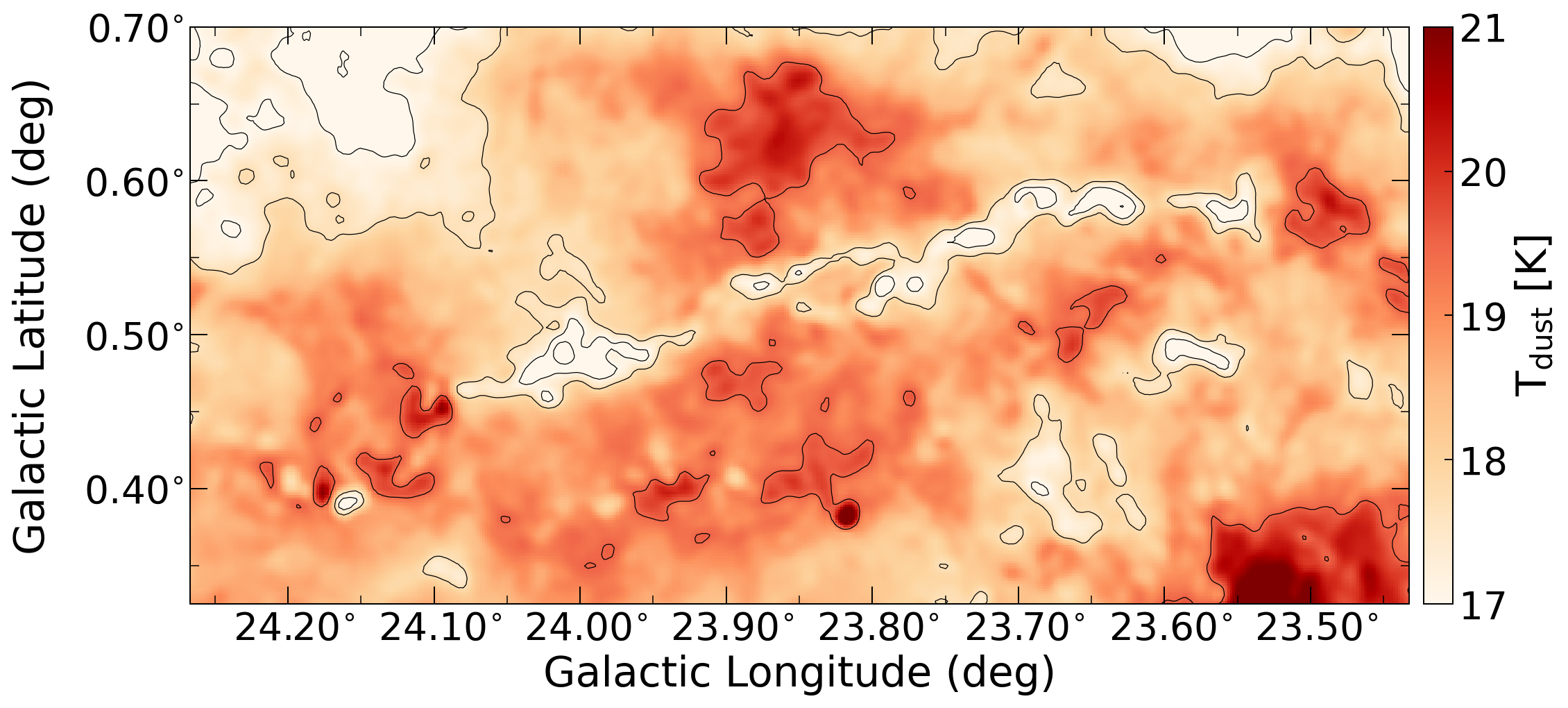}
 \caption{Dust temperature of G24 at a resolution of 36\arcsecuni, overlaid contours with levels of [16, 17, 17.7, 19.5]\,K.}
 \label{img: G24_NH2andTemp_1}
\end{figure}

The SED fitting procedure is performed with the Python library \texttt{CurveFit}. For the sake of revealing more details of the filament structure, a high-resolution (18.2$^{\prime\prime}$) column density map is created according to the method introduced in \cite{Palmeirim_2013AnA...550A..38P} Appendix~A. Column density maps with higher resolution (e.g., 11.5\,\arcsecuni) can be generated. \cite{Hill_2012AnA...548L...6H} Appendix~A describes similar steps but improves the resolution to 11.5\,\arcsecuni\ by including APEX/P-ArT\'eMiS 450\,\micron\ continuum observations. Only the $Herschel$ continuum dust data are used in our work. 
The dust temperature map with the resolution of 36\arcsecuni\ and the $\rm H_2$ column density map with the resolution of 18.2\arcsecuni\ are presented in Figure~\ref{img: G24_NH2andTemp_1} and Figure~\ref{img: column_density1} panel (c), respectively. 

\section{Background Subtraction}
\label{appendix: Background Subtraction}

\begin{figure}
 \centering 
 \includegraphics[width=0.8\columnwidth]{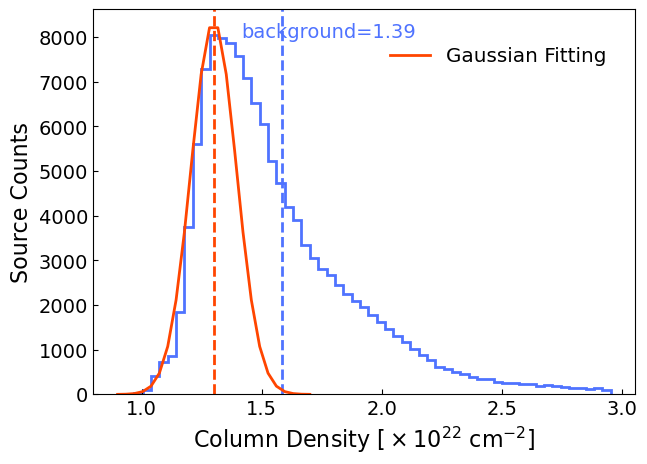}
 \caption{The PDF of column density map (Figure~\ref{img: column_density1} panel (c)). The red curve is the 1D Gaussian fitting, and the blue vertical dashed line is one standard deviation of the Gaussian fitting.}
 \label{img: Bg_sb_PDF}
\end{figure}

The background is subtracted following \cite{Peretto_2010AnA...518L..98P}. 
As presented in Figure~\ref{img: Bg_sb_PDF}, a probability density function (PDF) towards the column density map (Figure~\ref{img: column_density1} panel (c)) is generated to estimate the boundaries of G24. The PDF shows a log-normal distribution at low column density and an extended tail at higher column density. 
The log-normal distribution is fitted by a 1D Gaussian model. To avoid the deformation of the log-normal distribution caused by the extended tail, the data for Gaussian fitting is obtained by getting a symmetric shape by mirroring the part lower than the peak value of the log-normal distribution. 

\begin{figure}
 \centering 
 \includegraphics[width=1.0\columnwidth]{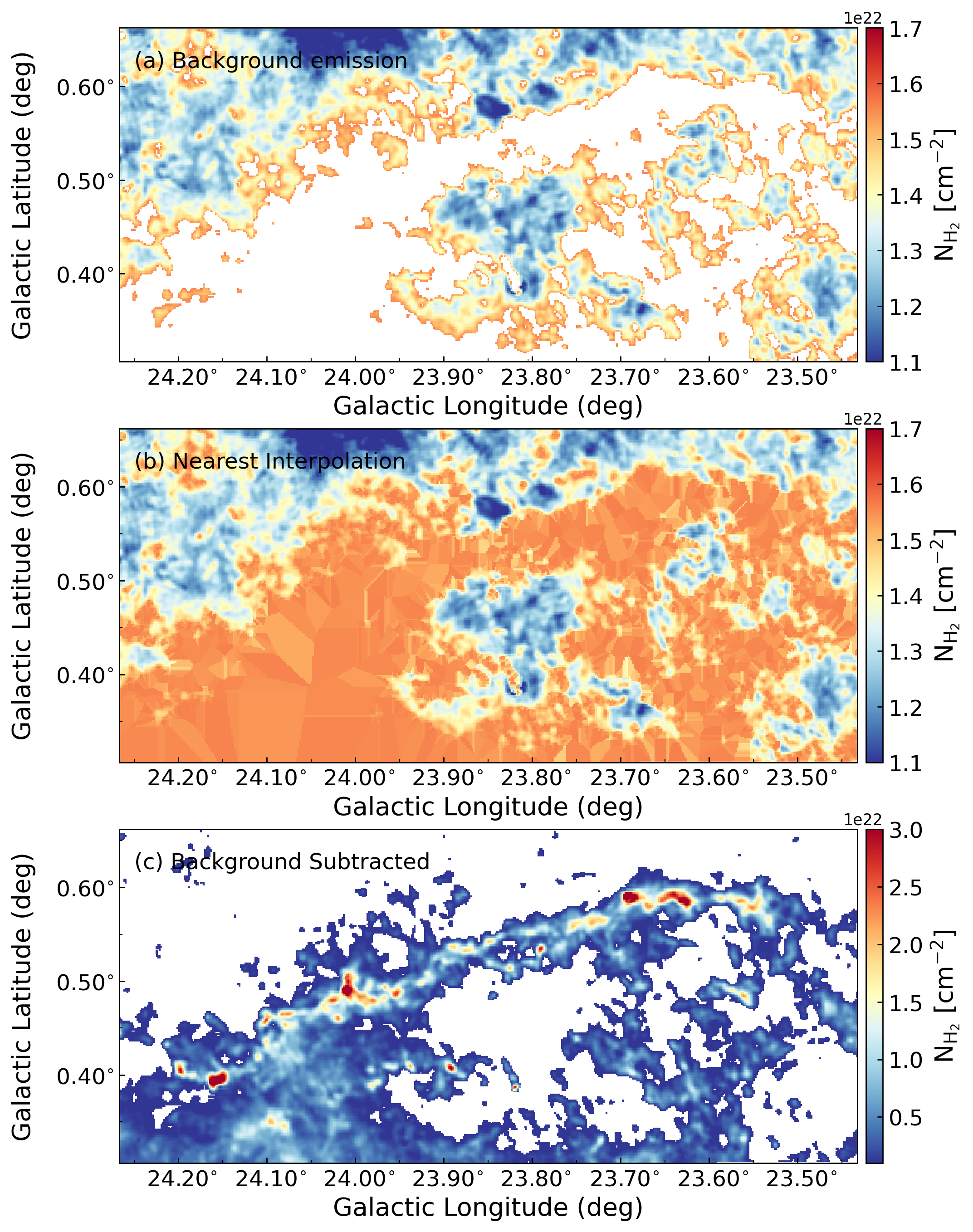}
 \caption{The demonstration of the background subtraction process. (a) The column density map of G24 with structures masked out. (b) The background map with the masked region interpolated by the nearest neighbour algorithm. (c) The final background-subtracted map of G24.}
 \label{img: Bg_sb_map}
\end{figure}

The one standard deviation (1$\sigma$, $\rm 1.39\,\times\,10^{22}\,cm^{-2}$) obtained from the Gaussian fitting is then used to distinguish the G24 structure from background. That is, pixels with column density $\geqslant 1\sigma$ is considered to be included in the structure of G24, otherwise the pixels are considered as background.  As shown in Figure~\ref{img: Bg_sb_map} panel (a), the structure is masked out from the original column density map.
 
The map with mask is interpolated by the nearest neighbour algorithm to obtain a background map as shown in Figure~\ref{img: Bg_sb_map} panel (b). In this way, the background fluctuations are taken into account, and the background maps are reconstructed. Finally, the structure of G24 is derived by subtracting the background map from the original map, as shown in Figure~\ref{img: Bg_sb_map} panel (c).

The average column density of G24 is $\rm \geqslant 2\times10^{22} \,cm^{-2}$. 

\begin{figure}
  \centering 
  \includegraphics[width=1.0\columnwidth]{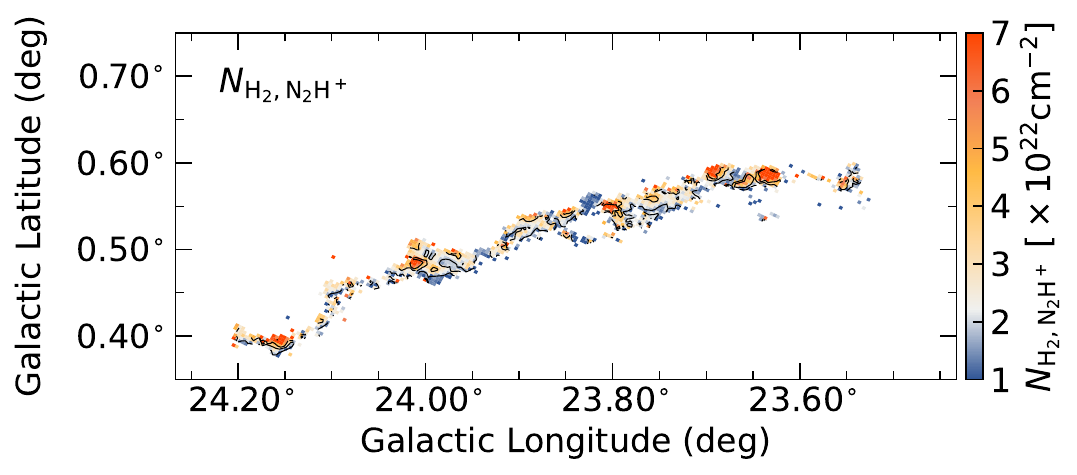}
  \caption{\ce{H2} Column density map derived from \ce{N2H+}\,(1--0) with contours of [1, 2, 3, 4, 5, 6, 7]$\rm \times 10^{22}$\,$\rm cm^{-2}$.}
  \label{img: column_density2}
\end{figure} 

\section{Fitting the Spectral Line of \texorpdfstring{\ce{^13CO}}{13CO}, \texorpdfstring{\ce{C^18O}}{C18O} and \texorpdfstring{\ce{N2H+}}{N2H+}}
\label{appendix: Spectral Line Fitting}

\begin{figure}
 \centering 
 \includegraphics[width=1.0\columnwidth]{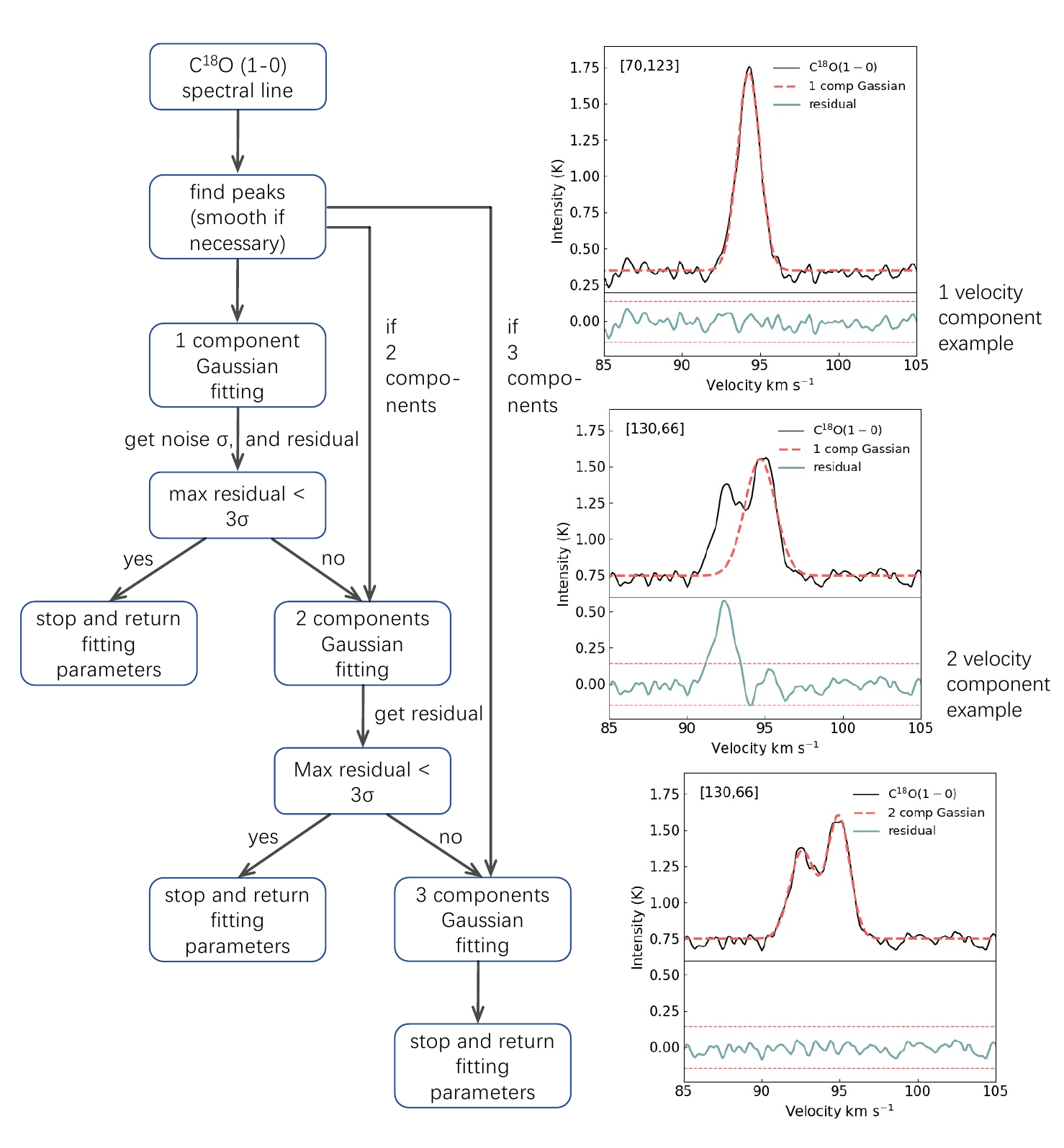}
 \caption{The flow chart of \ce{C^18O} multiple-component Gaussian fitting, with right panels presenting examples of a single component emission and a double component emission. In each panel, the spectral line is shown with a black line, and the red dashed line is the fitting result. The green line is the residual, and the horizontal red dashed lines are the $\rm \pm 3.5\times$standard deviation. When the residual exceeds the level, it is assumed that some signals are still not fitted, and the number of components increases. The pixel coordinates are shown in the upper left corner. }
 \label{img: C18O_flow_chat} 
\end{figure}

The data from both \ce{N2H+} and \ce{C^18O} are fitted pixel-by-pixel to obtain the maps of velocity dispersion. For each pixel, the spectral line is averaged within a beam-scale. That is, the spectral line is averaged over $2\times2$ pixels where the target pixel is located in the first row and first column. 
The \ce{C^18O} spectral lines appear to contain multiple emission peaks for some pixels of the \ce{C^18O}\,(1--0) data cube. Thus, we carry out multiple-component Gaussian fitting following the flow chart in Figure~\ref{img: C18O_flow_chat}. We remind the reader that here we artificially set the maximum velocity component to three. 

Each velocity component will obtain three parameters from the fitting: the antenna temperature at peak emission, centroid velocity $v_{\rm LSR, \ce{C^18O}}$, and velocity dispersion $\sigma_{\ce{C^18O}}$. The final values of the three parameters of each pixel are a weighted average value of that of each component, with the integrated intensity as the weight. The integral is integrated over the range of $v_{\rm LSR, \ce{C^18O}} \pm 2\sigma_{\ce{C^18O}}$. 

The line profile of \ce{N2H+} is fitted by a 1-dimensional hyperfine structure (HFS) model in the Python library \texttt{pyspeckit}. The output parameters are: excitation temperature $T_{\rm ex}$, optical depth $\tau_{\ce{N2H+}}$, centroid velocity $v_{\rm LSR}$, and velocity dispersion $\sigma_{\ce{N2H+}}$. 
 
We notice that for some spectra the velocity dispersion cannot be fitted well by a one-component HFS model, as the obtained optical depth is always with an excessively large uncertainty (an order of magnitude or more than the fitted $\tau_{\ce{N2H+}}$). In addition, when we try to fit \ce{N2H+} with two velocity components, one of the optical depths is always with excessive uncertainty.

In order to obtain reasonable results (parameters greater than at least 3 times their uncertainties), the parameter $\tau_{\ce{N2H+}}$ is fixed to a certain value of 1.30. The value is estimated by the following steps. 
We choose a circular area at a random position of the data that takes a random value from 3, 4, and 5 pixels as radius. The spectral line is averaged over these pixels to improve the signal-to-noise ratio. 
The line profile is then fitted by a single-component HFS model without any parameters fixed. If all four output parameters are reasonable (parameters greater than at least 3 times of their uncertainties), the $\tau_{\ce{N2H+}}$ is kept, otherwise it is discarded. This step is repeated 200 times to obtain a sample of well fitted $\tau_{\ce{N2H+}}$ and their uncertainties $\sigma_{\rm \tau, \ce{N2H+}}$. Then the final optical depth, 1.30, is obtained as the weighted average value of all $\tau_{\ce{N2H+}}$ from the sample with 1/$\sigma_{\rm \tau, \ce{N2H+}}^2$ as the weight. 
 
\section{The Radial Profile}
\label{appendix: The Radial Profile}
 
\begin{figure}[htbp]
  \centering 
  \includegraphics[width=1.0\columnwidth]{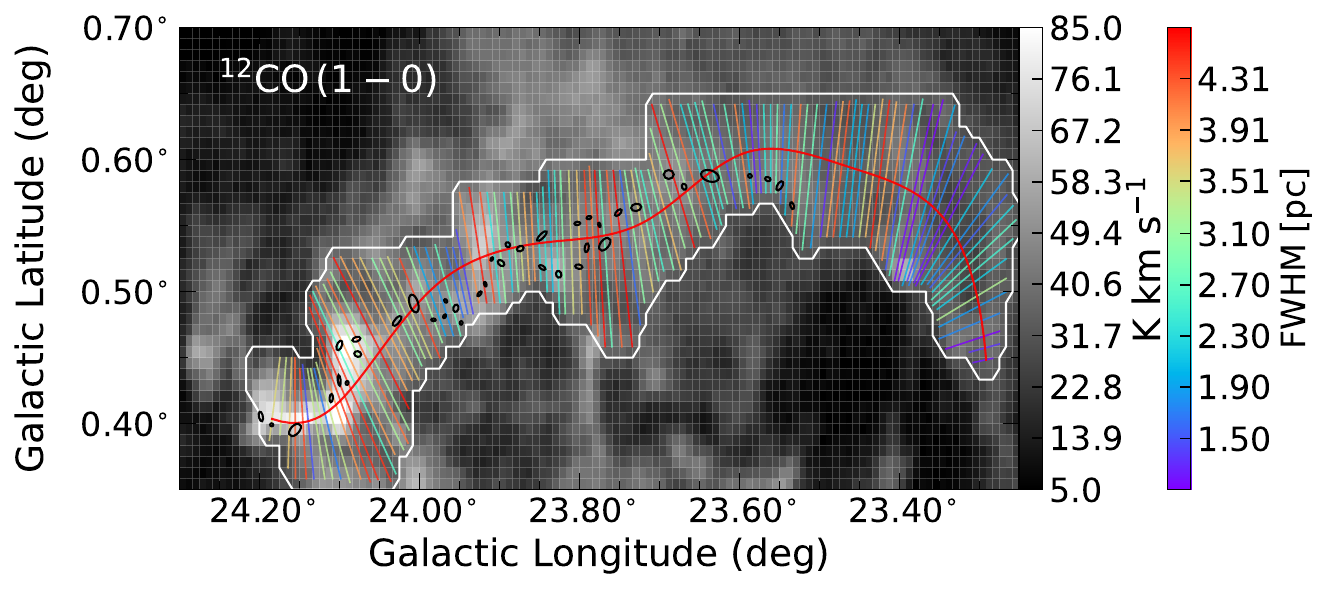} \\
  \includegraphics[width=0.9\columnwidth]{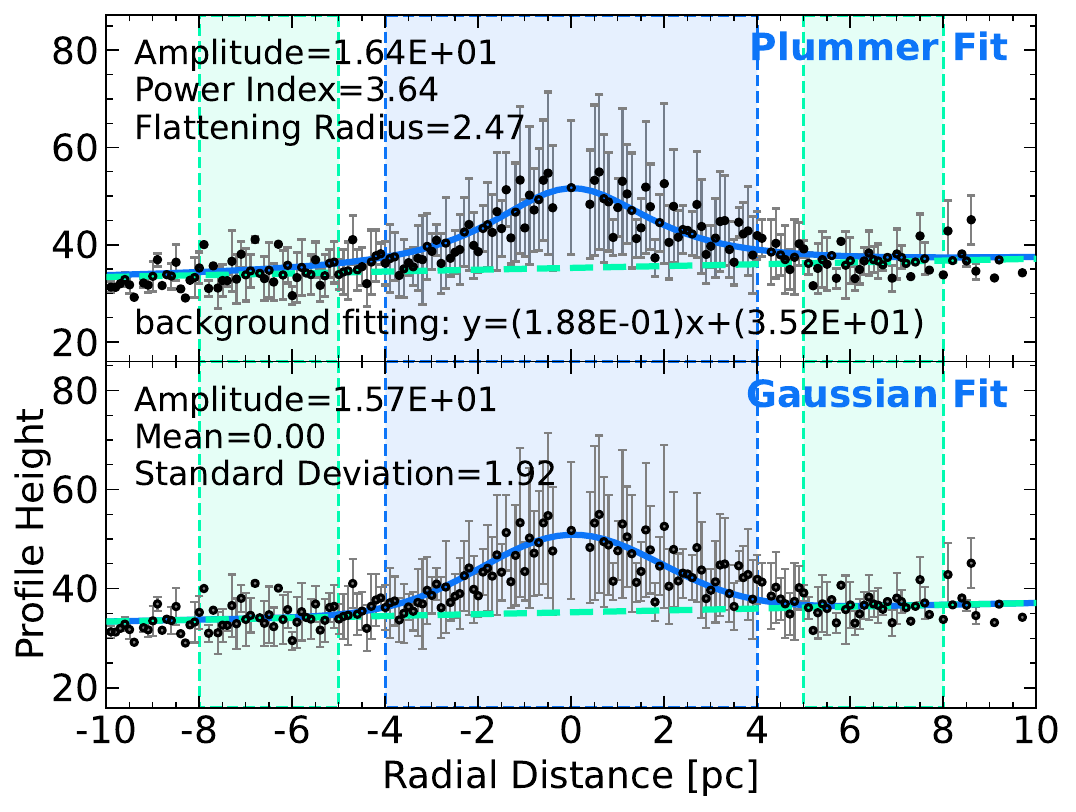}

  \caption{\textit{upper: }The skeleton, boundary of G24 and paths of transverse slices, colour-coded by the Gaussian FWHM. The background is the integrated intensity map of CO\,(1--0) with the integral range of 90.0-99.5\,\velouni. The curve of coloured dots at the centre of slices depicts the route of the skeleton. The white contour is the boundary of G24 obtained from \texttt{Radfil}.
  The positions of clumps are marked by black ellipses.
  \textit{lower: }The fit result of Plummer-like profile and Gaussian profile from \texttt{RadFil}. The x-axis is the physical distance from the peak emission. The data points are binned at intervals of 0.1\,pc. Each bin shows only the mean intensity with error bars, which is obtained from the standard deviation of intensities within that bin. The green translucent regions are the distance range for background fitting, and the green dashed line is the fitted background. The blue translucent region is the distance range for line profile fitting, and the blue solid line is the fitted profile. The results of the fitted parameters are marked in the upper left corners.}
 \label{img: radial_fil_profile_co1-0}
\end{figure}

\begin{figure}[htbp]
  \centering 
  \includegraphics[width=1.0\columnwidth]{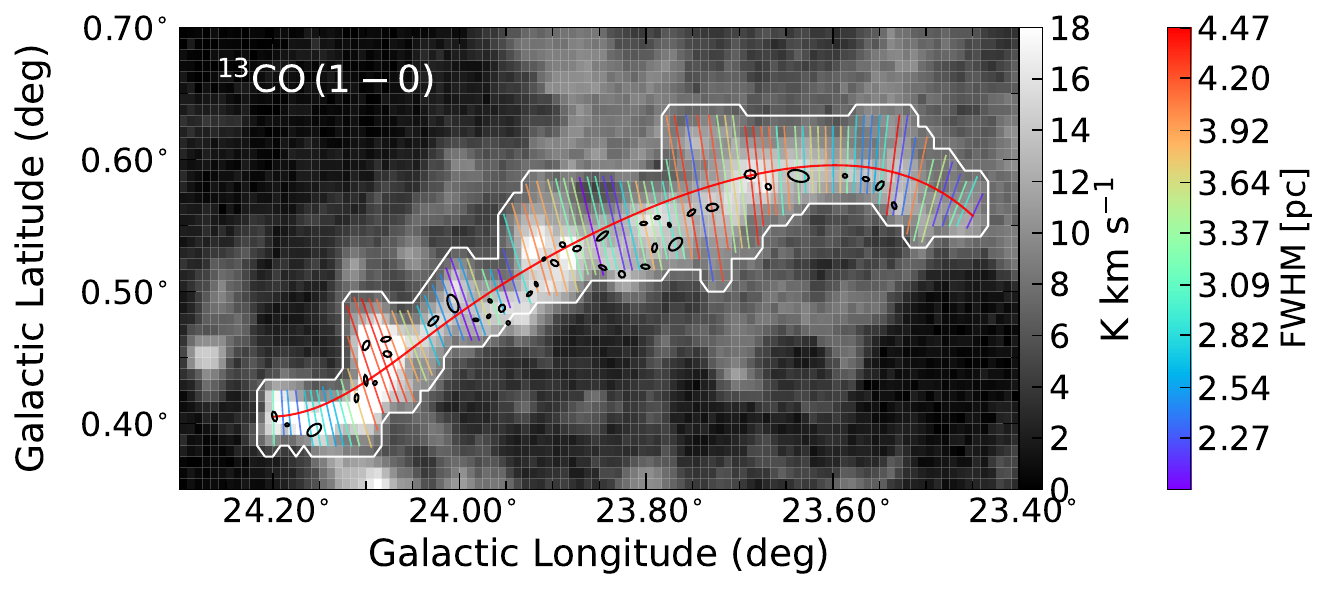} \\
  \includegraphics[width=0.9\columnwidth]{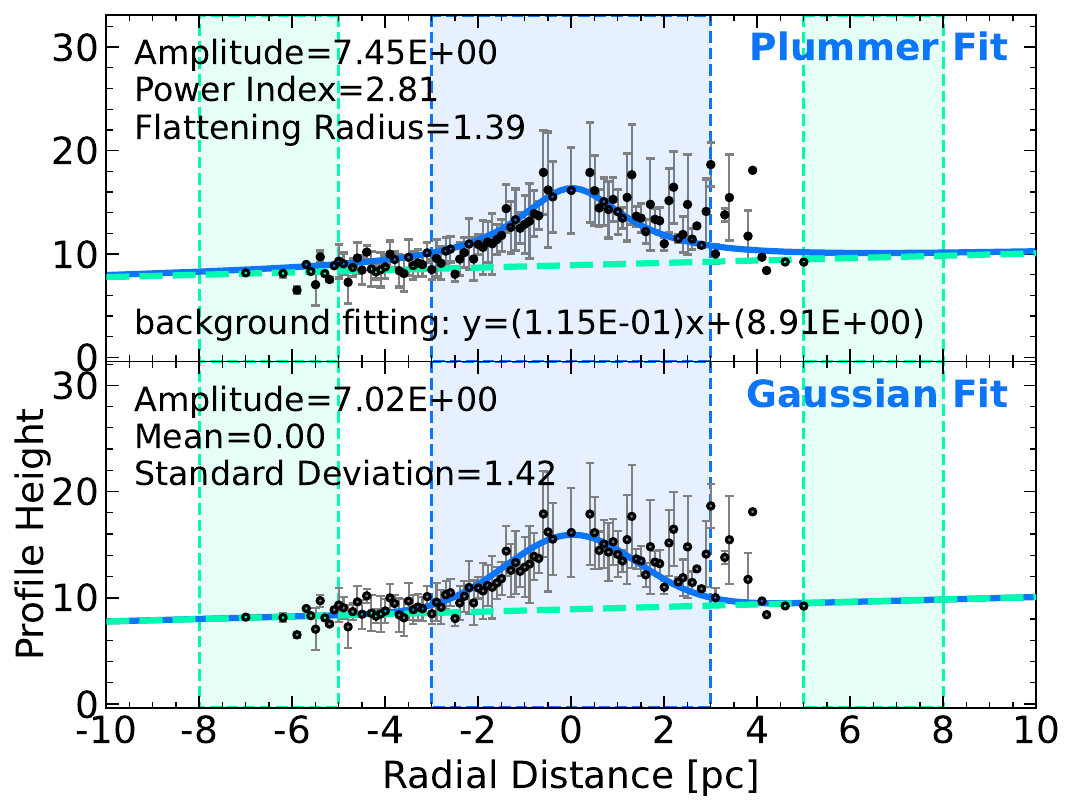}

  \caption{\textit{upper: }The skeleton, boundary of G24 and paths of transverse slices, colour-coded by the Gaussian FWHM. The background is the integrated intensity map of \ce{^13CO}\,(1--0) with the integral range of 90.0-99.5\,\velouni . \textit{lower: }The fit result of Plummer-like profile and Gaussian profile from \texttt{RadFil}. All of the markers and colour scheme are the same as Figure~\ref{img: radial_fil_profile_co1-0}. }
 \label{img: radial_fil_profile_13co1-0}
\end{figure}

\begin{figure}[htbp]
  \centering 
  \includegraphics[width=1.0\columnwidth]{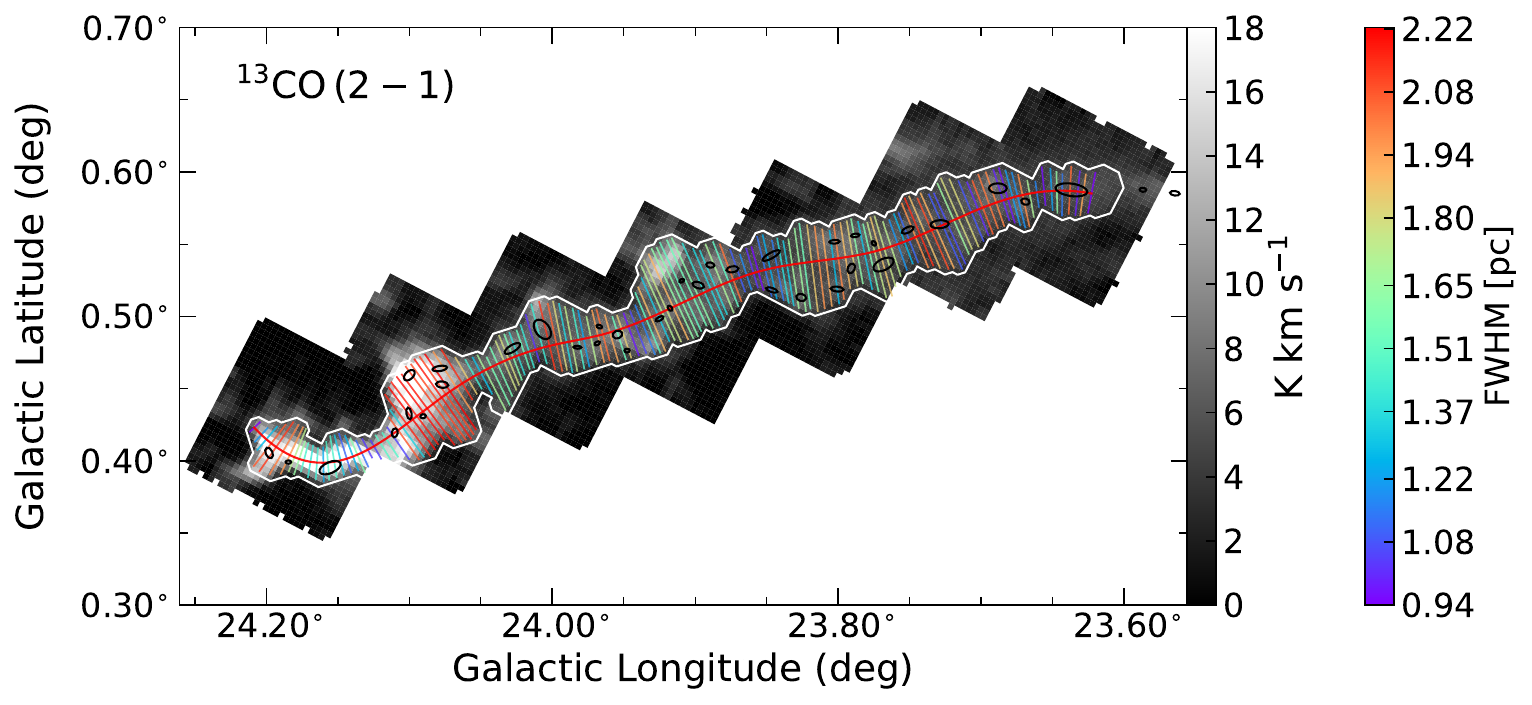} \\
  \includegraphics[width=0.9\columnwidth]{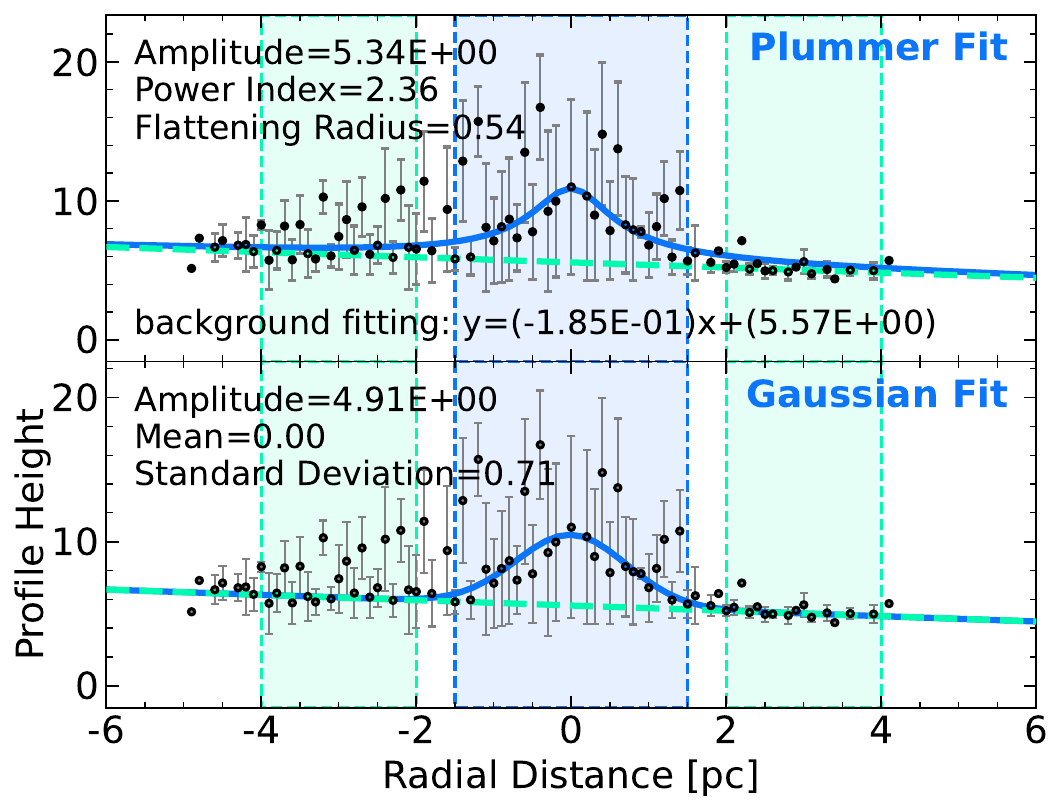}

  \caption{\textit{upper: }The skeleton, boundary of G24 and paths of transverse slices, colour-coded by the Gaussian FWHM. The background is the integrated intensity map of \ce{^13CO}\,(2--1) with the integral range of 90.0-98.0\,\velouni . All of the markers and colour scheme are the same as Figure~\ref{img: radial_fil_profile_co1-0}. \textit{lower: }The fit result of Plummer-like profile and Gaussian profile from \texttt{RadFil}. All of the markers and colour scheme are the same as Figure~\ref{img: radial_fil_profile_co1-0}. }
 \label{img: radial_fil_profile_13co2-1}
\end{figure}

\begin{figure}[htbp]
  \centering 
  \includegraphics[width=0.9\columnwidth]{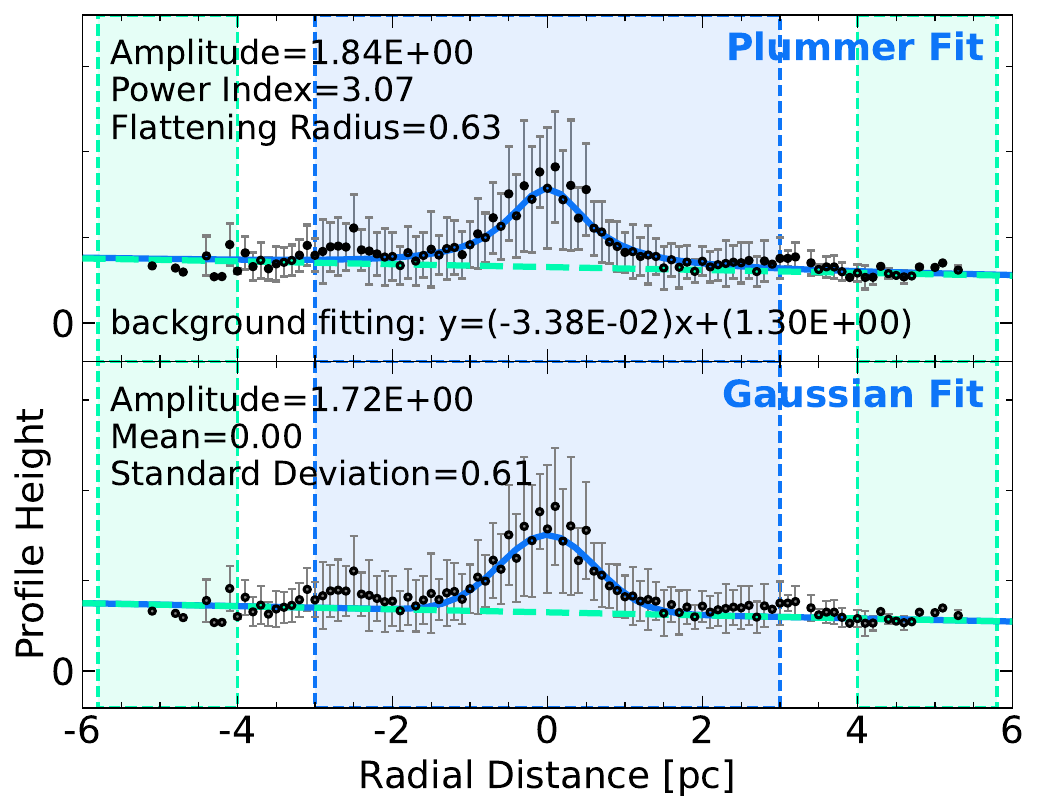}
  \caption{
  The fit result of Plummer-like profile and Gaussian profile from \texttt{RadFil} for the \ce{C^18O}\,(1--0) data. All of the markers and colour scheme are the same as Figure~\ref{img: radial_fil_profile_co1-0}. }
 \label{img: radial_fil_profile_c18o1-0_fitting}
\end{figure}

\begin{figure}[htbp]
  \centering 
  \includegraphics[width=1.0\columnwidth]{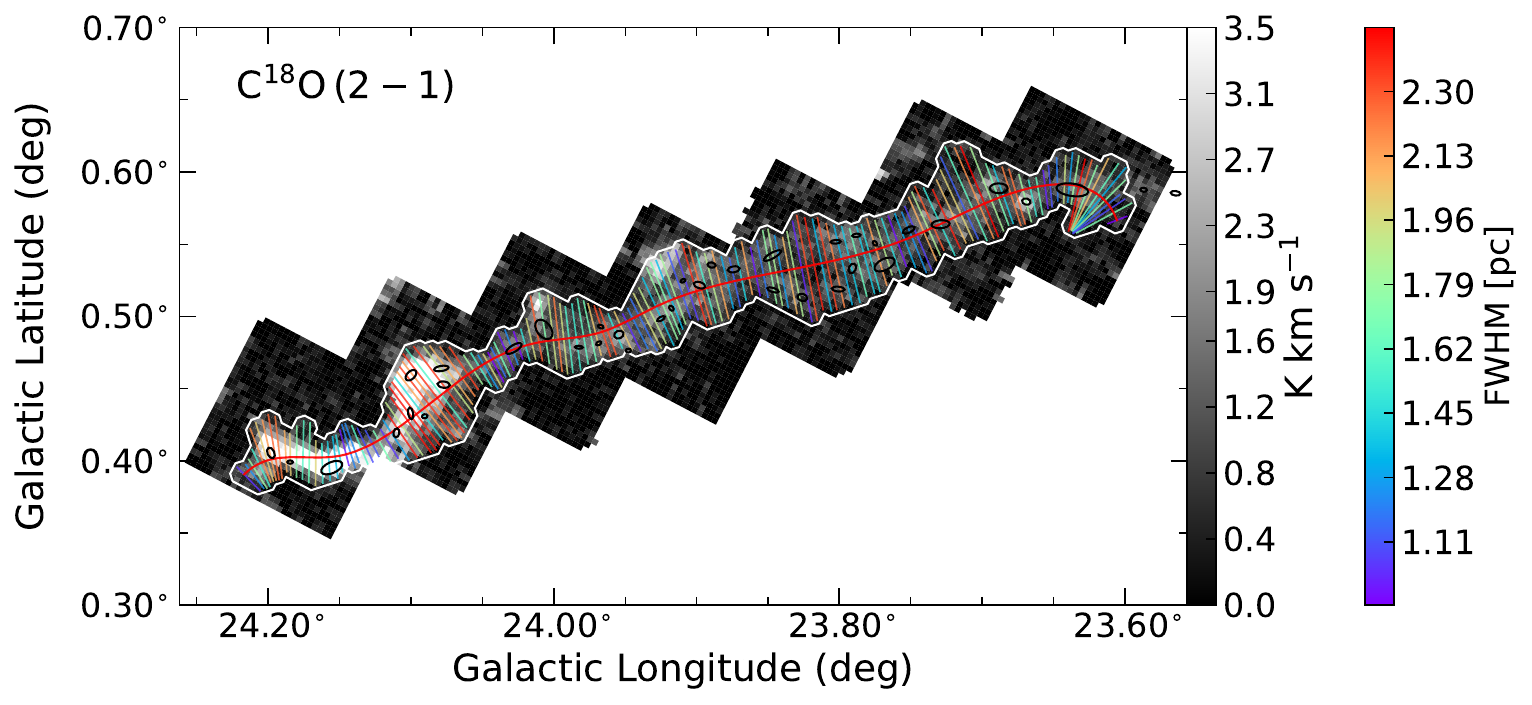} \\
  \includegraphics[width=0.9\columnwidth]{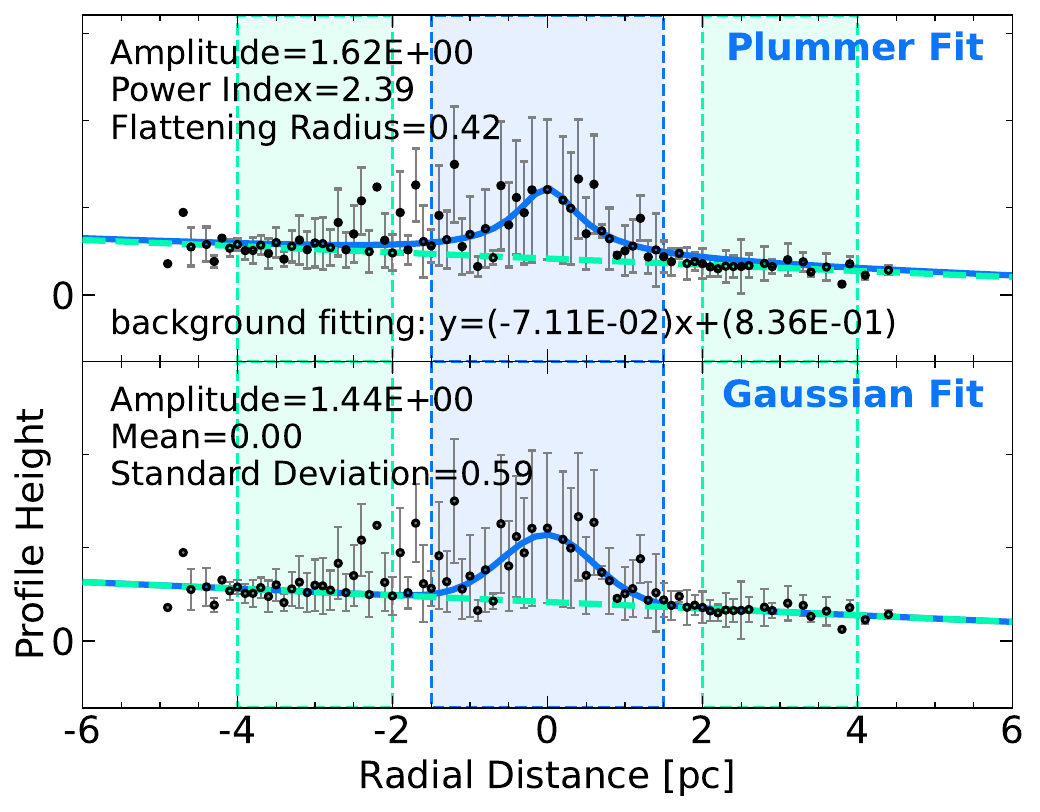}

  \caption{\textit{upper: }The skeleton, boundary of G24 and paths of transverse slices, colour-coded by the Gaussian FWHM. The background is the integrated intensity map of \ce{C^18O}\,(2--1) with the integral range of 90.0-98.0\,\velouni . All of the markers and colour scheme are the same as Figure~\ref{img: radial_fil_profile_co1-0}.
  \textit{lower: }The fit result of Plummer-like profile and Gaussian profile from \texttt{RadFil}. All of the markers and colour scheme are the same as Figure~\ref{img: radial_fil_profile_co1-0}. }
 \label{img: radial_fil_profile_c18o2-1}
\end{figure}

\begin{figure}[htbp]
  \centering 
  \includegraphics[width=1.0\columnwidth]{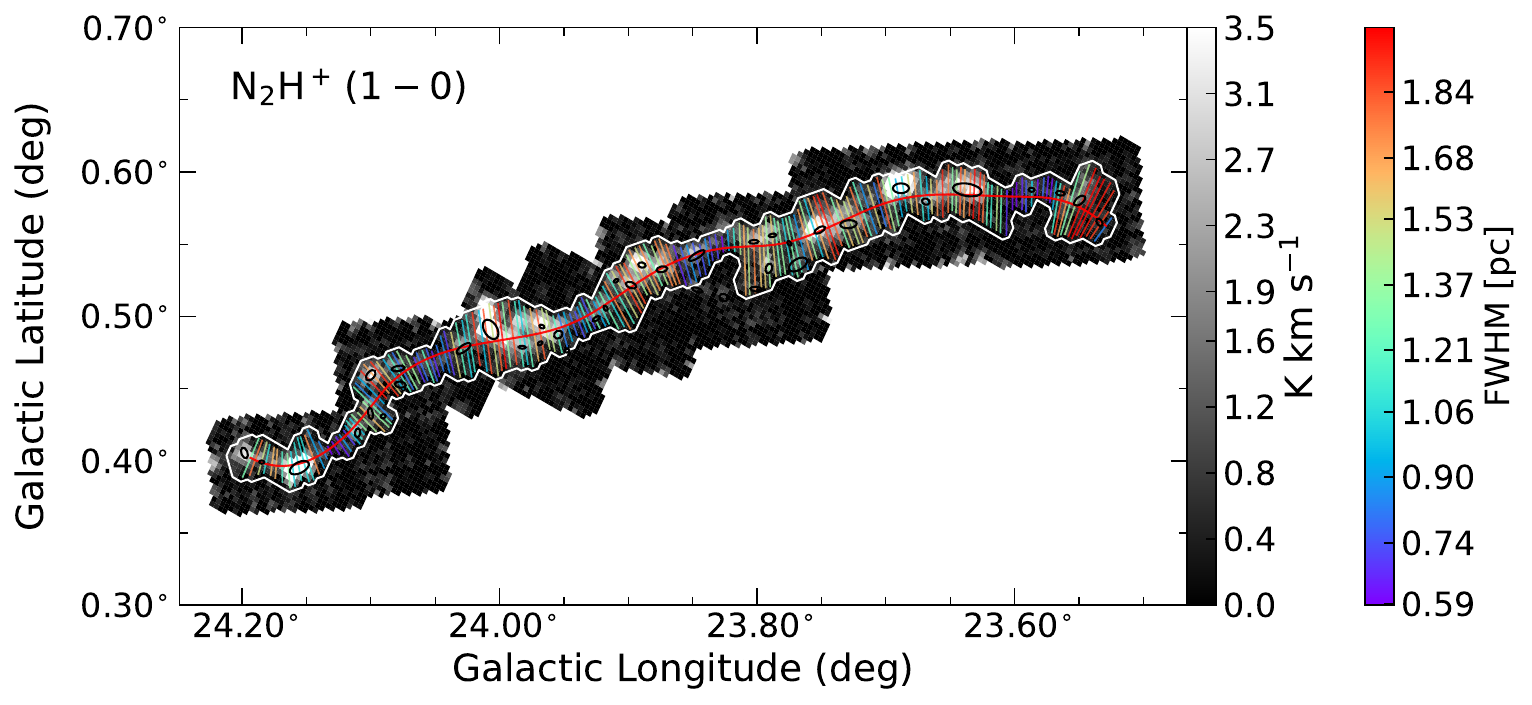} \\
  \includegraphics[width=0.9\columnwidth]{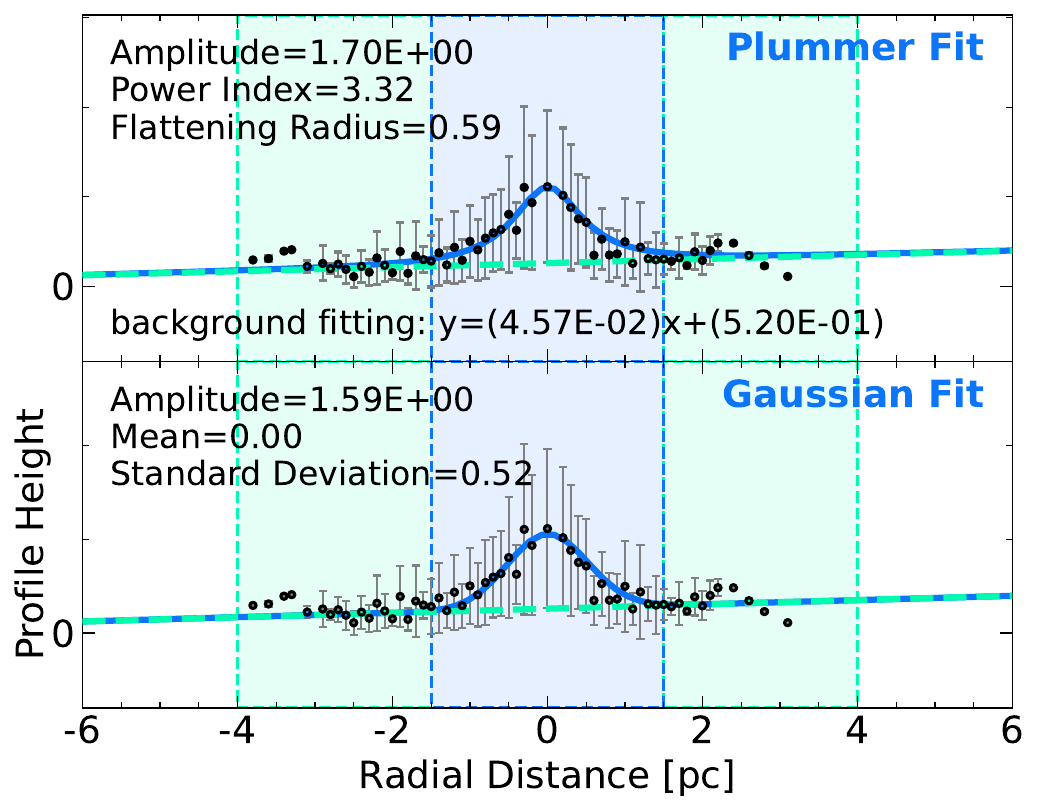}

  \caption{\textit{upper: }The skeleton, boundary of G24 and paths of transverse slices, colour-coded by the Gaussian FWHM. The background is the integrated intensity map of \ce{N2H+}\,(1--0) with the integral range of 84.0-104.5\,\velouni . All of the markers and colour scheme are the same as Figure~\ref{img: radial_fil_profile_co1-0}.
  \textit{lower: }The fit result of Plummer-like profile and Gaussian profile from \texttt{RadFil}. All of the markers and colour scheme are the same as Figure~\ref{img: radial_fil_profile_co1-0}. }
 \label{img: radial_fil_profile_n2hp}
\end{figure}

\begin{figure}[htbp]
  \centering 
  \includegraphics[width=1.0\columnwidth]{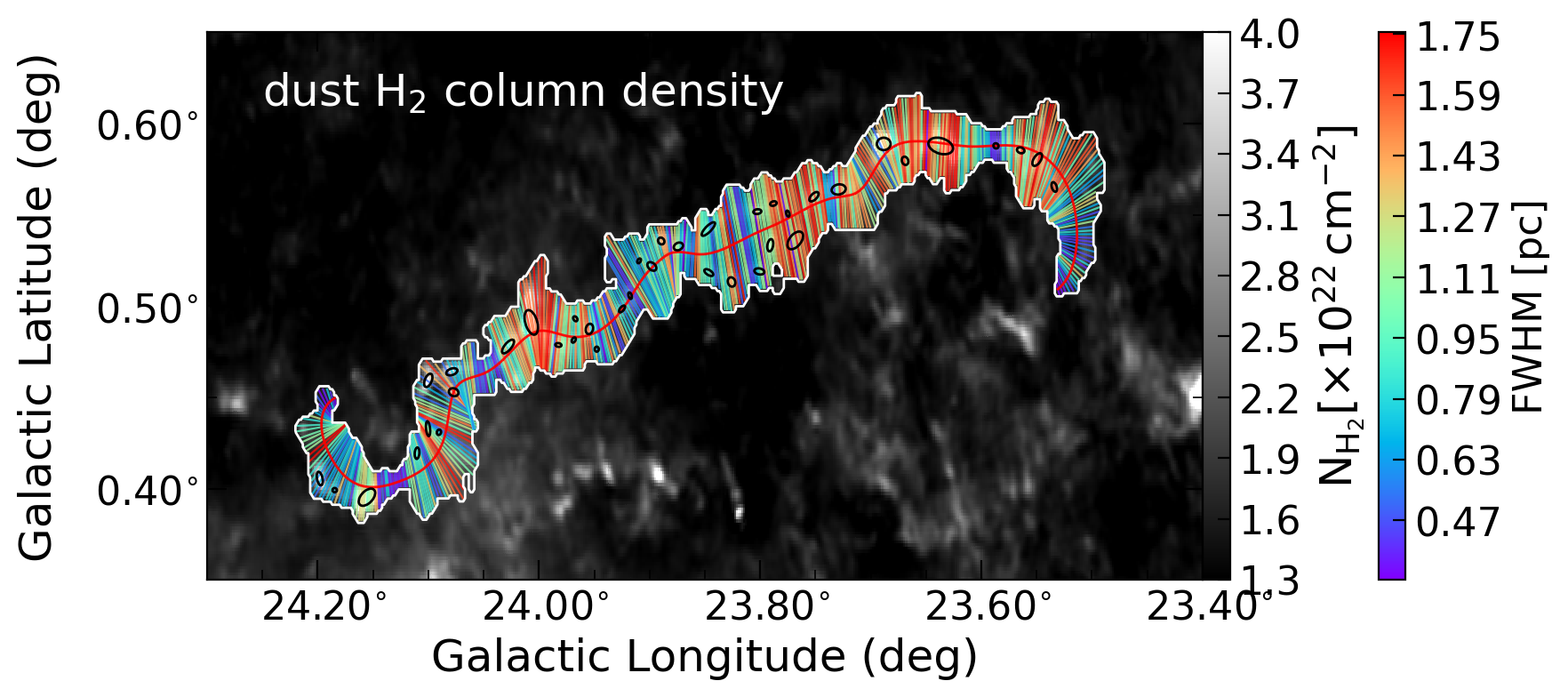} \\
  \includegraphics[width=0.9\columnwidth]{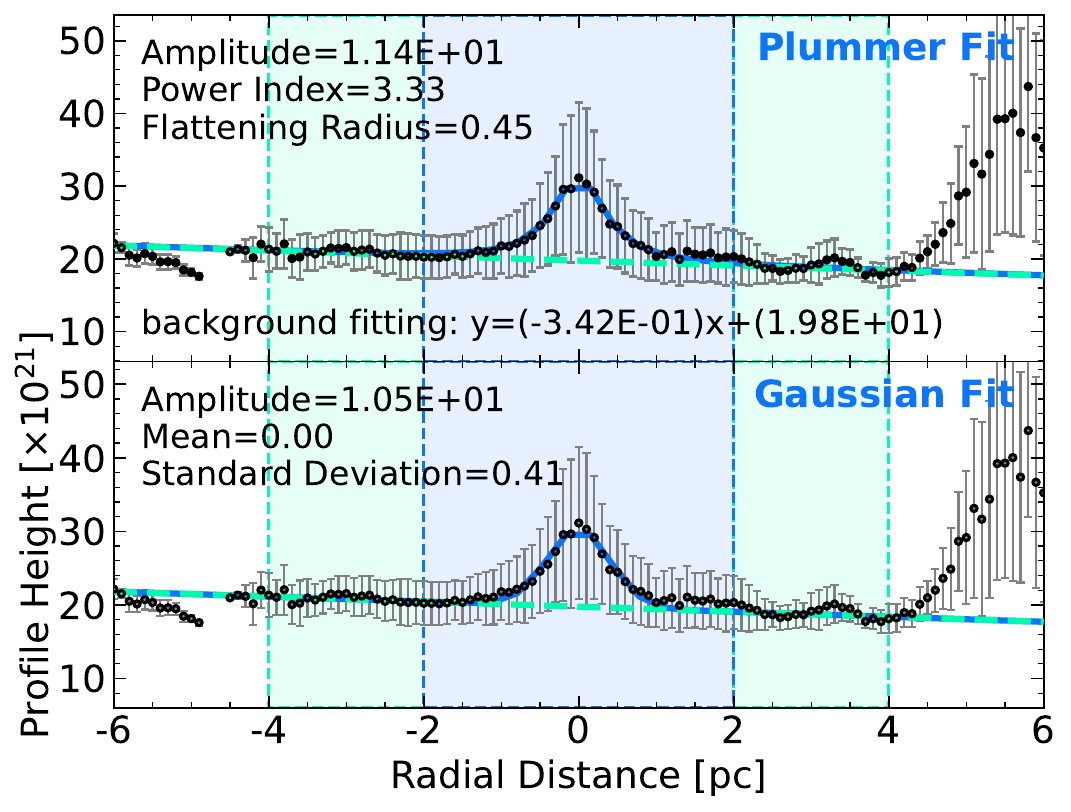}
  
  \caption{\textit{upper: }The skeleton, boundary of G24 and paths of transverse slices, colour-coded by the Gaussian FWHM. The background is the \textit{Herschel}-based \ce{H2} column density map. All of the markers and colour scheme are the same as Figure~\ref{img: radial_fil_profile_co1-0}.
  \textit{lower: }The fit result of Plummer-like profile and Gaussian profile from \texttt{RadFil}. All of the markers and colour scheme are the same as Figure~\ref{img: radial_fil_profile_co1-0}. }
 \label{img: radial_fil_profile_clmn}
\end{figure}

The radial profiles of the G24 filament obtained from CO\,(1--0), \ce{^13CO}\,(1--0), \ce{^13CO}\,(2--1), \ce{C^18O}\,(2--1), \ce{N2H+}\,(1--0) integrated intensity map and the $Herschel$-based \ce{H2} column density map are presented in this section from Figure~\ref{img: radial_fil_profile_co1-0} to Figure~\ref{img: radial_fil_profile_clmn}. 
 
In addition, for the \texttt{RadFil} fitting, the input parameter \texttt{fitdist} should be defined by the effective distance, which is an offset from the position of the maximum amplitude of a radial profile. It depends on the length of the radial profile and varies with the different tracers we used. Based on the spatial distribution of each molecular line used here, we provided different values of \texttt{fitdist}: 6\,pc for \ce{^12CO}\,(1--0), 3\,pc for \ce{^13CO} and \ce{C^18O}\,(1--0),  and 2\,pc for \ce{^13CO}\,(2--1), \ce{C^18O}\,(2--1), \ce{N2H+}\,(1--0), and $N$(\ce{H2})$_{Herschel}$. Another input parameter, \texttt{bgdist}, is used for fitting the background of the radial profile. It is a section offset from the peak emission, where no emissions are detected. We set \texttt{bgdist} to 5.0--8.0\,pc for \ce{^12CO}\,(1--0) and \ce{^13CO}\,(1--0), 4.0--6.0\,pc for \ce{C^18O}\,(1--0), 2.0--4.0\,pc for \ce{^13CO}, \ce{C^18O}\,$J=$2--1 and $N$(\ce{H2})$_{Herschel}$, and 1.5--4.0\,pc for \ce{N2H+}\,(1--0).

\section{The Size of G24 along the Line of Sight}
\label{appendix: Line Radiative Transfer}
The size of a filament along the line of sight provides a third dimension of scale, which is important in determining the morphology of the filament. 
To investigate the thickness of G24, which refers to the size along the line of sight (LOS), we need to estimate the \ce{H2} volume density ($n_{\rm \ce{H2}}$), which is determined with a line ratio of two different transition levels of the same molecular species and using the non-LTE radiative transfer code RADEX \citep{Van_der_Tak_2007AnA...468..627V}.  In this work we used $J$=1--0 and $J$=2--1 transition levels of \ce{C^18O}. 
To create the velocity-integrated intensity ratio map of the \ce{C^18O}\,(1--0)/\ce{C^18O}\,(2--1), the \ce{C^18O}\,(1--0) data with $\theta_{\rm beam}$ of $\sim 24^{\prime\prime}$ was smoothed by the $\theta_{\rm beam}$ of $\sim 30^{\prime\prime}$ of the \ce{C^18O}\,(2--1) data, and the smoothing was done by convolving the \ce{C^18O}\,(1--0) data with a 2D Gaussian kernel. After the beam convolution, the \ce{C^18O}\,(1--0) data is regridded to the same pixel scale of the \ce{C^18O}\,(2--1) data, and then the intensity integral line ratio is generated over the velocity range of the emission per pixel, which is the centroid velocity $\pm$3 times the standard deviation obtained from the Gaussian fitting with a single- or multiple-component Gaussian profile. 
  
\begin{table}[ht!]
  \centering  
  \caption{Global Input Parameters for RADEX.}
  \label{tab: Paramters for RADEX}
  \setlength{\tabcolsep}{1.0mm}{
  \begin{tabular}{l   c}
  \toprule[1pt]
  Free Input Parameters & Value \\
    \midrule[0.5pt]
  Sampling point number for $n_{\rm \ce{H2}}$ & 50 \\
  Sampling point number for $T_{\rm kin}$ & 56\\
  $T_{\rm kin, min}$ & 5\,K\\
  $T_{\rm kin, max}$ & 60\,K \\
  $n_{\rm H_2min}$ & $\rm 10^{3}$\,$\rm cm^{-3}$ \\
  $n_{\rm H_2max}$ & $\rm 10^{9}$\,$\rm cm^{-3}$\\
  \bottomrule[1pt]
  \end{tabular}}
\end{table}
\begin{figure}[ht!]
 \centering
  \includegraphics[width=0.5\textwidth]{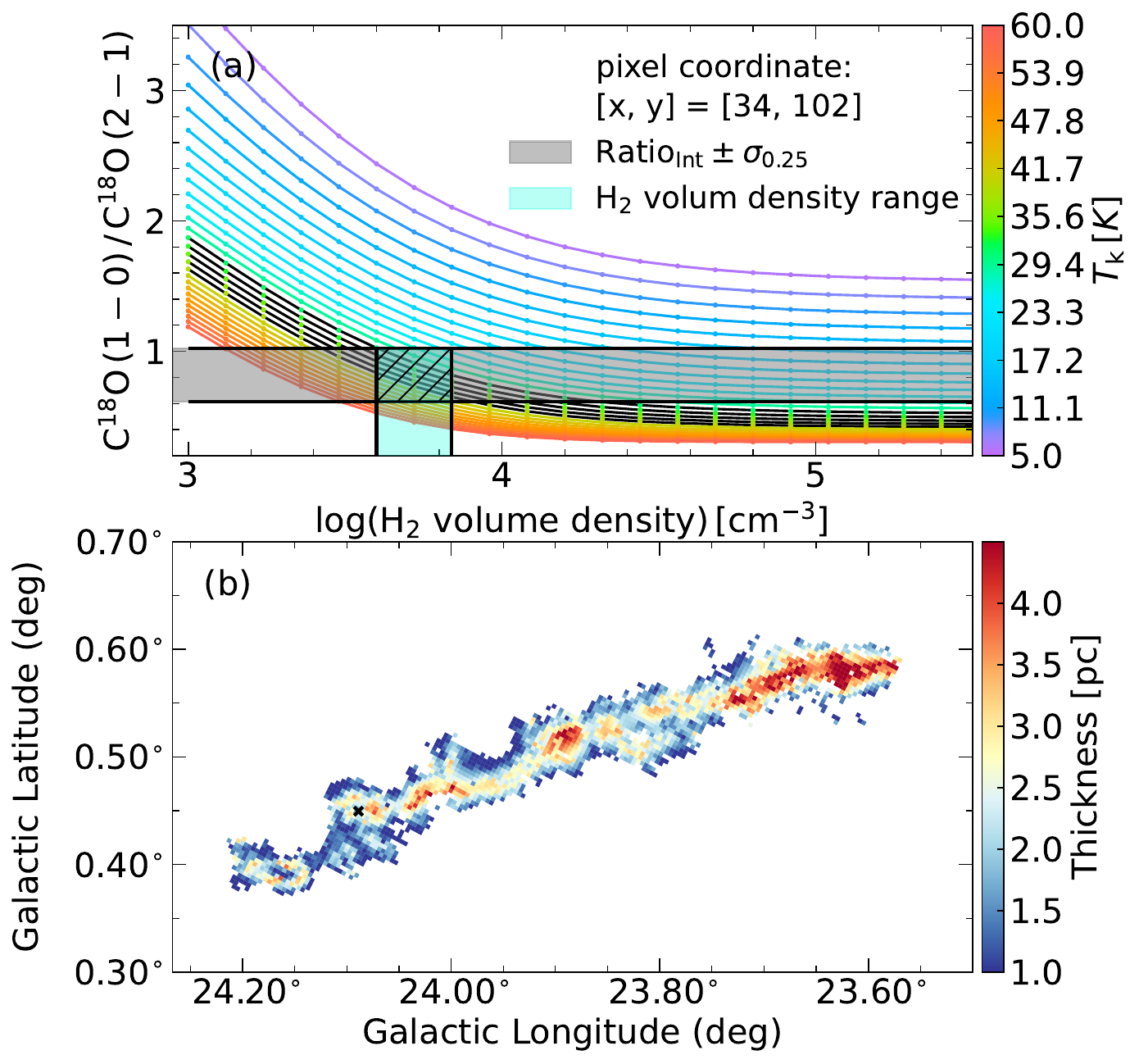}
  \caption{An example of \ce{H2} volume density estimation and the thickness of G24 along the line of sight.
  (a) An example of estimating the range of \ce{H2} volume density by \ce{C^18O}\,(1--0)/\ce{C^18O}\,(2--1) integrated intensity ratio at pixel [x, y]\,=\,[34, 102]. The ratio is a function of the logarithm of \ce{H2} volume density. The lines are colour-coded by the kinetic temperatures produced by the simulation. The black dashed curves are the dust temperature at pixel [x, y]\,=\,[34, 102] in Figure~\ref{img: G24_NH2andTemp_1} with $\pm$25\% uncertainty. The shaded grey area between the two horizontal black lines is the observed \ce{C^18O}\,(1--0)/\ce{C^18O}\,(2--1) integrated intensity ratio $\pm$25\% uncertainty. The vertical black lines indicate the intersections of the black curves and the horizontal lines. The range of \ce{H2} volume density is represented by translucent green and outlined with black hatched lines. (b) The thickness of G24 along the line of sight estimated by RADEX. The black cross is the pixel used as an example in panel (a).}
  \label{img: fil_thickness}
\end{figure}
 
Table~\ref{tab: Paramters for RADEX} lists the global input parameters for RADEX. The global input parameters apply to the entire map and they are the sampling point number in the kinetic temperature $T_{\rm kin}$, minimum and maximum kinetic temperatures ($T_{\rm kin, min}$, $T_{\rm kin, max}$), the sampling point number of \ce{H2} column density $n_{\ce{H2}}$, and minimum and maximum \ce{H2} volume density ($n_{\rm H_2min}$, $n_{\rm H_2max}$). Except for the global input parameters, we also provided additional input parameters that vary in pixels, which are column density $N$(\ce{C^18O}) and FWHM of velocity dispersion of \ce{C^18O}\,(1--0) (see Section~\ref{subsec: Clmndens}). The range of $T_{\rm kin}$ is limited from 5\,K to 60\,K and the range of $n_{\ce{H2}}$ $10^3 - 10^8$\,cm$^{-3}$, large enough to include the dust temperature and column density determined in Section~\ref{subsec: Clmndens}. We selected the sampling point numbers for \( T_{\rm kin} \) and \( n_{\ce{H2}} \) as 50 and 56, respectively.

The outputs of RADEX are kinetic temperature ($T_{\rm kin}$), excitation temperature ($T_{\rm ex}$), optical depth ($\tau_{\ce{C^18O}}$), integrated intensity ratio (\ce{C^18O}\,(1--0)/\ce{C^18O}\,(2--1)) and $n_{\ce{H2}}$, and the based on these outputs, RADEX creates a grid of 50 ($n_{\ce{H2}}$)~$\times$ 56 ($T_{\rm kin}$) for \ce{C^18O}\,(1--0)/\ce{C^18O}\,(2--1) line intensity ratio, $T_{\rm ex}$, and $\tau_{\ce{C^18O}}$. We compared our observational measurements (i.e., $T_{\rm kin}$, $T_{\rm ex}$, $\tau_{\ce{C^18O}}$ and integrated intensity ratio \ce{C^18O}\,(1--0)/\ce{C^18O}\,(2--1)) on the grid to constrain the range of $n_{\ce{H2}}$. For the observational values of $T_{\rm kin}$ and $T_{\rm ex}$, the $T_{\rm dust}$ is used as $T_{\rm kin}$, and the $T_{\rm ex}$ is derived from \ce{^12CO}\,(1--0). Consequently, these adopted temperatures may introduce additional uncertainties when comparing the RADEX grid with observational measurements. Each parameter mentioned above is assigned an uncertainty of 25\% to provide a reasonable range for the $n_{\ce{H2}}$ in the parameter space. The final volume density for a pixel is determined as the median value within the $n_{\ce{H2}}$ range.

The upper panel (a) in Figure~\ref{img: fil_thickness} shows an example of a grid of the $n_{\ce{H2}}$ range in the x-axis and the $T_{\rm kin}$ in the colour bar, extracted at a pixel position [x, y]\,=\,[34, 102]. The integrated intensity ratio of \ce{C^18O}\,(1--0) and \ce{C^18O}\,(2--1) is shown on the left y-axis, with a $\pm25\%$ uncertainty interval indicated by the horizontal shaded area. The $T_{\rm kin}$ with a $\pm25\%$ uncertainty derived from this work are represented by the black dashed curves. The optimal fitting area for the observed \ce{C^18O} intensity ratio and the observational $T_{\rm kin}$ is indicated by the hatched region, which corresponds to the range of $n_{\ce{H2}}$. Overall, the volume densities of \ce{H2} fitted throughout G24 consistently are in a range of $10^3$--$10^4\,{\rm cm}^{-3}$. 

Based on the obtained $N$(\ce{H2}) and $n_{\rm \ce{H2}}$, the mean and median thickness along the LOS are both equal to 2.2 pc. The panel (b) of Figure~\ref{img: fil_thickness} displays the distribution of the thickness of dense gas along the LOS. The thicknesses along the LOS of the high-density regions of G24 are 2.5-4.0 pc, which are 2-4 times larger than the \ce{C^18O} projected FWHM$_{\rm deconv}$ of 1.07\,pc (see Table~\ref{tab: FWHM_estimation}) and are $\sim40$ times shorter than the length of G24 in the plane of the sky. This rejects the possibility that G24 is a projection of flat structures, such as edge-on sheets, along the LOS. The greater values of the thickness compared to the filament widths may be due to the assumption of $T_{\rm kin}$ being equal to the $T_{\rm dust}$. The gas temperature is, in general, expected to be well coupled to the dust temperature at densities above $\rm \sim3\times10^4\,cm^{-3}$ \citep{Galli_2002AnA...394..275G}, which is higher than the typical $n_{\ce{H2}}$ found with the \ce{C^18O} line ratios. Thus, the actual kinetic temperature may be lower than the dust temperature, leading to a higher volume density than the value obtained under the previously mentioned assumption. 

\cite{Li-Goldsmith_2012ApJ...756...12L} used \ce{HC3N}\,$J$=2--1 and 10--9 transitions to estimate the \ce{H2} volume density and the LOS dimension of the high-density regions of the filament B213 in the Taurus molecular cloud based on results from RADEX. They demonstrated that a kinetic temperature of 10\,K seems appropriate for the optically thin \ce{HC3N} emission. We tried to estimate the volume densities for a $T_{\rm kin}$ of 10\,K. However, in some areas, the thickness values are unrealistically small. For example, the thickness of the shell around the \htworegion\ region (introduced in Section~\ref{sec: Analysis} and presented by arch-shaped radio emission white contours in Figure~\ref{img: RGB}) is on the order of $10^{-3}$\,pc. Establishing a uniform kinetic temperature is challenging and requires the use of dense gas tracers, such as ammonia. Additionally, this estimation does not consider the effect of the inclination angle on the thickness, leading to a potential overestimation of that thickness. 
 
\bibliography{main}{}

\begin{thebibliography}{}
\expandafter\ifx\csname natexlab\endcsname\relax\def\natexlab#1{#1}\fi
\providecommand{\url}[1]{\href{#1}{#1}}
\providecommand{\dodoi}[1]{doi:~\href{http://doi.org/#1}{\nolinkurl{#1}}}
\providecommand{\doeprint}[1]{\href{http://ascl.net/#1}{\nolinkurl{http://ascl.net/#1}}}
\providecommand{\doarXiv}[1]{\href{https://arxiv.org/abs/#1}{\nolinkurl{https://arxiv.org/abs/#1}}}

\bibitem[{Abreu-Vicente {et~al.}(2016)Abreu-Vicente, Ragan, Kainulainen, Henning, Beuther, \& Johnston}]{Abreu-Vicente_2016AnA...590A.131A}
Abreu-Vicente, J., Ragan, S., Kainulainen, J., {et~al.} 2016, A\&A, 590, A131, \dodoi{10.1051/0004-6361/201527674}

\bibitem[{Adams(1991)}]{Adams_1991ApJ...382..544A}
Adams, F.~C. 1991, ApJ, 382, 544, \dodoi{10.1086/170741}

\bibitem[{Anderson {et~al.}(2021)Anderson, Peretto, Ragan, Rigby, Avison, Duarte-Cabral, Fuller, Shirley, Traficante, \& Williams}]{Anderson_2021MNRAS.508.2964A}
Anderson, M., Peretto, N., Ragan, S.~E., {et~al.} 2021, MNRAS, 508, 2964, \dodoi{10.1093/mnras/stab2674}

\bibitem[{Andr{\'{e}} {et~al.}(2014)Andr{\'{e}}, {Di Francesco}, Ward-Thompson, Inutsuka, Pudritz, \& Pineda}]{André_2014prpl.conf...27A}
Andr{\'{e}}, P., {Di Francesco}, J., Ward-Thompson, D., {et~al.} 2014, in Protostars and Planets VI, ed. H.~Beuther, R.~S. Klessen, C.~P. Dullemond, \& T.~Henning, 27--51, \dodoi{10.2458/azu_uapress_9780816531240-ch002}

\bibitem[{Andr{\'{e}} {et~al.}(2022)Andr{\'{e}}, Palmeirim, \& Arzoumanian}]{André_2022AnA...667L...1A}
Andr{\'{e}}, P., Palmeirim, P., \& Arzoumanian, D. 2022, A\&A, 667, L1, \dodoi{10.1051/0004-6361/202244541}

\bibitem[{Andr{\'{e}} {et~al.}(2010)Andr{\'{e}}, Men'shchikov, Bontemps, K{\"{o}}nyves, Motte, Schneider, Didelon, Minier, Saraceno, Ward-Thompson, di~Francesco, White, Molinari, Testi, Abergel, Griffin, Henning, Royer, Mer\'\in, Vavrek, Attard, Arzoumanian, Wilson, Ade, Aussel, Baluteau, Benedettini, Bernard, Blommaert, Cambr{\'{e}}sy, Cox, di~Giorgio, Hargrave, Hennemann, Huang, Kirk, Krause, Launhardt, Leeks, {Le Pennec}, Li, Martin, Maury, Olofsson, Omont, Peretto, Pezzuto, Prusti, Roussel, Russeil, Sauvage, Sibthorpe, Sicilia-Aguilar, Spinoglio, Waelkens, Woodcraft, \& Zavagno}]{André_2010AnA...518L.102A}
Andr{\'{e}}, P., Men'shchikov, A., Bontemps, S., {et~al.} 2010, A\&A, 518, L102, \dodoi{10.1051/0004-6361/201014666}

\bibitem[{Aniano {et~al.}(2011)Aniano, Draine, Gordon, \& Sandstrom}]{Aniano_2011PASP..123.1218A}
Aniano, G., Draine, B., Gordon, K., \& Sandstrom, K. 2011, PASP, 123, 1218, \dodoi{10.1086/662219}

\bibitem[{Arzoumanian {et~al.}(2013)Arzoumanian, Andr{\'{e}}, Peretto, \& K{\"{o}}nyves}]{Arzoumanian_2013AnA...553A.119A}
Arzoumanian, D., Andr{\'{e}}, P., Peretto, N., \& K{\"{o}}nyves, V. 2013, A\&A, 553, A119, \dodoi{10.1051/0004-6361/201220822}

\bibitem[{Arzoumanian {et~al.}(2011)Arzoumanian, Andr{\'{e}}, Didelon, K{\"{o}}nyves, Schneider, Men'shchikov, Sousbie, Zavagno, Bontemps, di~Francesco, Griffin, Hennemann, Hill, Kirk, Martin, Minier, Molinari, Motte, Peretto, Pezzuto, Spinoglio, Ward-Thompson, White, \& Wilson}]{Arzoumanian_2011AnA...529L...6A}
Arzoumanian, D., Andr{\'{e}}, P., Didelon, P., {et~al.} 2011, A\&A, 529, L6, \dodoi{10.1051/0004-6361/201116596}

\bibitem[{Arzoumanian {et~al.}(2019)Arzoumanian, Andr{\'{e}}, K{\"{o}}nyves, Palmeirim, Roy, Schneider, Benedettini, Didelon, {Di Francesco}, Kirk, \& Ladjelate}]{Arzoumanian_2019AnA...621A..42A}
Arzoumanian, D., Andr{\'{e}}, P., K{\"{o}}nyves, V., {et~al.} 2019, A\&A, 621, A42, \dodoi{10.1051/0004-6361/201832725}

\bibitem[{Baluev(2008)}]{Baluev_2008MNRAS.385.1279B}
Baluev, R. 2008, MNRAS, 385, 1279, \dodoi{10.1111/j.1365-2966.2008.12689.x}

\bibitem[{Battersby \& Bally(2014)}]{Battersby_2014ASSP...36..417B}
Battersby, C., \& Bally, J. 2014, in Astrophysics and Space Science Proceedings, Vol.~36, The Labyrinth of Star Formation, 417, \dodoi{10.1007/978-3-319-03041-8_82}

\bibitem[{Battersby {et~al.}(2014)Battersby, Ginsburg, Bally, Longmore, Dunham, \& Darling}]{Battersby_2014ApJ...787..113B}
Battersby, C., Ginsburg, A., Bally, J., {et~al.} 2014, ApJ, 787, 113, \dodoi{10.1088/0004-637X/787/2/113}

\bibitem[{Beaumont {et~al.}(2013)Beaumont, Offner, Shetty, Glover, \& Goodman}]{Beaumont_2013ApJ...777..173B}
Beaumont, C.~N., Offner, S.~S., Shetty, R., Glover, S.~C., \& Goodman, A.~A. 2013, ApJ, 777, 173, \dodoi{10.1088/0004-637X/777/2/173}

\bibitem[{Benjamin {et~al.}(2003)Benjamin, Churchwell, Babler, Bania, Clemens, Cohen, Dickey, Indebetouw, Jackson, Kobulnicky, Lazarian, Marston, Mathis, Meade, Seager, Stolovy, Watson, Whitney, Wolff, \& Wolfire}]{Benjamin_2003PASP..115..953B}
Benjamin, R.~A., Churchwell, E., Babler, B.~L., {et~al.} 2003, PASP, 115, 953, \dodoi{10.1086/376696}

\bibitem[{Beuther {et~al.}(2010)Beuther, Henning, Linz, Krause, Nielbock, \& Steinacker}]{Beuther_2010AnA...518L..78B}
Beuther, H., Henning, T., Linz, H., {et~al.} 2010, A\&A, 518, L78, \dodoi{10.1051/0004-6361/201014532}

\bibitem[{Bourke {et~al.}(1997)Bourke, Garay, Lehtinen, K{\"{o}}hnenkamp, Launhardt, Nyman, May, Robinson, \& Hyland}]{Bourke_1997ApJ...476..781B}
Bourke, T.~L., Garay, G., Lehtinen, K.~K., {et~al.} 1997, ApJ, 476, 781, \dodoi{10.1086/303642}

\bibitem[{Carey {et~al.}(2009)Carey, Noriega-Crespo, Mizuno, Shenoy, Paladini, Kraemer, Price, Flagey, Ryan, Ingalls, Kuchar, {Pinheiro Gon{\c{c}}alves}, Indebetouw, Billot, Marleau, Padgett, Rebull, Bressert, Ali, Molinari, Martin, Berriman, Boulanger, Latter, Miville-Deschenes, Shipman, \& Testi}]{Carey_2009PASP..121...76C}
Carey, S., Noriega-Crespo, A., Mizuno, D., {et~al.} 2009, PASP, 121, 76, \dodoi{10.1086/596581}

\bibitem[{Carter {et~al.}(2012)Carter, Lazareff, Maier, Chenu, Fontana, Bortolotti, Boucher, Navarrini, Blanchet, Greve, John, Kramer, Morel, Navarro, Pe{\~{n}}alver, Schuster, \& Thum}]{Carter_2012AnA...538A..89C}
Carter, M., Lazareff, B., Maier, D., {et~al.} 2012, A\&A, 538, A89, \dodoi{10.1051/0004-6361/201118452}

\bibitem[{Caselli {et~al.}(2002)Caselli, Walmsley, Zucconi, Tafalla, Dore, \& Myers}]{Caselli_2002ApJ...565..344C}
Caselli, P., Walmsley, C., Zucconi, A., {et~al.} 2002, ApJ, 565, 344, \dodoi{10.1086/324302}

\bibitem[{Clarke {et~al.}(2019)Clarke, Williams, Ib{\'{a}}{\~{n}}ez-Mej{\'{i}}a, \& Walch}]{Clarke_2019MNRAS.484.4024C}
Clarke, S., Williams, G., Ib{\'{a}}{\~{n}}ez-Mej{\'{i}}a, J., \& Walch, S. 2019, MNRAS, 484, 4024, \dodoi{10.1093/mnras/stz248}

\bibitem[{Colombo {et~al.}(2021)Colombo, K{\"{o}}nig, Urquhart, Wyrowski, Mattern, Menten, Lee, Brand, Wienen, Mazumdar, Schuller, \& Leurini}]{Colombo_2021AnA...655L...2C}
Colombo, D., K{\"{o}}nig, C., Urquhart, J., {et~al.} 2021, A\&A, 655, L2, \dodoi{10.1051/0004-6361/202142182}

\bibitem[{Cox {et~al.}(2016)Cox, Arzoumanian, Andr{\'{e}}, Rygl, Prusti, Men'shchikov, Royer, K{\'{o}}sp{\'{a}}l, Palmeirim, Ribas, K{\"{o}}nyves, Bernard, Schneider, Bontemps, Merin, Vavrek, {Alves de Oliveira}, Didelon, Pilbratt, \& Waelkens}]{Cox_2016AnA...590A.110C}
Cox, N., Arzoumanian, D., Andr{\'{e}}, P., {et~al.} 2016, A\&A, 590, A110, \dodoi{10.1051/0004-6361/201527068}

\bibitem[{Dewangan {et~al.}(2019)Dewangan, Pirogov, Ryabukhina, Ojha, \& Zinchenko}]{Dewangan_2019ApJ...877....1D}
Dewangan, L., Pirogov, L., Ryabukhina, O., Ojha, D., \& Zinchenko, I. 2019, ApJ, 877, 1, \dodoi{10.3847/1538-4357/ab1aa6}

\bibitem[{Fiege \& Pudritz(2000)}]{Fiege_2000MNRAS.311..105F}
Fiege, J.~D., \& Pudritz, R.~E. 2000, MNRAS, 311, 105, \dodoi{10.1046/j.1365-8711.2000.03067.x}

\bibitem[{Frerking {et~al.}(1982)Frerking, Langer, \& Wilson}]{Ferking_1982ApJ...262..590F}
Frerking, M., Langer, W., \& Wilson, R. 1982, ApJ, 262, 590, \dodoi{10.1086/160451}

\bibitem[{Galli {et~al.}(2002)Galli, Walmsley, \& Gon{\c{c}}alves}]{Galli_2002AnA...394..275G}
Galli, D., Walmsley, M., \& Gon{\c{c}}alves, J. 2002, A\&A, 394, 275, \dodoi{10.1051/0004-6361:20021125}

\bibitem[{Garden {et~al.}(1991)Garden, Hayashi, Gatley, Hasegawa, \& Kaifu}]{Garden_1991ApJ...374..540G}
Garden, R., Hayashi, M., Gatley, I., Hasegawa, T., \& Kaifu, N. 1991, ApJ, 374, 540, \dodoi{10.1086/170143}

\bibitem[{Ge \& Wang(2022)}]{Ge_2022ApJS..259...36G}
Ge, Y., \& Wang, K. 2022, ApJS, 259, 36, \dodoi{10.3847/1538-4365/ac4a76}

\bibitem[{Ge {et~al.}(2023)Ge, Wang, Duarte-Cabral, Pettitt, Dobbs, S{\'{a}}nchez-Monge, Neralwar, Urquhart, Colombo, Dur{\'{a}}n-Camacho, Beuther, Bronfman, Rigby, Eden, Neupane, Barnes, Henning, \& Yang}]{Ge_2023AnA...675A.119G}
Ge, Y., Wang, K., Duarte-Cabral, A., {et~al.} 2023, A\&A, 675, A119, \dodoi{10.1051/0004-6361/202245784}

\bibitem[{Goodman {et~al.}(2014)Goodman, Alves, Beaumont, Benjamin, Borkin, Burkert, Dame, Jackson, Kauffmann, Robitaille, \& Smith}]{Goodman_2014ApJ...797...53G}
Goodman, A.~A., Alves, J., Beaumont, C.~N., {et~al.} 2014, ApJ, 797, 53, \dodoi{10.1088/0004-637X/797/1/53}

\bibitem[{Griffin {et~al.}(2010)Griffin, Abergel, Abreu, Ade, Andr{\'{e}}, Augueres, Babbedge, Bae, Baillie, Baluteau, Barlow, Bendo, Benielli, Bock, Bonhomme, Brisbin, Brockley-Blatt, Caldwell, Cara, Castro-Rodriguez, Cerulli, Chanial, Chen, Clark, Clements, Clerc, Coker, Communal, Conversi, Cox, Crumb, Cunningham, Daly, Davis, de~Antoni, Delderfield, Devin, di~Giorgio, Didschuns, Dohlen, Donati, Dowell, Dowell, Duband, Dumaye, Emery, Ferlet, Ferrand, Fontignie, Fox, Franceschini, Frerking, Fulton, Garcia, Gastaud, Gear, Glenn, Goizel, Griffin, Grundy, Guest, Guillemet, Hargrave, Harwit, Hastings, Hatziminaoglou, Herman, Hinde, Hristov, Huang, Imhof, Isaak, Israelsson, Ivison, Jennings, Kiernan, King, Lange, Latter, Laurent, Laurent, Leeks, Lellouch, Levenson, Li, Li, Lilienthal, Lim, Liu, Lu, Madden, Mainetti, Marliani, McKay, Mercier, Molinari, Morris, Moseley, Mulder, Mur, Naylor, Nguyen, O'Halloran, Oliver, Olofsson, Olofsson, Orfei, Page, Pain, Panuzzo, Papageorgiou, Parks, Parr-Burman, Pearce, Pearson,
  P{\'{e}}rez-Fournon, Pinsard, Pisano, Podosek, Pohlen, Polehampton, Pouliquen, Rigopoulou, Rizzo, Roseboom, Roussel, Rowan-Robinson, Rownd, Saraceno, Sauvage, Savage, Savini, Sawyer, Scharmberg, Schmitt, Schneider, Schulz, Schwartz, Shafer, Shupe, Sibthorpe, Sidher, Smith, Smith, Smith, Spencer, Stobie, Sudiwala, Sukhatme, Surace, Stevens, Swinyard, Trichas, Tourette, Triou, Tseng, Tucker, Turner, Vaccari, Valtchanov, Vigroux, Virique, Voellmer, Walker, Ward, Waskett, Weilert, Wesson, White, Whitehouse, Wilson, Winter, Woodcraft, Wright, Xu, Zavagno, Zemcov, Zhang, \& Zonca}]{Griffin_2010AnA...518L...3G}
Griffin, M., Abergel, A., Abreu, A., {et~al.} 2010, A\&A, 518, L3, \dodoi{10.1051/0004-6361/201014519}

\bibitem[{Guo {et~al.}(2021)Guo, Chen, Feng, Sun, Wang, Su, Sun, Ao, Zhang, Zhou, Yuan, \& Yang}]{Guo_2021ApJ...921...23G}
Guo, W., Chen, X., Feng, J., {et~al.} 2021, ApJ, 921, 23, \dodoi{10.3847/1538-4357/ac15fe}

\bibitem[{G{\"{u}}sten {et~al.}(2006)G{\"{u}}sten, Nyman, Schilke, Menten, Cesarsky, \& Booth}]{Güsten_2006AnA...454L..13G}
G{\"{u}}sten, R., Nyman, L.~r., Schilke, P., {et~al.} 2006, A\&A, 454, L13, \dodoi{10.1051/0004-6361:20065420}

\bibitem[{Hacar {et~al.}(2017)Hacar, Alves, Tafalla, \& Goicoechea}]{Hacar_2017A&A...602L...2H}
Hacar, A., Alves, J., Tafalla, M., \& Goicoechea, J. 2017, A\&A, 602, L2, \dodoi{10.1051/0004-6361/201730732}

\bibitem[{Hacar {et~al.}(2023)Hacar, Clark, Heitsch, Kainulainen, Panopoulou, Seifried, \& Smith}]{Hacar_2023ASPC..534..153H}
Hacar, A., Clark, S., Heitsch, F., {et~al.} 2023, in Astronomical Society of the Pacific Conference Series, Vol. 534, Protostars and Planets VII, ed. S.~Inutsuka, Y.~Aikawa, T.~Muto, K.~Tomida, \& M.~Tamura, 153, \dodoi{10.48550/arXiv.2203.09562}

\bibitem[{Hacar {et~al.}(2018)Hacar, Tafalla, Forbrich, Alves, Meingast, Grossschedl, \& Teixeira}]{Hacar_2018AnA...610A..77H}
Hacar, A., Tafalla, M., Forbrich, J., {et~al.} 2018, A\&A, 610, A77, \dodoi{10.1051/0004-6361/201731894}

\bibitem[{Hennemann {et~al.}(2012)Hennemann, Motte, Schneider, Didelon, Hill, Arzoumanian, Bontemps, Csengeri, Andr{\'{e}}, Konyves, Louvet, Marston, Men'shchikov, Minier, {Nguyen Luong}, Palmeirim, Peretto, Sauvage, Zavagno, Anderson, Bernard, {Di Francesco}, Elia, Li, Martin, Molinari, Pezzuto, Russeil, Rygl, Schisano, Spinoglio, Sousbie, Ward-Thompson, \& White}]{Hennemann_2012AnA...543L...3H}
Hennemann, M., Motte, F., Schneider, N., {et~al.} 2012, A\&A, 543, L3, \dodoi{10.1051/0004-6361/201219429}

\bibitem[{Hill {et~al.}(2012)Hill, Andr{\'{e}}, Arzoumanian, Motte, Minier, Men'shchikov, Didelon, Hennemann, K{\"{o}}nyves, Nguyen-Luong, Palmeirim, Peretto, Schneider, Bontemps, Louvet, Elia, Giannini, Rev{\'{e}}ret, {Le Pennec}, Rodriguez, Boulade, Doumayrou, Dubreuil, Gallais, Lortholary, Martignac, Talvard, \& {De Breuck}}]{Hill_2012AnA...548L...6H}
Hill, T., Andr{\'{e}}, P., Arzoumanian, D., {et~al.} 2012, A\&A, 548, L6, \dodoi{10.1051/0004-6361/201220504}

\bibitem[{Inutsuka \& Miyama(1992)}]{Inutsuka_1992ApJ...388..392I}
Inutsuka, S.-I., \& Miyama, S.~M. 1992, ApJ, 388, 392, \dodoi{10.1086/171162}

\bibitem[{Jackson {et~al.}(2006)Jackson, Rathborne, Shah, Simon, Bania, Clemens, Chambers, Johnson, Dormody, Lavoie, \& Heyer}]{Jackson_2006ApJS..163..145J}
Jackson, J., Rathborne, J., Shah, R., {et~al.} 2006, ApJS, 163, 145, \dodoi{10.1086/500091}

\bibitem[{Jackson {et~al.}(2010)Jackson, Finn, Chambers, Rathborne, \& Simon}]{Jackson_2010ApJ...719L.185J}
Jackson, J.~M., Finn, S.~C., Chambers, E.~T., Rathborne, J.~M., \& Simon, R. 2010, ApJL, 719, L185, \dodoi{10.1088/2041-8205/719/2/L185}

\bibitem[{Kainulainen {et~al.}(2016)Kainulainen, Hacar, Alves, Beuther, Bouy, \& Tafalla}]{Kainulainen_2016AnA...586A..27K}
Kainulainen, J., Hacar, A., Alves, J., {et~al.} 2016, A\&A, 586, A27, \dodoi{10.1051/0004-6361/201526017}

\bibitem[{Kauffmann {et~al.}(2008)Kauffmann, Bertoldi, Bourke, {Evans II}, \& Lee}]{Kauffmann_2008AnA...487..993K}
Kauffmann, J., Bertoldi, F., Bourke, T., {Evans II}, N., \& Lee, C. 2008, A\&A, 487, 993, \dodoi{10.1051/0004-6361:200809481}

\bibitem[{Kirk {et~al.}(2013)Kirk, Myers, Bourke, Gutermuth, Hedden, \& Wilson}]{Kirk_2013ApJ...766..115K}
Kirk, H., Myers, P.~C., Bourke, T.~L., {et~al.} 2013, ApJ, 766, 115, \dodoi{10.1088/0004-637X/766/2/115}

\bibitem[{K{\"{o}}nyves {et~al.}(2015)K{\"{o}}nyves, Andr{\'{e}}, Men'shchikov, Palmeirim, Arzoumanian, Schneider, Roy, Didelon, Maury, Shimajiri, {Di Francesco}, Bontemps, Peretto, Benedettini, Bernard, Elia, Griffin, Hill, Kirk, Ladjelate, Marsh, Martin, Motte, {Nguy{\^{e}}n Luong}, Pezzuto, Roussel, Rygl, Sadavoy, Schisano, Spinoglio, Ward-Thompson, \& White}]{Könyves_2015AnA...584A..91K}
K{\"{o}}nyves, V., Andr{\'{e}}, P., Men'shchikov, A., {et~al.} 2015, A\&A, 584, A91, \dodoi{10.1051/0004-6361/201525861}

\bibitem[{Kumar {et~al.}(2020)Kumar, Palmeirim, Arzoumanian, \& Inutsuka}]{Kumar_2020AnA...642A..87K}
Kumar, M., Palmeirim, P., Arzoumanian, D., \& Inutsuka, S. 2020, A\&A, 642, A87, \dodoi{10.1051/0004-6361/202038232}

\bibitem[{Ladd {et~al.}(1991)Ladd, Adams, Casey, Davidson, Fuller, Harper, Myers, \& Padman}]{Ladd_1991ApJ...382..555L}
Ladd, E., Adams, F.~C., Casey, S., {et~al.} 1991, ApJ, 382, 555, \dodoi{10.1086/170742}

\bibitem[{Li \& Goldsmith(2012)}]{Li-Goldsmith_2012ApJ...756...12L}
Li, D., \& Goldsmith, P.~F. 2012, ApJ, 756, 12, \dodoi{10.1088/0004-637X/756/1/12}

\bibitem[{Li {et~al.}(2016)Li, Urquhart, Leurini, Csengeri, Wyrowski, Menten, \& Schuller}]{Li_2016AnA...591A...5L}
Li, G.~X., Urquhart, J.~S., Leurini, S., {et~al.} 2016, A\&A, 591, A5, \dodoi{10.1051/0004-6361/201527468}

\bibitem[{Li {et~al.}(2013)Li, Wyrowski, Menten, \& Belloche}]{Li_2013AnA...559A..34L}
Li, G.~X., Wyrowski, F., Menten, K., \& Belloche, A. 2013, A\&A, 559, A34, \dodoi{10.1051/0004-6361/201322411}

\bibitem[{Lin {et~al.}(2016)Lin, Liu, Li, Zhang, Ginsburg, Pineda, Qian, Galv{\'{a}}n-Madrid, McLeod, Rosolowsky, Dale, Immer, Koch, Longmore, Walker, \& Testi}]{Lin_2016ApJ...828...32L}
Lin, Y., Liu, H.~B., Li, D., {et~al.} 2016, ApJ, 828, 32, \dodoi{10.3847/0004-637X/828/1/32}

\bibitem[{Liu {et~al.}(2012)Liu, Quintana-Lacaci, Wang, Ho, Li, Zhang, \& Zhang}]{Liu_2012ApJ...745...61L}
Liu, H.~B., Quintana-Lacaci, G., Wang, K., {et~al.} 2012, ApJ, 745, 61, \dodoi{10.1088/0004-637X/745/1/61}

\bibitem[{Liu {et~al.}(2019)Liu, Stutz, \& Yuan}]{Liu_2019MNRAS.487.1259L}
Liu, H.-L., Stutz, A., \& Yuan, J.-H. 2019, MNRAS, 487, 1259, \dodoi{10.1093/mnras/stz1340}

\bibitem[{Liu {et~al.}(2015)Liu, Wu, Li, Yuan, Liu, \& Dong}]{Liu_2015ApJ...798...30L}
Liu, H.-L., Wu, Y., Li, J., {et~al.} 2015, ApJ, 798, 30, \dodoi{10.1088/0004-637X/798/1/30}

\bibitem[{Mangum \& Shirley(2016)}]{Mangum-Shirley_2016PASP..128b9201M}
Mangum, J.~G., \& Shirley, Y.~L. 2016, {Erratam to: How to Calculate Molecular Column Density, (Publications of the Astronomical Society of the Pacific, (2015) 127, 949)}, Publications of the Astronomical Society of the Pacific, Volume 128, Issue 960, pp. 029201 (2016)., \dodoi{10.1088/1538-3873/128/960/029201}

\bibitem[{Mattern {et~al.}(2018)Mattern, Kauffmann, Csengeri, Urquhart, Leurini, Wyrowski, Giannetti, Barnes, Beuther, Bronfman, Duarte-Cabral, Henning, Kainulainen, Menten, Schisano, \& Schuller}]{Mattern_2018AnA...619A.166M}
Mattern, M., Kauffmann, J., Csengeri, T., {et~al.} 2018, A\&A, 619, A166, \dodoi{10.1051/0004-6361/201833406}

\bibitem[{Molinari {et~al.}(2010{\natexlab{a}})Molinari, Swinyard, Bally, Barlow, Bernard, Martin, Moore, Noriega-Crespo, Plume, Testi, Zavagno, Abergel, Ali, Andr{\'{e}}, Baluteau, Benedettini, Bern{\'{e}}, Billot, Blommaert, Bontemps, Boulanger, Brand, Brunt, Burton, Campeggio, Carey, Caselli, Cesaroni, Cernicharo, Chakrabarti, Chrysostomou, Codella, Cohen, Compiegne, Davis, de~Bernardis, de~Gasperis, {Di Francesco}, di~Giorgio, Elia, Faustini, Fischera, Fukui, Fuller, Ganga, Garcia-Lario, Giard, Giardino, Glenn, Goldsmith, Griffin, Hoare, Huang, Jiang, Joblin, Joncas, Juvela, Kirk, Lagache, Li, Lim, Lord, Lucas, Maiolo, Marengo, Marshall, Masi, Massi, Matsuura, Meny, Minier, Miville-Desch{\^{e}}nes, Montier, Motte, M{\"{u}}ller, Natoli, Neves, Olmi, Paladini, Paradis, Pestalozzi, Pezzuto, Piacentini, Pomar{\`{e}}s, Popescu, Reach, Richer, Ristorcelli, Roy, Royer, Russeil, Saraceno, Sauvage, Schilke, Schneider-Bontemps, Schuller, Schultz, Shepherd, Sibthorpe, Smith, Smith, Spinoglio, Stamatellos, Strafella,
  Stringfellow, Sturm, Taylor, Thompson, Tuffs, Umana, Valenziano, Vavrek, Viti, Waelkens, Ward-Thompson, White, Wyrowski, Yorke, \& Zhang}]{Molinari_2010PASP..122..314M}
Molinari, S., Swinyard, B., Bally, J., {et~al.} 2010{\natexlab{a}}, PASP, 122, 314, \dodoi{10.1086/651314}

\bibitem[{Molinari {et~al.}(2010{\natexlab{b}})Molinari, Swinyard, Bally, Barlow, Bernard, Martin, Moore, Noriega-Crespo, Plume, Testi, Zavagno, Abergel, Ali, Anderson, Andr{\'{e}}, Baluteau, Battersby, Beltr{\'{a}}n, Benedettini, Billot, Blommaert, Bontemps, Boulanger, Brand, Brunt, Burton, Calzoletti, Carey, Caselli, Cesaroni, Cernicharo, Chakrabarti, Chrysostomou, Cohen, Compiegne, de~Bernardis, de~Gasperis, di~Giorgio, Elia, Faustini, Flagey, Fukui, Fuller, Ganga, Garcia-Lario, Glenn, Goldsmith, Griffin, Hoare, Huang, Ikhenaode, Joblin, Joncas, Juvela, Kirk, Lagache, Li, Lim, Lord, Marengo, Marshall, Masi, Massi, Matsuura, Minier, Miville-Desch{\^{e}}nes, Montier, Morgan, Motte, Mottram, M{\"{u}}ller, Natoli, Neves, Olmi, Paladini, Paradis, Parsons, Peretto, Pestalozzi, Pezzuto, Piacentini, Piazzo, Polychroni, Pomar{\`{e}}s, Popescu, Reach, Ristorcelli, Robitaille, Robitaille, Rod{\'{o}}n, Roy, Royer, Russeil, Saraceno, Sauvage, Schilke, Schisano, Schneider, Schuller, Schulz, Sibthorpe, Smith, Smith,
  Spinoglio, Stamatellos, Strafella, Stringfellow, Sturm, Taylor, Thompson, Traficante, Tuffs, Umana, Valenziano, Vavrek, Veneziani, Viti, Waelkens, Ward-Thompson, White, Wilcock, Wyrowski, Yorke, \& Zhang}]{Molinari_2010AnA...518L.100M}
---. 2010{\natexlab{b}}, A\&A, 518, L100, \dodoi{10.1051/0004-6361/201014659}

\bibitem[{Motte {et~al.}(2018)Motte, Bontemps, \& Louvet}]{Motte_2018ARAnA..56...41M}
Motte, F., Bontemps, S., \& Louvet, F. 2018, ARA\&A, 56, 41, \dodoi{10.1146/annurev-astro-091916-055235}

\bibitem[{Myers(2009)}]{Myers_2009ApJ...700.1609M}
Myers, P.~C. 2009, ApJ, 700, 1609, \dodoi{10.1088/0004-637X/700/2/1609}

\bibitem[{Nagasawa(1987)}]{Nagasawa_1987PThPh..77..635N}
Nagasawa, M. 1987, Progress of Theoretical Physics, 77, 635, \dodoi{10.1143/PTP.77.635}

\bibitem[{Nakamura \& Li(2008)}]{Nakamura_2008ApJ...687..354N}
Nakamura, F., \& Li, Z.-Y. 2008, ApJ, 687, 354, \dodoi{10.1086/591641}

\bibitem[{{Nguyen Luong} {et~al.}(2011){Nguyen Luong}, Motte, Hennemann, Hill, Rygl, Schneider, Bontemps, Men'shchikov, Andr{\'{e}}, Peretto, Anderson, Arzoumanian, Deharveng, Didelon, di~Francesco, Griffin, Kirk, K{\"{o}}nyves, Martin, Maury, Minier, Molinari, Pestalozzi, Pezzuto, Reid, Roussel, Sauvage, Schuller, Testi, Ward-Thompson, White, \& Zavagno}]{Nguyen_Luong_2011AnA...535A..76N}
{Nguyen Luong}, Q., Motte, F., Hennemann, M., {et~al.} 2011, A\&A, 535, A76, \dodoi{10.1051/0004-6361/201117831}

\bibitem[{Ostriker(1964)}]{Ostriker_1964ApJ...140.1056O}
Ostriker, J. 1964, ApJ, 140, 1056, \dodoi{10.1086/148005}

\bibitem[{Ott(2010)}]{Ott_2010ASPC..434..139O}
Ott, S. 2010, in Astronomical Society of the Pacific Conference Series, Vol. 434, Astronomical Data Analysis Software and Systems XIX, ed. Y.~Mizumoto, K.~I. Morita, \& M.~Ohishi, 139, \dodoi{10.48550/arXiv.1011.1209}

\bibitem[{Palmeirim {et~al.}(2013)Palmeirim, Andr{\'{e}}, Kirk, Ward-Thompson, Arzoumanian, K{\"{o}}nyves, Didelon, Schneider, Benedettini, Bontemps, {Di Francesco}, Elia, Griffin, Hennemann, Hill, Martin, Men'shchikov, Molinari, Motte, {Nguyen Luong}, Nutter, Peretto, Pezzuto, Roy, Rygl, Spinoglio, \& White}]{Palmeirim_2013AnA...550A..38P}
Palmeirim, P., Andr{\'{e}}, P., Kirk, J., {et~al.} 2013, A\&A, 550, A38, \dodoi{10.1051/0004-6361/201220500}

\bibitem[{Panopoulou {et~al.}(2022)Panopoulou, Clark, Hacar, Heitsch, Kainulainen, Ntormousi, Seifried, \& Smith}]{Panopoulou_2022AnA...657L..13P}
Panopoulou, G., Clark, S., Hacar, A., {et~al.} 2022, A\&A, 657, L13, \dodoi{10.1051/0004-6361/202142281}

\bibitem[{Peretto \& Fuller(2009)}]{Peretto_2009AnA...505..405P}
Peretto, N., \& Fuller, G. 2009, A\&A, 505, 405, \dodoi{10.1051/0004-6361/200912127}

\bibitem[{Peretto {et~al.}(2010)Peretto, Fuller, Plume, Anderson, Bally, Battersby, Beltran, Bernard, Calzoletti, Digiorgio, Faustini, Kirk, Lenfestey, Marshall, Martin, Molinari, Montier, Motte, Ristorcelli, Rod{\'{o}}n, Smith, Traficante, Veneziani, Ward-Thompson, \& Wilcock}]{Peretto_2010AnA...518L..98P}
Peretto, N., Fuller, G., Plume, R., {et~al.} 2010, A\&A, 518, L98, \dodoi{10.1051/0004-6361/201014652}

\bibitem[{Peretto {et~al.}(2013)Peretto, Fuller, Duarte-Cabral, Avison, Hennebelle, Pineda, Andr{\'{e}}, Bontemps, Motte, Schneider, \& Molinari}]{Peretto_2013AnA...555A.112P}
Peretto, N., Fuller, G., Duarte-Cabral, A., {et~al.} 2013, A\&A, 555, A112, \dodoi{10.1051/0004-6361/201321318}

\bibitem[{Peretto {et~al.}(2014)Peretto, Fuller, Andr{\'{e}}, Arzoumanian, Rivilla, Bardeau, {Duarte Puertas}, {Guzman Fernandez}, Lenfestey, Li, Olguin, R{\"{o}}ck, de~Villiers, \& Williams}]{Peretto_2014AnA...561A..83P}
Peretto, N., Fuller, G., Andr{\'{e}}, P., {et~al.} 2014, A\&A, 561, A83, \dodoi{10.1051/0004-6361/201322172}

\bibitem[{Pety(2005)}]{Pety_2005sf2a.conf..721P}
Pety, J. 2005, in SF2A-2005: Semaine de l'Astrophysique Francaise, ed. F.~Casoli, T.~Contini, J.~Hameury, \& L.~Pagani, 721.
\newblock \url{https://ui.adsabs.harvard.edu/abs/2005sf2a.conf..721P}

\bibitem[{Poglitsch {et~al.}(2010)Poglitsch, Waelkens, Geis, Feuchtgruber, Vandenbussche, Rodriguez, Krause, Renotte, van Hoof, Saraceno, Cepa, Kerschbaum, Agn{\`{e}}se, Ali, Altieri, Andreani, Augueres, Balog, Barl, Bauer, Belbachir, Benedettini, Billot, Boulade, Bischof, Blommaert, Callut, Cara, Cerulli, Cesarsky, Contursi, Creten, {De Meester}, Doublier, Doumayrou, Duband, Exter, Genzel, Gillis, Gr{\"{o}}zinger, Henning, Herreros, Huygen, Inguscio, Jakob, Jamar, Jean, de~Jong, Katterloher, Kiss, Klaas, Lemke, Lutz, Madden, Marquet, Martignac, Mazy, Merken, Montfort, Morbidelli, M{\"{u}}ller, Nielbock, Okumura, Orfei, Ottensamer, Pezzuto, Popesso, Putzeys, Regibo, Reveret, Royer, Sauvage, Schreiber, Stegmaier, Schmitt, Schubert, Sturm, Thiel, Tofani, Vavrek, Wetzstein, Wieprecht, \& Wiezorrek}]{Poglitsch_2010AnA...518L...2P}
Poglitsch, A., Waelkens, C., Geis, N., {et~al.} 2010, A\&A, 518, L2, \dodoi{10.1051/0004-6361/201014535}

\bibitem[{Polychroni {et~al.}(2013)Polychroni, Schisano, Elia, Roy, Molinari, Martin, Andr{\'{e}}, Turrini, Rygl, {Di Francesco}, Benedettini, Busquet, di~Giorgio, Pestalozzi, Pezzuto, Arzoumanian, Bontemps, Hennemann, Hill, K{\"{o}}nyves, Men'shchikov, Motte, Nguyen-Luong, Peretto, Schneider, \& White}]{Polychroni_2013ApJ...777L..33P}
Polychroni, D., Schisano, E., Elia, D., {et~al.} 2013, ApJL, 777, L33, \dodoi{10.1088/2041-8205/777/2/L33}

\bibitem[{Preibisch {et~al.}(1993)Preibisch, Ossenkopf, Yorke, \& Henning}]{Preibisch_1993AnA...279..577P}
Preibisch, T., Ossenkopf, V., Yorke, H., \& Henning, T. 1993, A\&A, 279, 577.
\newblock \url{https://ui.adsabs.harvard.edu/abs/1993A&A...279..577P}

\bibitem[{Priestley \& Whitworth(2020)}]{Priestley-Whitworth_2020_MNRAS.499.3728P}
Priestley, F., \& Whitworth, A. 2020, MNRAS, 499, 3728, \dodoi{10.1093/mnras/staa3111}

\bibitem[{Ragan {et~al.}(2014)Ragan, Henning, Tackenberg, Beuther, Johnston, Kainulainen, \& Linz}]{Ragan_2014AnA...568A..73R}
Ragan, S., Henning, T., Tackenberg, J., {et~al.} 2014, A\&A, 568, A73, \dodoi{10.1051/0004-6361/201423401}

\bibitem[{Rosolowsky {et~al.}(2008)Rosolowsky, Pineda, Kauffmann, \& Goodman}]{Rosolowsky_2008ApJ...679.1338R}
Rosolowsky, E., Pineda, J., Kauffmann, J., \& Goodman, A. 2008, ApJ, 679, 1338, \dodoi{10.1086/587685}

\bibitem[{Schisano {et~al.}(2014)Schisano, Rygl, Molinari, Busquet, Elia, Pestalozzi, Polychroni, Billot, Carey, Paladini, Noriega-Crespo, Moore, Plume, Glover, \& V{\'{a}}zquez-Semadeni}]{Schisano_2014ApJ...791...27S}
Schisano, E., Rygl, K., Molinari, S., {et~al.} 2014, ApJ, 791, 27, \dodoi{10.1088/0004-637X/791/1/27}

\bibitem[{Schisano {et~al.}(2020)Schisano, Molinari, Elia, Benedettini, Olmi, Pezzuto, Traficante, Brescia, Cavuoti, di~Giorgio, Liu, Moore, Noriega-Crespo, Riccio, Baldeschi, Becciani, Peretto, Merello, Vitello, Zavagno, Beltr{\'{a}}n, Cambr{\'{e}}sy, Eden, {Li Causi}, Molinaro, Palmeirim, Sciacca, Testi, Umana, \& Whitworth}]{Schisano_2020MNRAS.492.5420S}
Schisano, E., Molinari, S., Elia, D., {et~al.} 2020, MNRAS, 492, 5420, \dodoi{10.1093/mnras/stz3466}

\bibitem[{Schneider {et~al.}(2010)Schneider, Csengeri, Bontemps, Motte, Simon, Hennebelle, Federrath, \& Klessen}]{Schneider_2010AnA...520A..49S}
Schneider, N., Csengeri, T., Bontemps, S., {et~al.} 2010, A\&A, 520, A49, \dodoi{10.1051/0004-6361/201014481}

\bibitem[{Schneider {et~al.}(2012)Schneider, Csengeri, Hennemann, Motte, Didelon, Federrath, Bontemps, {Di Francesco}, Arzoumanian, Minier, Andr{\'{e}}, Hill, Zavagno, Nguyen-Luong, Attard, Bernard, Elia, Fallscheer, Griffin, Kirk, Klessen, K{\"{o}}nyves, Martin, Men'shchikov, Palmeirim, Peretto, Pestalozzi, Russeil, Sadavoy, Sousbie, Testi, Tremblin, Ward-Thompson, \& White}]{Schneider_2012AnA...540L..11S}
Schneider, N., Csengeri, T., Hennemann, M., {et~al.} 2012, A\&A, 540, L11, \dodoi{10.1051/0004-6361/201118566}

\bibitem[{Schneider \& Elmegreen(1979)}]{Schneider_1979ApJS...41...87S}
Schneider, S., \& Elmegreen, B. 1979, ApJS, 41, 87, \dodoi{10.1086/190609}

\bibitem[{Shan {et~al.}(2012)Shan, Yang, Shi, Yao, Zuo, Lin, Chen, Zhang, Duan, Cao, Li, Li, Liu, \& Zhong}]{Shan_2012ITTST...2..593S}
Shan, W., Yang, J., Shi, S., {et~al.} 2012, IEEE Transactions on Terahertz Science and Technology, 2, 593, \dodoi{10.1109/TTHZ.2012.2213818}

\bibitem[{Shimajiri {et~al.}(2023)Shimajiri, Andr{\'{e}}, Peretto, Arzoumanian, Ntormousi, \& K{\"{o}}nyves}]{Shimajiri_2023AnA...672A.133S}
Shimajiri, Y., Andr{\'{e}}, P., Peretto, N., {et~al.} 2023, A\&A, 672, A133, \dodoi{10.1051/0004-6361/202140857}

\bibitem[{Stil {et~al.}(2006)Stil, Taylor, Dickey, Kavars, Martin, Rothwell, Boothroyd, Lockman, \& McClure-Griffiths}]{Stil_2006AJ....132.1158S}
Stil, J., Taylor, A., Dickey, J., {et~al.} 2006, AJ, 132, 1158, \dodoi{10.1086/505940}

\bibitem[{Stod{\'{o}}lkiewicz(1963)}]{Stodólkiewicz_1963AcA....13...30S}
Stod{\'{o}}lkiewicz, J. 1963, ACTAA, 13, 30.
\newblock \url{https://ui.adsabs.harvard.edu/abs/1963AcA....13...30S}

\bibitem[{Stutz \& Gould(2016)}]{Stutz_2016AnA...590A...2S}
Stutz, A.~M., \& Gould, A. 2016, A\&A, 590, A2, \dodoi{10.1051/0004-6361/201527979}

\bibitem[{Su {et~al.}(2019)Su, Yang, Zhang, Gong, Wang, Zhou, Wang, Chen, Sun, Chen, Xu, \& Jiang}]{Su_2019ApJS..240....9S}
Su, Y., Yang, J., Zhang, S., {et~al.} 2019, ApJS, 240, 9, \dodoi{10.3847/1538-4365/aaf1c8}

\bibitem[{Sun {et~al.}(2020)Sun, Yang, Xu, Zhang, Su, Wang, Chen, Lu, Sun, Ju, Zhang, Zhou, \& Jiang}]{Sun_2020ApJS..246....7S}
Sun, Y., Yang, J., Xu, Y., {et~al.} 2020, ApJS, 246, 7, \dodoi{10.3847/1538-4365/ab5b97}

\bibitem[{Tackenberg {et~al.}(2013)Tackenberg, Beuther, Plume, Henning, Stil, Walmsley, Schuller, \& Schmiedeke}]{Tackenberg_2013AnA...550A.116T}
Tackenberg, J., Beuther, H., Plume, R., {et~al.} 2013, A\&A, 550, A116, \dodoi{10.1051/0004-6361/201220140}

\bibitem[{Tackenberg {et~al.}(2014)Tackenberg, Beuther, Henning, Linz, Sakai, Ragan, Krause, Nielbock, Hennemann, Pitann, \& Schmiedeke}]{Tackenberg_2014AnA...565A.101T}
Tackenberg, J., Beuther, H., Henning, T., {et~al.} 2014, A\&A, 565, A101, \dodoi{10.1051/0004-6361/201321555}

\bibitem[{Traficante {et~al.}(2017)Traficante, Fuller, Billot, Duarte-Cabral, Merello, Molinari, Peretto, \& Schisano}]{Traficante_2017MNRAS.470.3882T}
Traficante, A., Fuller, G., Billot, N., {et~al.} 2017, MNRAS, 470, 3882, \dodoi{10.1093/mnras/stx1375}

\bibitem[{Traficante {et~al.}(2020)Traficante, Fuller, Duarte-Cabral, Elia, Heyer, Molinari, Peretto, \& Schisano}]{Traficante_2020MNRAS.491.4310T}
Traficante, A., Fuller, G., Duarte-Cabral, A., {et~al.} 2020, MNRAS, 491, 4310, \dodoi{10.1093/mnras/stz3344}

\bibitem[{Traficante {et~al.}(2018)Traficante, Fuller, Smith, Billot, Duarte-Cabral, Peretto, Molinari, \& Pineda}]{Traficante_2018MNRAS.473.4975T}
Traficante, A., Fuller, G., Smith, R., {et~al.} 2018, MNRAS, 473, 4975, \dodoi{10.1093/mnras/stx2672}

\bibitem[{Traficante {et~al.}(2023)Traficante, Jones, Avison, Fuller, Benedettini, Elia, Molinari, Peretto, Pezzuto, Pillai, Rygl, Schisano, \& Smith}]{Traficante_2023MNRAS.520.2306T}
Traficante, A., Jones, B., Avison, A., {et~al.} 2023, MNRAS, 520, 2306, \dodoi{10.1093/mnras/stad272}

\bibitem[{Trevi{\~{n}}o-Morales {et~al.}(2019)Trevi{\~{n}}o-Morales, Fuente, S{\'{a}}nchez-Monge, Kainulainen, Didelon, Suri, Schneider, Ballesteros-Paredes, Lee, Hennebelle, Pilleri, Gonz{\'{a}}lez-Garc{\'{i}}a, Kramer, Garc{\'{i}}a-Burillo, Luna, Goicoechea, Tremblin, \& Geen}]{Trevino-Morales_2019AnA...629A..81T}
Trevi{\~{n}}o-Morales, S., Fuente, A., S{\'{a}}nchez-Monge, {\'{A}}., {et~al.} 2019, A\&A, 629, A81, \dodoi{10.1051/0004-6361/201935260}

\bibitem[{Ungerechts \& Thaddeus(1987)}]{Ungerechts_1987ApJS...63..645U}
Ungerechts, H., \& Thaddeus, P. 1987, ApJS, 63, 645, \dodoi{10.1086/191176}

\bibitem[{van~der Tak {et~al.}(2007)van~der Tak, Black, Sch{\"{o}}ier, Jansen, \& van Dishoeck}]{Van_der_Tak_2007AnA...468..627V}
van~der Tak, F., Black, J., Sch{\"{o}}ier, F., Jansen, D., \& van Dishoeck, E. 2007, A\&A, 468, 627, \dodoi{10.1051/0004-6361:20066820}

\bibitem[{Veena {et~al.}(2021)Veena, Schilke, S{\'{a}}nchez-Monge, Sormani, Klessen, Schuller, Colombo, Csengeri, Mattern, \& Urquhart}]{Veena_2021ApJ...921L..42V}
Veena, V., Schilke, P., S{\'{a}}nchez-Monge, {\'{A}}., {et~al.} 2021, ApJL, 921, L42, \dodoi{10.3847/2041-8213/ac341f}

\bibitem[{Wang {et~al.}(2024)Wang, Ge, \& Baug}]{Wang_2024AnA...686L..11W}
Wang, K., Ge, Y., \& Baug, T. 2024, A\&A, 686, L11, \dodoi{10.1051/0004-6361/202450296}

\bibitem[{Wang {et~al.}(2016)Wang, Testi, Burkert, Walmsley, Beuther, \& Henning}]{Wang_2016ApJS..226....9W}
Wang, K., Testi, L., Burkert, A., {et~al.} 2016, ApJS, 226, 9, \dodoi{10.3847/0067-0049/226/1/9}

\bibitem[{Wang {et~al.}(2015)Wang, Testi, Ginsburg, Walmsley, Molinari, \& Schisano}]{Wang_2015MNRAS.450.4043W}
Wang, K., Testi, L., Ginsburg, A., {et~al.} 2015, MNRAS, 450, 4043, \dodoi{10.1093/mnras/stv735}

\bibitem[{Wang {et~al.}(2012)Wang, Zhang, Wu, Li, \& Zhang}]{Wang_2012ApJ...745L..30W}
Wang, K., Zhang, Q., Wu, Y., Li, H.-b., \& Zhang, H. 2012, ApJL, 745, L30, \dodoi{10.1088/2041-8205/745/2/L30}

\bibitem[{Wang {et~al.}(2011)Wang, Zhang, Wu, \& Zhang}]{Wang_2011ApJ...735...64W}
Wang, K., Zhang, Q., Wu, Y., \& Zhang, H. 2011, ApJ, 735, 64, \dodoi{10.1088/0004-637X/735/1/64}

\bibitem[{Wang {et~al.}(2014)Wang, Zhang, Testi, van~der Tak, Wu, Zhang, Pillai, Wyrowski, Carey, Ragan, \& Henning}]{Wang_2014MNRAS.439.3275W}
Wang, K., Zhang, Q., Testi, L., {et~al.} 2014, MNRAS, 439, 3275, \dodoi{Wang_10.1093/mnras/stu127}

\bibitem[{Xie {et~al.}(2021)Xie, Fuller, Li, Chen, Ren, Wu, Duan, Wang, Li, Peretto, Liu, \& Shen}]{Xie_2021SCPMA..6479511X}
Xie, J., Fuller, G.~A., Li, D., {et~al.} 2021, Science China Physics, Mechanics, and Astronomy, 64, 279511, \dodoi{10.1007/s11433-021-1695-0}

\bibitem[{Xu {et~al.}(2018)Xu, Xu, Zhang, Liu, Yu, Ning, \& Ju}]{Xu_2018AnA...609A..43X}
Xu, J.-L., Xu, Y., Zhang, C.-P., {et~al.} 2018, A\&A, 609, A43, \dodoi{10.1051/0004-6361/201629189}

\bibitem[{Yuan {et~al.}(2018)Yuan, Li, Wu, Ellingsen, Henkel, Wang, Liu, Liu, Zavagno, Ren, \& Huang}]{Yuan_2018ApJ...852...12Y}
Yuan, J., Li, J.-Z., Wu, Y., {et~al.} 2018, ApJ, 852, 12, \dodoi{10.3847/1538-4357/aa9d40}

\bibitem[{Zhang {et~al.}(2020)Zhang, Ren, Wu, Li, Zhu, Zhang, Mardones, Wang, Shi, Yue, Luo, Xie, Jiao, Liu, Xu, \& Wang}]{Zhang_2020MNRAS.497..793Z}
Zhang, C., Ren, Z., Wu, J., {et~al.} 2020, MNRAS, 497, 793, \dodoi{10.1093/mnras/staa1958}

\bibitem[{Zhou {et~al.}(2022)Zhou, Liu, Evans, Garay, Goldsmith, G{\'{o}}mez, V{\'{a}}zquez-Semadeni, Liu, Stutz, Wang, Juvela, He, Li, Bronfman, Liu, Xu, Tej, Dewangan, Li, Zhang, Zhang, Ren, Tatematsu, {Shing Li}, {Won Lee}, Baug, Qin, Wu, Peng, Zhang, Liu, Luo, Ge, Saha, Chakali, Zhang, Kim, Ristorcelli, Shen, \& Li}]{Zhou_2022MNRAS.514.6038Z}
Zhou, J.-W., Liu, T., Evans, N.~J., {et~al.} 2022, MNRAS, 514, 6038, \dodoi{10.1093/mnras/stac1735}

\bibitem[{Zucker {et~al.}(2015)Zucker, Battersby, \& Goodman}]{Zucker_2015ApJ...815...23Z}
Zucker, C., Battersby, C., \& Goodman, A. 2015, ApJ, 815, 23, \dodoi{10.1088/0004-637X/815/1/23}

\bibitem[{Zucker {et~al.}(2018)Zucker, Battersby, \& Goodman}]{Zucker_2018ApJ...864..153Z}
---. 2018, ApJ, 864, 153, \dodoi{10.3847/1538-4357/aacc66}

\bibitem[{Zucker \& Chen(2018)}]{Zucker_2018ApJ...864..152Z}
Zucker, C., \& Chen, H. H.-H. 2018, ApJ, 864, 152, \dodoi{10.3847/1538-4357/aad3b5}

\end{thebibliography}
\bibliographystyle{aasjournal}



\end{document}